\documentclass[aps,prd, floatfixpreprintnumbers,superscriptaddress,showpacs,notitlepage,nofootinbib]{revtex4-1}

\usepackage[colorlinks=true,linkcolor=black, citecolor=black,
urlcolor=black]{hyperref}

\usepackage{multirow,graphics}
\usepackage{amsmath}
\usepackage{amsfonts}
\usepackage{amssymb}
\usepackage{slashed}
\usepackage{slashed}
\usepackage{bm}
\usepackage{graphicx}%
\usepackage[T1]{fontenc} 
\usepackage{mathrsfs}
\usepackage{yfonts}

    \newcommand{\beq}{\begin{equation}}
    \newcommand{\eeq}{\end{equation}}
    \newcommand\beqa{\begin{eqnarray}}
    \newcommand\eeqa{\end{eqnarray}}

\usepackage{amstext}
\usepackage{amssymb}
\usepackage{amsmath}
\usepackage{graphicx}
\usepackage{color}
\usepackage{slashed}

\newcommand{\dhd}{{\textstyle d}
\lower.03ex\hbox{\kern-0.38em$^{\scriptstyle-}$}\kern-0.05em{}}
\newcommand{\dbar}{{\textstyle \delta}
\lower.03ex\hbox{\kern-0.38em$^{\scriptstyle-}$}\kern-0.05em{}}
\newcommand{\half}{{1\over 2}}

\newcommand{\bart}{{\bar t}}
\newcommand{\baru}{{\bar u}}
\newcommand{\barv}{{\bar v}}

\newcommand{\bfi}{{\bar \phi}}
\newcommand{\bsi}{{\bar \psi}}
\newcommand{\Bsi}{{\bar \Psi}}

\newcommand{\cala}{{\cal A}}

\newcommand{\cald}{{\cal D}}  
\newcommand{\calf}{{\cal F}}

\newcommand{\calo}{{\cal O}}    
  
\newcommand{\calp}{{\cal P}}

\newcommand{\vro}{{\varrho}}
\newcommand{\ie}{{i\epsilon}}

\newcommand{\notp}{{\not\! p}}

\newcommand{\ve}{{\varepsilon}}
\newcommand{\vfi}{{\varphi}}

\newcommand{\rmT}{{\rm T}}

\newcommand{\tbe}{\tilde b}
\newcommand{\tmu}{\tilde {\mu}}

\newcommand{\Telta}{\tilde {\Delta}}

\newcommand{\scrf}{\mathscr{F}}

\newcommand{\slP}{\slashed{P}}

\begin{document}

\newcommand{\IM}{{\rm Im}\,}
\newcommand{\card}{\#}
\newcommand{\la}[1]{\label{#1}}
\newcommand{\eq}[1]{(\ref{#1})}
\newcommand{\figref}[1]{Fig. \ref{#1}}
\newcommand{\abs}[1]{\left|#1\right|}
\newcommand{\comD}[1]{{\color{red}#1\color{black}}}

\makeatletter
     \@ifundefined{usebibtex}{\newcommand{\ifbibtexelse}[2]{#2}} {\newcommand{\ifbibtexelse}[2]{#1}}
\makeatother

\preprint{JLAB-THY-??-????}

\newcommand{\footnoteab}[2]{\ifbibtexelse{%
\footnotetext{#1}%
\footnotetext{#2}%
\cite{Note1,Note2}%
}{%
\newcommand{\textfootnotea}{#1}%
\newcommand{\textfootnoteab}{#2}%
\cite{thefootnotea,thefootnoteab}}}

\def\e{\epsilon}
     \def\bT{{\bf T}}
    \def\bQ{{\bf Q}}
    \def\wT{{\mathbb{T}}}
    \def\wQ{{\mathbb{Q}}}
    \def\ttQ{{\bar Q}}
    \def\tQ{{\tilde \bP}}
        \def\bP{{\bf P}}
    \def\CF{{\cal F}}
    \def\cC{\CF}
     \def\Tr{\text{Tr}}
     \def\l{\lambda}
\def\hbZ{{\widehat{ Z}}}
\def\bZ{{\resizebox{0.28cm}{0.33cm}{$\hspace{0.03cm}\check {\hspace{-0.03cm}\resizebox{0.14cm}{0.18cm}{$Z$}}$}}}
\newcommand{\rb}{\right)}
\newcommand{\lb}{\left(}

\newcommand{\gT}{T}\newcommand{\gQ}{Q}

\title{Rapidity evolution of TMDs  with running coupling}

\author{Ian Balitsky }\email{balitsky@jlab.org}
\affiliation{
	Physics Dept., ODU, Norfolk VA 23529,  and\\
	Theory Group, Jlab, 12000 Jefferson Ave, Newport News, VA 23606}

\author{
	Giovanni A. Chirilli}\email{giovanni.chirilli@ur.de}
\address{Institut f\"ur Theoretische Physik, Universit\"at Regensburg,\\ D-93040 Regensburg, Germany}

\begin{abstract}
The scale of a coupling constant for rapidity-only evolution of transverse-momentum dependent (TMD)
operators  in the Sudakov kinematic region is
calculated using the Brodsky-Lepage-Mackenzie (BLM) optimal scale setting \cite{Brodsky:1982gc}. The effective argument of a coupling constant
is halfway in the logarithmical scale between the transverse momentum and energy of TMD distribution. 
The resulting rapidity-only evolution equation is solved for quark and gluon TMDs. 
\end{abstract}

 \maketitle

\tableofcontents

\section{Introduction}
The  transverse-momentum dependent parton distributions (TMDs) 
\cite{Collins:1981uw,Collins:1984kg,Ji:2004wu,GarciaEchevarria:2011rb}  have been widely used in the analysis of 
 processes such as semi-inclusive deep inelastic scattering 
or particle production in hadron-hadron collisions (for a review, see Ref. \cite{Collins:2011zzd}).  
The typical kinematics of TMD applications corresponds to the case of Bjorken $x\sim 1$. However,
in recent years there is a surge of interest in a possible extension of TMD formalism to small-$x$
processes. Moreover, the future EIC accelerator will study particle production in the whole region 
of kinematics between moderate $x$ and small $x$. To this end, it is desirable to have an adequate TMD formalism 
that smoothly interpolates between those regions. Unfortunately, the classical  
Collins-Soper-Sterman (CSS)
approach cannot be extended to low $x$ since it was designed to describe the fixed-angle rather than the Regge
limit of large momenta. 

In a series of recent papers \cite{Balitsky:2015qba,Balitsky:2016dgz,Balitsky:2019ayf} the evolution of TMDs was studied by small-$x$ methods. It was demonstrated that
using a small-$x$-inspired rapidity-only cutoff for TMD operators one can obtain an evolution 
equation that smoothly interpolates between the linear case at moderate $x$ and nonlinear evolution at small $x$. 
The obtained rapidity evolution equation correctly reproduces
 three different limits: Dokshitzer-Gribov-Lipatov-Altarelli-Parisi (DGLAP), 
Balitsky-Fadin-Kuraev-Lipatov (BFKL), and Sudakov evolutions, but unfortunately in the intermediate region this 
 equation is very complicated and not very practical. 
Another disadvantage of rapidity-only evolution is that the argument of the coupling constant is not fixed by
the leading-order equation and can be obtained only after next-to-leading order (NLO) calculation.  
This is quite in contrast 
with the usual DGLAP evolution (or CSS one) where the argument of the coupling constant is assigned by 
renormgroup even in the leading order. 
   The rapidity-only evolution of TMDs in the Sudakov region was obtained in Refs. \cite{Balitsky:2015qba,Balitsky:2016dgz}  and studied in Ref. \cite{Balitsky:2019ayf} where it was demonstrated that the leading-order TMD evolution was 
conformally invariant given the proper choice of rapidity cutoff. To use this equation in real QCD one needs to fix
somehow the argument of the coupling constant.
In this paper, the argument of the coupling constant is determined by the BLM approach \cite{Brodsky:1982gc}
(see also \cite{Beneke:1994qe} for higher-order analysis and \cite{Brodsky:1998kn} for small-$x$ application 
similar to what is considered here).
The essence of the BLM approach is to calculate the small part of the NLO result, namely the quark loop contribution to a gluon 
propagator, and promote $-{1\over 6\pi}n_f$ to the full $b_0={11\over 12\pi}N_c-{1\over 6\pi}n_f$. 
This procedure was successfully used for studies of small-$x$ evolution of color dipoles 
where the argument of the coupling constant was fixed using NLO calculation and
renormalon/BLM considerations, (see Refs. \cite{Balitsky:2006wa,Kovchegov:2006vj}).

The paper is organized as follows. Section \ref{rapidityEvolqTMD} is devoted to the leading-order calculation of rapidity evolution of quark TMDs 
and discusses the choice of rapidity-only cutoff. In Section \ref{qloopq} we obtain the quark loop correction to this evolution. 
Section \ref{EvolutionPlusInf} is
about the TMDs with gauge links out to $+\infty$.  We derive the one-loop evolution for gluon TMDs in Section \ref{rapiditygTMD} and 
discuss conclusions in Section \ref{Conclusions}. The necessary technical details and sidelined explanations are presented in the appendices.

\section{Rapidity evolution of quark TMDs \label{rapidityEvolqTMD}}

We will start with the discussion of the evolution of quark TMD operators. For definiteness,  we consider
quark TMDs with gauge links going to $-\infty$  in the ``+'' direction which appear in the description of particle production in hadron-hadron collisions. The typical example is the Drell-Yan (DY) process of production of the $\mu^+\mu^-$ pair in the so-called
Sudakov region where the invariant mass of $\mu^+\mu^-$ pair $Q$ is much greater 
than the sum of their transverse momenta $q_\perp$.
In that region the DY hadronic tensor $W_{\mu\nu}(q)$   can be represented in a standard
 TMD-factorized way \cite{Collins:2011zzd, Collins:2014jpa},
\begin{eqnarray}
&&\hspace{-2mm}
W_{\mu\nu}(q)~
=~\sum_{\rm flavors}e_f^2\!\int\! d^2k_\perp
\cald_{f/A}^{(i)}(x_A,k_\perp)\cald_{f/B}^{(i)}(x_B,q_\perp-k_\perp)
C_{\mu\nu}(q,k_\perp)~+~{\rm power ~corrections}~+~{\rm Y-terms}
\label{TMDfakt}
\end{eqnarray}
where $\cald_{f/A}(x_A,k_\perp)$ is the TMD density of  a quark $f$  in hadron $A$
with fraction of momentum $x_A$ and transverse momentum $k_\perp$, $\cald_{f/B}(x_B,q_\perp-k_\perp)$ 
is a similar quantity for hadron $B$, and  
 coefficient functions $C_i(q,k)$ are determined by the cross section $\sigma(ff\rightarrow \mu^+\mu^-)$  of production of DY pair of invariant mass $q^2$ in the scattering of two quarks. 
 
 The TMD densities $\cald_{f/A}(x_A,k_\perp)$ and  $\cald_{f/B}(x_B,k_\perp)$are 
 defined by quark-antiquark operators with gauge links going to $-\infty$.
For example, the TMD $f_1$ responsible for the total DY cross section for unpolarized hadrons is defined by 
\begin{eqnarray}
&&\hspace{-1mm}
f_1^f(x_B,k_\perp)~=~{1\over 16\pi^3}\!\int\!dz_+ d^2z_\perp~e^{-ix_B z^+\sqrt{s\over 2} +i(k,z)_\perp}
\langle p_N|\bsi_f(z_+,z_\perp)[z,z-\infty n]\slashed{n}\psi_f(0)|p_N\rangle
\label{fdef}
\end{eqnarray}
where $|p_N\rangle$ is an unpolarized nucleon with momentum $p_N\simeq p_N^-$  and $n=({1\over\sqrt{2}},0,0,{1\over\sqrt{2}})$ 
is a lightlike vector in the ``+'' direction (almost) collinear to vector $p_A$. 
Hereafter we use the notation
\begin{equation}
~[x,y]~\equiv~{\rm P}e^{ig\int\! du~(x-y)^\mu A_\mu(ux+(1-u)y)}
\label{defu}
\end{equation}
for a straight-line gauge link connecting points $x$ and $y$. The infinite  lightlike gauge 
links are sometimes called Wilson lines and we will use this terminology. Note also that the operator in 
the right-hand side (RHS) of Eq (\ref{fdef}) is not time ordered.

 In this paper we will study the rapidity-only evolution of the operators 
\begin{equation}
\hspace{-1mm}
\bsi(x^+,x_\perp)[x,x\pm\infty n]~[\pm\infty n+x_\perp,\pm\infty n+y_\perp]\Gamma[\pm\infty n+y,y]\psi(y^+,y_\perp)
\label{qtmdop}
\end{equation}
for quark TMDs, and 
\begin{equation}
\hspace{-1mm}
F^{-i}(x^+,x_\perp)[x,x\pm\infty n][\pm\infty n+x_\perp,\pm\infty n+y_\perp][\pm\infty n+y,y]F^{-j}(y^+,y_\perp)
\label{gtmdop}
\end{equation}
for the gluon ones.
Here $\Gamma$ is one of the matrices $\gamma^-,\gamma^-\gamma_5,\gamma^-\gamma_\perp$ so we single out ``good'' projections
in the light-cone language. 
Note that we do not multiply operators (\ref{qtmdop}) by the square root of the soft factor so, strictly speaking, our  operators (\ref{qtmdop})
enter the ``old version'' of TMD factorization \cite{Collins:1984kg,Collins:2011zzd} such as
\begin{eqnarray}
&&\hspace{-2mm}
W_{\mu\nu}(q)~
=~\sum_{\rm flavors}e_f^2\!\int\! d^2k_\perp ~S(q_\perp,k_\perp)
\tilde\cald_{f/A}^{(i)}(x_A,k_\perp)\tilde\cald_{f/B}^{(i)}(x_B,q_\perp-k_\perp)
C_{\mu\nu}(q,k_\perp)~+~\dots
\label{TMDfaktold}
\end{eqnarray}
where $S(q_\perp,k_\perp)$ is a soft factor. After assigning the square root of a soft factor to each TMD one gets the 
``new'' version \cite{Collins:2003fm} of TMD factorization (\ref{TMDfakt}). 
Since the soft factor is a correlation function of semi-infinite Wilson lines, it will be affected by using rapidity-only
cutoffs. We postpone the calculation of soft (and hard) factors in Eq. (\ref{TMDfakt}) until future publication 
and right now concentrate on the rapidity-only evolution of the operator (\ref{qtmdop}) {\it per se}.

 \subsection{Leading-order evolution of quark TMDs}
 We will start with TMD operators (\ref{qtmdop}) with gauge links going to $-\infty$. 
 It is well known that TMDs (\ref{qtmdop}) exhibit rapidity divergences due to infinitely long gauge links.
 The rapidity-only cutoff corresponds to restricting the $+$ component of gluons emitted by Wilson lines,
\begin{eqnarray}
&&\hspace{-0mm} 
A^\sigma_\mu(x)~=~\int\!{d^4 k\over 16\pi^4} ~\theta\big(\sigma\vro-|k^+|\big)e^{-ik\cdot x} A_\mu(k)
\label{cutoff}
\end{eqnarray}
where we use the notation $\varrho\equiv\sqrt{s\over 2}$.
(Actually, as we will see below, it is more convenient to use smooth cutoff in $|k^+|$ instead of a rigid one imposed by $\theta$-function). 
As mentioned in the Introduction, the goal of this paper is to find the  evolution of the TMD operator (\ref{qtmdop}) 
with respect to the rapidity cutoff $\sigma$
in the ``Sudakov region''  $\sigma x_B s\gg k_\perp^2\sim q_\perp^2$. 

As usual, to find the evolution kernel we need to integrate over gluons  
with $\sigma>k^+/\vro>\sigma'$ and temporarily freeze the fields with $k^+/\vro<\sigma'$. The result will be some kernel
multiplied by TMD operators with rapidity cutoff $\sigma'$. To get the evolution kernel in the leading order,
we need to calculate one-loop diagrams  for the ``matrix element'' of the operator (\ref{qtmdop})
 in the background fields 
\begin{equation}
\hspace{-1mm}
\langle \bsi(x^+,x_\perp)[x^+,-\infty ^+]_x[x_\perp-\infty^+,y_\perp-\infty ^+][-\infty^+,y^+]_y\Gamma\psi(y^+,y_\perp)\rangle_{\Psi,A}
\label{tmdmael}
\end{equation}
where $\Psi$ and $A$ are quarks and gluons with small $k^+<\sigma\vro$.  Hereafter we denote lightlike gauge links by
\begin{equation}
[x^+,y^+]_z~\equiv~[x^+ +z_\perp,y^+ +z_\perp]
\end{equation}
for brevity.
As discussed in Refs. \cite{Balitsky:2015qba,Balitsky:2016dgz,Balitsky:2017flc,Balitsky:2017gis}, in the leading order one can take 
$\Psi$ and $A$  fields
with $k^+=0$ which means background fields $\Psi(x^+,x_\perp)$ and $A(x^+,x_\perp)$. Also, it is convenient to use the
$A^-=0$ gauge for background fields. 

Since the operator in Eq. (\ref{tmdmael}) is not time ordered we need to insert a full set of states at $t=\infty$ 
so the matrix element (\ref{tmdmael}) will be represented as a double functional integral for ``cut diagrams'' in these background fields.
The self-consistency condition is that the background field should be the 
same to the left of the cut and to the right of the cut. Indeed, summation over the full set of intermediate states 
corresponds to the boundary conditions that the fields to the left and to the right of the cut coincide at $t=\infty$. Since 
the background fields do not depend on $x^-$, if they coincide at $x^+=\infty$, they have to  be equal 
everywhere (see the discussion in Refs. \cite{Balitsky:2017flc,Balitsky:2017gis}).
We choose the $A^-=0$ gauge for background gluon fields 
so an extra background gluon line would mean an extra  $F_{\mu\nu}$. 
This gives a higher-twist contribution which we neglect in this paper (see the discussion in Refs.\cite{Balitsky:2017flc,Balitsky:2020jzt}). 
\begin{figure}[htb]
\begin{center}
\includegraphics[width=131mm]{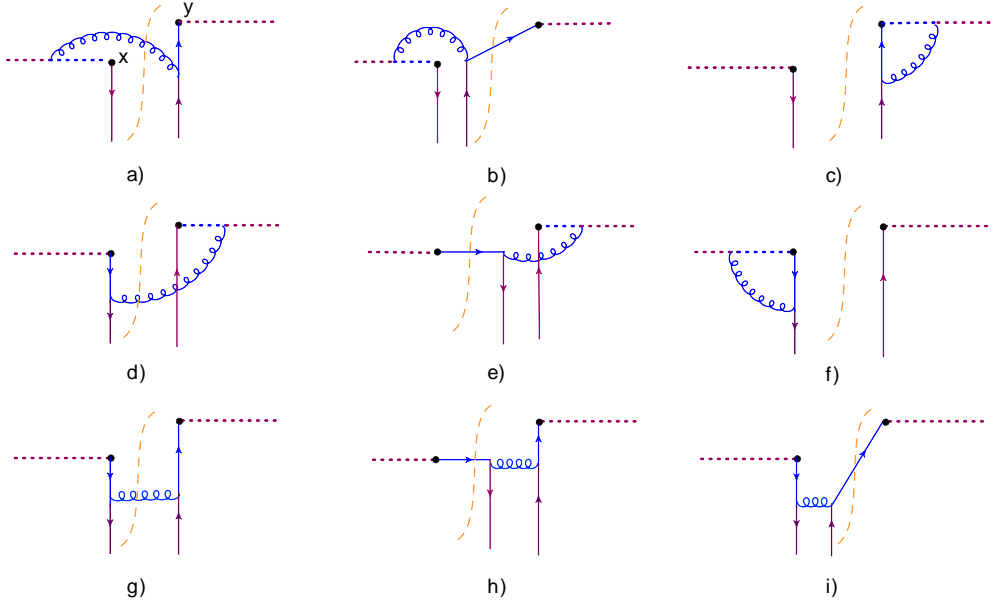}
\end{center}
\caption{One-loop diagrams for TMD operator (\ref{qtmdop}) in the background quark field. The dashed lines denote gauge links.\label{fig:kvdiagrams}}
\end{figure}
For quantum gluons, we use the background-Feynman gauge 
which reduces to the usual Feynman gauge in diagrams without background gluons. It is well-known that in such a 
gauge the contribution of the gauge link at infinity $[x_\perp-\infty n,y_\perp-\infty n]$ 
can be neglected, and we get diagrams shown in Fig. \ref{fig:kvdiagrams}. 

We will use Sudakov variables  
$\alpha\equiv p^+/\vro$ and $\beta\equiv p^-/\vro$ so that $p=\alpha p_1+\beta p_2+p_\perp$ 
where $p_1=n\vro$ and $p_2$ is a lightlike vector close to $p_B$ so that $p_B=x_Bp_B^+ +p_{B_\perp}$. 
In these variables $p\cdot q~=~(\alpha_p\beta_q+\alpha_q\beta_p){s\over 2}-(p,q)_\perp$ where $(p,q)_\perp\equiv -p_iq^i$. 
Throughout the paper, the sum over the Latin indices $i$, $j, \dots$, runs over the two transverse 
components while the sum over Greek indices runs over the four components as usual.

It is convenient to define Fourier transforms of the background fields $\Psi$,
\begin{equation}
\hspace{-0mm}
\Psi(\beta_B,p_{B_\perp})~=~\varrho\!\int\! dz^+ dz_\perp~\Psi(z^+,z_\perp)e^{i\varrho\beta_Bz^+ -i(p_B,z)_\perp},~~~
\Bsi(\beta'_B,p'_{B_\perp})~=~\varrho\!\int\! dz^+ dz_\perp~\Bsi(z^+,z_\perp)e^{i\varrho\beta'_Bz^+ -i(p'_B,z)_\perp}
\label{psis}
\end{equation}
Hereafter we will use the notation $\beta_B\equiv x_B$ since we will calculate integrals using Sudakov variables.

Note that as discussed in Refs. \cite{Balitsky:2015qba,Balitsky:2016dgz,Balitsky:2017flc,Balitsky:2017gis}, in a general gauge one should replace 
\begin{equation}
\hspace{-0mm}
\Psi(z^+,z_\perp)~\rightarrow~[-\infty^+,z^+]_z\Psi(z^+,z_\perp), ~~~~~~\Bsi(z^+,z_\perp)~\rightarrow~\Bsi(z^+,z_\perp)[z^+,-\infty^+]_z
\end{equation}
in the case of evolution equations for the operator (\ref{tmdmael}), and 
\begin{equation}
\hspace{-0mm}
\Psi(z^+,z_\perp)~\rightarrow~[\infty^+,z^+]_z\Psi(z^+,z_\perp), ~~~~~~\Bsi(z^+,z_\perp)~\rightarrow~\Bsi(z^+,z_\perp)[z^+,\infty^+]_z
\end{equation}
for evolution equations of operators (\ref{qtmdop}) with gauge links out to $+\infty$.

\subsection{Diagrams in Figs. \ref{fig:kvdiagrams}(a)-\ref{fig:kvdiagrams}(c) \label{sect:diagsac}}

\subsubsection{ A choice of rapidity cutoff \label{sect:choicecut}}
Let us start with the diagram in Fig. \ref{fig:kvdiagrams}(c) where all propagators are 
of Feynman type.  Note that a possible diagram with Fig.  \ref{fig:kvdiagrams}(c) topology and with a three-gluon 
vertex in the left sector vanishes
since the background field $\Psi(\beta_B,p_{B_\perp})$ cannot produce any real particles. Also, we do not draw 
the diagrams with self-energy insertions in quark tails since they are not relevant for rapidity evolution. 
Simple calculation yields
\footnote{Throughout the paper we distinguish between $\vfi(x)\vfi(y)$, ${\rm T}\{\vfi(x)\vfi(y)\}$, 
and $\tilde{\rm T}\{\vfi(x)\vfi(y)\}\equiv \theta(y_0-x_0)\vfi(x)\vfi(y)+\theta(x_0-y_0)\vfi(y)\vfi(x)$ so
the notation $\langle\vfi(x)\vfi(y)\rangle$ is used only for Wightman-type Green functions.}
\begin{eqnarray}
&&\hspace{-1mm}
\langle{\rm T}\{[-\infty,y^+]_y\Gamma\psi(y^+,y_\perp)\}\rangle_\Psi^{\rm Fig. \ref{fig:kvdiagrams}c}~=~-ig^2c_F\!\int\! \dhd\beta_B\dhd p_{B_\perp}
\nonumber\\
&&\hspace{-1mm}
\times~e^{-ip_By}\!\int\!\dhd\alpha\dhd\beta\dhd p_\perp{1\over \beta+\ie}{\theta(\sigma-|\alpha|)\over \alpha\beta s-p_\perp^2+\ie}
{s(\beta-\beta_B)\over \alpha(\beta-\beta_B)s-(p-p_B)_\perp^2+\ie}\Gamma\Psi(\beta_B,p_{B_\perp})
\nonumber\\
&&\hspace{-1mm}
=~-g^2c_F\!\int\! \dhd\beta'_B\dhd p_{B_\perp}e^{-ip_By}\!\int_0^\sigma\!\dhd\alpha\!\int\!\dhd p_\perp
{\beta_Bs\over p_\perp^2[\alpha\beta_Bs-(p-p_B)_\perp^2+\ie]}\Gamma\Psi(\beta_B,p_{B_\perp})
\label{virtual}
\end{eqnarray}
Hereafter we use space-saving $\hbar$-inspired notations
$\dhd^n p\equiv {d^n p\over(2\pi)^n}$.  
Note that the integral in the RHS of Eq. (\ref{virtual}) diverges as $p_\perp\rightarrow 0$,
but one should expect that this divergence cancels with the contribution of diagrams in Figs. \ref{fig:kvdiagrams}a,b.  

Next we calculate the diagrams in Figs. \ref{fig:kvdiagrams}(a) and \ref{fig:kvdiagrams}(b) with a combination
of Feynman, complex conjugate, and cut propagators. One obtains
\begin{eqnarray}
&&\hspace{-1mm}
\langle[x^+,-\infty]_x\Gamma\psi(y^+,y_\perp)\rangle_\Psi^{\rm Fig. \ref{fig:kvdiagrams}a,b}
\nonumber\\
&&\hspace{-1mm}
=~g^2c_F\!\int\! \dhd\beta'_B\dhd p_{B_\perp}e^{-i(p_B,y)}\!\int\!\dhd\alpha\dhd\beta\dhd p_\perp\Big[2\pi\delta(\alpha(\beta-\beta_B)s-(p-p_B)_\perp^2)
(\beta-\beta_B)s\theta(\alpha) 
{1\over \alpha\beta s-p_\perp^2-\ie}
\nonumber\\
&&\hspace{22mm}
+~{(\beta-\beta_B)s\over \alpha(\beta-\beta_B)s-(p-p_B)_\perp^2+\ie}
2\pi\delta( \alpha\beta s-p_\perp^2)\theta(\alpha)\Big]{\theta(\sigma-|\alpha|)\over \beta+\ie}e^{-i\beta\vro\Delta^++i(p,\Delta)_\perp}\Gamma\Psi(\beta_B,p_{B_\perp})
\nonumber\\
&&\hspace{-1mm}
=~g^2c_F\!\int\!\dhd\beta_B\dhd p_{B_\perp}\psi(\beta_B,p_{B_\perp}) e^{-ip_By}
\!\int_0^\sigma\!{\dhd\alpha\over\alpha}\!\int\!\dhd p_\perp e^{i(p,\Delta)_\perp}
\Big[{(\alpha\beta_Bs-p_\perp^2)e^{-i{p_\perp^2\over\alpha s}\vro\Delta^+}
\over p_\perp^2[\alpha\beta_Bs+(p-p_B)_\perp^2-p_\perp^2-\ie]}
\nonumber\\
&&\hspace{22mm}
+~{(p-p_B)_\perp^2e^{-i(\beta_B+{(p-p_B)_\perp^2\over\alpha s})\vro\Delta^+}
\over [\alpha\beta_B s+(p-p_B)_\perp^2+\ie][\alpha\beta_B s+(p-p_B)_\perp^2-p_\perp^2-\ie]}\Big]\Gamma\Psi(\beta_B,p_{B_\perp})
\label{reals}
\end{eqnarray}
Hereafter we use the notation $\Delta\equiv x-y$ for brevity. The dimension of the transverse space is $d-2=2+2\ve$ (or
$d=2$ if we do not need dimensional regularization).

It is convenient to rewrite Eq. (\ref{reals}) as a sum of two terms
\begin{eqnarray}
&&\hspace{-1mm}
\langle[x^+,-\infty]_x\Gamma\psi(y^+,y_\perp)\rangle_{\Psi}~=~
\langle[x^+,-\infty]_x\Gamma\psi(y^+,y_\perp)\rangle_{\Psi}^{(1)}+\langle[x^+,-\infty]_x\Gamma\psi(y^+,y_\perp)\rangle_{\Psi}^{(2)} 
\end{eqnarray}
where 
\begin{eqnarray}
&&\hspace{-0mm}
\langle[x^+,-\infty]_x\Gamma\psi(y^+,y_\perp)\rangle_{\Psi}^{(1)}
~
\nonumber\\
&&\hspace{22mm}
=~g^2c_F\!\int\!\dhd\beta_B\dhd p_{B_\perp}\Gamma\Psi(\beta_B,p_{B_\perp}) e^{-ip_By}
\!\int_0^\sigma\!\dhd\alpha\!\int\!{\dhd p_\perp \over p_\perp^2}
{\beta_Bse^{-i{p_\perp^2\over\alpha s}\vro\Delta^+ +i(p,\Delta)}
\over \alpha\beta_Bs+(p-p_B)_\perp^2+\ie},
\label{realpart1}\\
&&\hspace{0mm}
\langle[x^+,-\infty]_x\Gamma\psi(y^+,y_\perp)\rangle_{\Psi}^{(2)}~=~g^2c_F\!\int\!\dhd\beta_B\dhd p_{B_\perp}
\Gamma\Psi(\beta_B,p_{B_\perp})e^{-ip_By}
\nonumber\\
&&\hspace{22mm}
\times~\!\int_0^\sigma\!{\dhd\alpha\over\alpha}\!\int\!\dhd p_\perp
{(p-p_B)_\perp^2e^{i(p,\Delta)_\perp}\big[e^{-i(\beta_B+{(p-p_B)_\perp^2\over\alpha s})\vro\Delta^+}-e^{-i{p_\perp^2\over\alpha s}\vro\Delta^+}\big]
\over [\alpha\beta_B s+(p-p_B)_\perp^2+\ie][\alpha\beta_B s+(p-p_B)_\perp^2-p_\perp^2]}
\label{realpart2}
\end{eqnarray}
The integral in the RHS of Eq. (\ref{realpart2}) is convergent while the one in the RHS of Eq. (\ref{realpart1}) diverges as $p_\perp\rightarrow 0$.
As we mentioned above, one should expect that this divergence cancels with the contribution (\ref{virtual}) of diagram in Figs. \ref{fig:kvdiagrams}(c). 
Indeed, this divergence comes from the infinite length of gauge links in Eqs. (\ref{psis}).
As $p_\perp\rightarrow 0$ the integral (\ref{reals}) behaves in the same way as such an integral at $x_\perp=y_\perp$ so the contributions of infinite gauge links should cancel 
\begin{equation}
\langle[x^+,-\infty^+]_y[-\infty^+,y^+]_y\Gamma\psi(y^+,y_\perp)\rangle_\Psi~=~\langle[x^+,y^+]_y\Gamma\psi(y^+,y_\perp)\rangle_\Psi
\end{equation}

Unfortunately, ``rigid'' cutoff $\sigma>|\alpha|$ does not provide this property--the sum of Eqs. (\ref{virtual}) and (\ref{realpart1}) 
is still divergent as $p_\perp\rightarrow 0$. To ensure IR cancellations, we use a ``smooth'' cutoff in $\alpha$ imposed by point-splitting 
regularization $\psi(y^+,y_\perp)\rightarrow \psi(y^+,y_\perp,y^-)$. We get then
\begin{eqnarray}
&&\hspace{-1mm}
\langle{\rm T}\{[-\infty,y^+]_y\Gamma\psi(y^+,y_\perp,-\delta^-)\}\rangle_{\Psi}^{\rm Fig. \ref{fig:kvdiagrams}c}
~=~-ig^2c_F\!\int\! \dhd\beta_B\dhd p_{B_\perp}
\nonumber\\
&&\hspace{-1mm}
\times~e^{-ip_By}\!\int\!\dhd\alpha\dhd\beta\dhd p_\perp{1\over \beta+\ie}{e^{-i\alpha\vro\delta^-}\over \alpha\beta s-p_\perp^2+\ie}
{s(\beta-\beta_B)\over \alpha(\beta-\beta_B)s-(p-p_B)_\perp^2+\ie}\Gamma\Psi(\beta_B,p_{B_\perp})
\nonumber\\
&&\hspace{-1mm}
=~g^2c_F\!\int\! \dhd\beta_B\dhd p_{B_\perp}e^{-ip_By}\!\int_{-\infty}^0\!\dhd\alpha\!\int\!\dhd p_\perp
{\beta_Bs\over p_\perp^2[\alpha\beta_Bs+(p-p_B)_\perp^2-\ie]}\Gamma\Psi(\beta_B,p_{B_\perp})e^{-i{\alpha\over\sigma}}
\nonumber\\
&&\hspace{-1mm}
=~-g^2c_F\!\int\! \dhd\beta_B\dhd p_{B_\perp}e^{-ip_By}\Gamma\Psi(\beta_B,p_{B_\perp})\!\int_0^{\infty}\!\dhd\alpha\!\int\!\dhd p_\perp
{\beta_Bs\over p_\perp^2[\alpha\beta_Bs+(p-p_B)_\perp^2+\ie]}e^{-i{\alpha\over\sigma}}
\label{virt}
\end{eqnarray}
where 
\begin{equation}
\sigma\equiv {1\over \vro\delta^-}>0.
\label{sigma}
\end{equation}
Note that to get the last line in Eq. (\ref{virt}), we turned the contour of integration over $\alpha$ 
on angle $\pi$ in the lower half-plane of complex $\alpha$.
At $\beta_B>0$ the  singularity at $\alpha={(p-p_B)_\perp^2\over\beta_Bs}+\ie$ does not affect the rotation, while at $\beta_B<0$
the rotation pushes the singularity over $\alpha$ up to $+\ie$.

For the diagrams in Figs. \ref{fig:kvdiagrams}(a) and \ref{fig:kvdiagrams}(b) with point splitting one obtains
\begin{eqnarray}
&&\hspace{-0mm}
\langle[x^+,-\infty]_x\Gamma\psi(y^+,y_\perp,-\delta^-)\rangle_{\Psi}^{(1)}
\nonumber\\
&&\hspace{22mm}
=~g^2c_F\!\int\!\dhd\beta_B\dhd p_{B_\perp}e^{-ip_By}\Gamma\Psi(\beta_B,p_{B_\perp})
\!\int_0^\infty\!\dhd\alpha\!\int\!{\dhd p_\perp \over p_\perp^2}
{\beta_Bse^{-i{p_\perp^2\over\alpha s}\vro\Delta^+ +i(p,\Delta)}
\over \alpha\beta_Bs+(p-p_B)_\perp^2+\ie}e^{-i{\alpha\over\sigma}},
\label{repart1}\\
&&\hspace{0mm}
\langle[x^+,-\infty]_x\Gamma\psi(y^+,y_\perp,-\delta^-)\rangle_\Psi^{(2)}~=~g^2c_F\!\int\!\dhd\beta_B\dhd p_{B_\perp}\Gamma\Psi(\beta_B,p_{B_\perp})
 e^{-ip_By}
\nonumber\\
&&\hspace{22mm}
\times~\!\int_0^\infty\!{\dhd\alpha\over\alpha}\!\int\!\dhd p_\perp
{(p-p_B)_\perp^2e^{i(p,\Delta)_\perp}\big[e^{-i(\beta_B+{(p-p_B)_\perp^2\over\alpha s})\vro\Delta^+}-e^{-i{p_\perp^2\over\alpha s}\vro\Delta^+}\big]
\over [\alpha\beta_B s+(p-p_B)_\perp^2+\ie][\alpha\beta_B s+(p-p_B)_\perp^2-p_\perp^2]}e^{-i{\alpha\over\sigma}}
\label{repart2}
\end{eqnarray}
Now we see that the sum of Eqs. (\ref{virt}), (\ref{repart1}), and (\ref{repart2}) 
\begin{eqnarray}
&&\hspace{-0mm}
\langle[x^+,-\infty]_x[-\infty,y^+]_y\Gamma\psi(y^+,y_\perp,-\delta^-)\rangle_{\Psi}^{\rm Fig. \ref{fig:kvdiagrams}a-c}
\nonumber\\
&&\hspace{22mm}
=~g^2c_F\!\int\!\dhd\beta_B\dhd p_{B_\perp}e^{-ip_By}\Gamma\Psi(\beta_B,p_{B_\perp})
\!\int_0^\infty\!\dhd\alpha~e^{-i{\alpha\over\sigma}}\!\int\!\dhd p_\perp ~\bigg(
{\beta_Bs\big(e^{-i{p_\perp^2\over\alpha s}\vro\Delta^+ +i(p,\Delta)_\perp}-1\big)
\over  p_\perp^2[\alpha\beta_Bs+(p-p_B)_\perp^2+\ie]}
\nonumber\\
&&\hspace{44mm}
+~{(p-p_B)_\perp^2e^{i(p,\Delta)_\perp}\big[e^{-i(\beta_B+{(p-p_B)_\perp^2\over\alpha s})\vro\Delta^+}-e^{-i{p_\perp^2\over\alpha s}\vro\Delta^+}\big]
\over \alpha[\alpha\beta_B s+(p-p_B)_\perp^2+\ie][\alpha\beta_B s+(p-p_B)_\perp^2-p_\perp^2]}\bigg)
\label{napravo}
\end{eqnarray}
is given by a convergent integral.  It should be emphasized that $\delta^-<0$;
otherwise, we would not be able to make the rotation of the contour in the last line in 
Eq. (\ref{virt}) and the cancellation of IR divergences would not happen. 
The reason for that is that all quantum operators in $[-\infty,y^+]_y\psi(y^+,y_\perp)$ 
commute since they are on the light ray, and to preserve this commutation property 
[which is necessary for using Feynman propagators in Eq. (\ref{virt})] we should shift 
$\psi(y^+,y_\perp)$ to a point separated by a spacelike distance from operators in the gauge link 
$[-\infty,y^+]_y$ (see the discussion in Appendix \ref{app:cutoff}).

\begin{figure}[htb]
\begin{center}
\includegraphics[width=131mm]{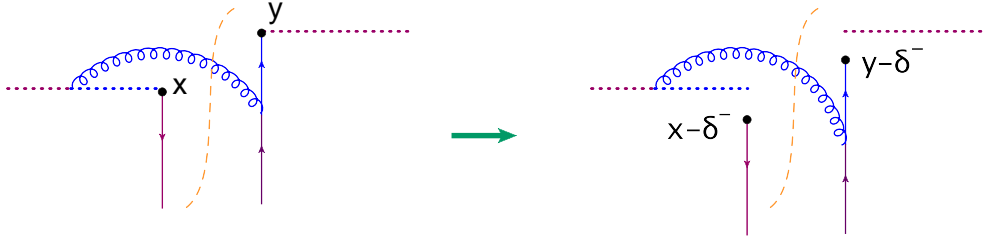}
\end{center}
\caption{Point-splitting regularization of rapidity divergence. \label{fig:split}}
\end{figure}

It should be emphasized that we do not suggest the nonperturbative studies of TMDs with our ``point splitting'' and the reason is that  
objects such as
\begin{equation}     
\langle p_N|\bsi\big(x^+,x_\perp,-{1\over\vro\sigma'}\big)[x^+,-\infty]_x[-\infty,y^+]_y\Gamma\psi\big(y^+,y_\perp,-{1\over\vro\sigma}\big)
|p_N\rangle
\label{psmael}
\end{equation}
are meaningless since the operator is not gauge invariant. Our message is that the longitudinal integrals in the perturbative diagrams for TMDs 
\begin{equation}     
\langle p_N|\bsi\big(x^+,x_\perp\big)[x^+,-\infty]_x[-\infty,y^+]_y\Gamma\psi(y^+,y_\perp)
|p_N\rangle
\label{goodmael}
\end{equation}
should be cut from above in a smooth way respecting unitarity and causality, and a mnemonic rule how to choose the proper
sign of the cutoff $e^{\pm i{\alpha/\sigma}}$ is to consider the point-splitting (\ref{psmael}). Also, the point splitting representation 
of rapidity cutoff in $\alpha$ helps to visualize  the coordinate space approximations that we make (see Fig.
\ref{fig:split}).
Thus, we will be using expressions such as (\ref{psmael})  and defining rapidity-regularized operators by
\begin{eqnarray}
&&\hspace{-0mm}
\bsi^\sigma(x^+,x_\perp) ~\equiv~\bsi\big(x^+,x_\perp,-{1\over\rho\sigma}\big)[x^+,-\infty]_x,~~~~
\psi^\sigma(y^+,y_\perp) ~\equiv~[-\infty,y^+]_y\psi\big(y^+,y_\perp,-{1\over\rho\sigma}\big),
\label{Psidef}
\end{eqnarray}
but only for the perturbative calculations.  \footnote{
Alternatively, one may use the classical 
cutoff with off-light-cone gauge links, but from experience with rapidity evolution of color dipoles we know that 
using off-light-cone gauge links  enormously complicates the NLO calculations 
(see  Refs. \cite{Balitsky:2007feb,Balitsky:2009xg} and  especially Appendix B to Ref. \cite{Balitsky:2007feb}).
}
In this paper we perform calculations in the (background) Feynman gauge, but  
in the Appendix \ref{app:lorentz} we demonstrate that for other gauges like Landau gauge the extra terms in gluon propagator 
lead to power corrections $\sim {q_\perp^2\over\beta_B\sigma_t s}\ll 1$. 

It should be mentioned that the standard regularization of TMD operator (\ref{goodmael}) at moderate $x$ 
is a combination of UV and rapidity cutoffs, (see Ref. \cite{Collins:2011zzd}).  We discuss the relation 
of that regularization to our rapidity-only cutoff in Appendix \ref{sec:compare}.

\subsubsection{Rapidity evolution of diagrams in Figs. \ref{fig:kvdiagrams}(a)-\ref{fig:kvdiagrams}(c)}

In this section we will calculate the $\sigma$ dependence of the integral (\ref{napravo})
\begin{equation}
\hspace{-0mm}
\sigma{d\over d\sigma}\!\int_0^\infty\!\dhd\alpha~e^{-i{\alpha\over\sigma}}\!\int\!\dhd p_\perp \bigg(
{\beta_Bs\big(e^{-i{p_\perp^2\over\alpha s}\vro\Delta^+ +i(p,\Delta)_\perp}-1\big)
\over  p_\perp^2[\alpha\beta_Bs+(p-p_B)_\perp^2+\ie]}
+~{(p-p_B)_\perp^2e^{i(p,\Delta)_\perp}\big[e^{-i(\beta_B+{(p-p_B)_\perp^2\over\alpha s})\vro\Delta^+}-e^{-i{p_\perp^2\over\alpha s}\vro\Delta^+}\big]
\over \alpha[\alpha\beta_B s+(p-p_B)_\perp^2+\ie][\alpha\beta_B s+(p-p_B)_\perp^2-p_\perp^2]}\bigg)\,.
\label{napravo1}
\end{equation}
First, note that the second term in the RHS does not contribute to the evolution. Indeed, characteristic $\alpha$'s in that term are $\sim {m_\perp^2\over\beta_Bs}$ 
where $m_\perp^2\sim \Delta_\perp^{-2}\sim  p_{B_\perp}^2$ so we can expand $e^{-i{\alpha\over\sigma}}$ and get approximately
\begin{eqnarray}
&&\hspace{-0mm}
\!\int_0^\infty\!\dhd\alpha~\!\int\!\dhd p_\perp ~\Big[1-i{\alpha\over\sigma}\theta(\sigma-\alpha)\Big]
{(p-p_B)_\perp^2e^{i(p,\Delta)_\perp}\big[e^{-i(\beta_B+{(p-p_B)_\perp^2\over\alpha s})\vro\Delta^+}-e^{-i{p_\perp^2\over\alpha s}\vro\Delta^+}\big]
\over \alpha[\alpha\beta_B s+(p-p_B)_\perp^2+\ie][\alpha\beta_B s+(p-p_B)_\perp^2-p_\perp^2]}\,.
\label{napravomalo1}
\end{eqnarray}
The first term is a convergent integral independent of $\sigma$ while the second is of order of ${m_\perp^2\over \sigma\beta_Bs}\ln{\sigma\beta_Bs\over m_\perp^2}$ 
so it is a power correction that we neglect.

Next, we study the dependence of the first term in the RHS of Eq. (\ref{napravo1}) on $\Delta^+$. From TMD factorization (\ref{TMDfakt}) and definition (\ref{fdef}) we see that we need operator 
(\ref{qtmdop}) in the region $\Delta^+\vro\beta_B\sim 1$. Let us demonstrate that in this region one can neglect $\Delta^+$. Indeed, 
\begin{eqnarray}
&&\hspace{-0mm}
\!\int\!\dhd p_\perp ~e^{i(p,\Delta)_\perp}
\bigg[\!\int_0^{{p_\perp^2\over s}\vro|\Delta^+|}\! d\alpha+\!\int_{{p_\perp^2\over s}\vro|\Delta^+|}^\infty\! d\alpha\bigg]e^{-i{\alpha\over\sigma}}
{\beta_Bs\big(e^{-i{p_\perp^2\over\alpha s}\vro\Delta^+ }-1\big)
\over  p_\perp^2[\alpha\beta_Bs+(p-p_B)_\perp^2+\ie]}
\label{napravomalo2}\\
&&\hspace{-0mm}
\simeq~\!\int\!\dhd p_\perp e^{i(p,\Delta)_\perp}\bigg[\!\int_0^{{p_\perp^2\over s}\vro|\Delta^+|}\! d\alpha {\beta_Bs\big(e^{-i{p_\perp^2\over\alpha s}\vro\Delta^+ }-1\big)
\over  p_\perp^2[\alpha\beta_Bs+(p-p_B)_\perp^2+\ie]}
-\!\int_{{p_\perp^2\over s}\vro|\Delta^+|}^\infty\! {d\alpha\over\alpha}{i\beta_B\rho\Delta^+\over  \alpha\beta_Bs+(p-p_B)_\perp^2+\ie}\bigg]
~+~O\big({m_\perp^2\over\sigma\beta_Bs}\big)
\nonumber
\end{eqnarray}
which is a sum of terms independent of $\sigma$ and power corrections.

Thus, we  need to consider only
\begin{eqnarray}
&&\hspace{-0mm}
\!\int_0^\infty\!\dhd\alpha\!\int\!{\dhd p_\perp\over  p_\perp^2}
{\beta_Bse^{-i{\alpha\over\sigma}}\big(e^{i(p,\Delta)_\perp}-1\big)
\over \alpha\beta_Bs+(p-p_B)_\perp^2+\ie}
=~\!\int_0^\infty\!\dhd\alpha~e^{-i{\alpha\over\sigma}}\!\int\!{\dhd p_\perp\over  p_\perp^2}
{\beta_Bs\big(e^{i(p,\Delta)_\perp}-1\big)\over  \alpha\beta_Bs+p_\perp^2+\ie}
\Big[1-{p_\perp^2-(p-p_B)_\perp^2\over \alpha\beta_Bs+(p-p_B)_\perp^2+\ie}\Big]
\nonumber\\
&&\hspace{-0mm}
=~\!\int_0^\infty\!\dhd\alpha~e^{-i{\alpha\over\sigma}}\!\int\!{\dhd p_\perp\over  p_\perp^2}
{\beta_Bs\big(e^{i(p,\Delta)_\perp}-1\big)\over  \alpha\beta_Bs+p_\perp^2+\ie}
~+~\!\int\!{\dhd p_\perp\over  p_\perp^2}\big(e^{i(p,\Delta)_\perp}-1\big)\ln{p_\perp^2\over(p-p_B)_\perp^2}
~+~O\big({m_\perp^2\over\sigma\beta_Bs}\Big)
\label{napravo2}
\end{eqnarray}
where we neglected the $e^{-i{\alpha\over\sigma}}$ cutoff in the second integral since it converges at $\alpha\sim{m_\perp^2\over \sigma|\beta_B|s}$. 

The $\sigma$ dependence comes only from the first term in the RHS of Eq. (\ref{napravo2}) 
calculated in Appendix \ref{app:integrals} [see Eq. (\ref{eiknapravotvet})] so from Eq. (\ref{evolint}) we get 
\begin{eqnarray}
&&\hspace{-0mm}
\sigma{d\over d\sigma}\langle[x^+,-\infty]_x[-\infty,y^+]_y\Gamma\psi(y^+,y_\perp,-{1\over\vro\sigma})\rangle_{\Psi}
\nonumber\\
&&\hspace{2mm}
=~-{g^2\over 8\pi^2}c_F\!\int\!\dhd\beta_B\dhd p_{B_\perp}\Gamma\Psi(\beta_B,y_\perp) e^{-ip_By}\ln\Big(-{i\over 4}(\beta_B+\ie)\sigma s\Delta_\perp^2e^\gamma\Big)
~+~O\big({m_\perp^2\over\beta_B\sigma s}\big)
\nonumber\\
&&\hspace{2mm}
=~-{g^2\over 8\pi^2}c_F\!\int\!\dhd\beta_B\Gamma\Psi(\beta_B,y_\perp) e^{-i\beta_B\vro y^+}\ln\Big(-{i\over 4}(\beta_B+\ie)\sigma s\Delta_\perp^2e^\gamma\Big)
~+~O\big({m_\perp^2\over\beta_B\sigma s}\big)
\label{napravotvet}
\end{eqnarray}
where $\gamma\simeq 0.577$ is the Euler constant and $\Psi(\beta_B,y_\perp)=\vro\int\! dy^+ e^{i\beta_B\vro y^+}\Psi(y^+,y_\perp)$, see Eq. (\ref{psis}).

\subsection{Diagrams in Figs. \ref{fig:kvdiagrams}(d)-\ref{fig:kvdiagrams}(i)}

The calculation of diagrams in Fig. \ref{fig:kvdiagrams}(d)-\ref{fig:kvdiagrams}(f)
repeats that of Fig. \ref{fig:kvdiagrams}(a)-\ref{fig:kvdiagrams}(c) with minimal changes.
Let us start with the diagram shown in Fig. \ref{fig:kvdiagrams}f
\begin{eqnarray}
&&\hspace{-1mm}
\langle\tilde{\rm T}\{\bsi(x^+,x_\perp,-\delta^-)\Gamma[x^+,-\infty]_x\}\rangle_{\Bsi}^{\rm Fig. \ref{fig:kvdiagrams}f}
~=~ig^2c_F\!\int\! \dhd\beta_B\dhd p_{B_\perp}\Bsi(\beta_B,p_{B_\perp})\Gamma
\nonumber\\
&&\hspace{-1mm}
\times~e^{-ip_Bx}\!\int\!\dhd\alpha\dhd\beta\dhd p_\perp{1\over \beta-\ie}{e^{-i\alpha\vro\delta^-}\over \alpha\beta s-p_\perp^2-\ie}
{s(\beta+\beta_B)\over \alpha(\beta+\beta_B)s-(p+p_B)_\perp^2-\ie}
\nonumber\\
&&\hspace{-1mm}
=~g^2c_F\!\int\! \dhd\beta_B\dhd p_{B_\perp}e^{-ip_By}
\Bsi(\beta_B,p_{B_\perp})\Gamma\!\int_{-\infty}^0\!\dhd\alpha\!\int\!\dhd p_\perp
{\beta_Bs\over p_\perp^2[\alpha\beta_Bs-(p+p_B)_\perp^2-\ie]}e^{ - i{\alpha\over\sigma}}
\nonumber\\
&&\hspace{-1mm}
=~-g^2c_F\!\int\! \dhd\beta_B\dhd p_{B_\perp}e^{-ip_By}\Bsi(\beta_B,p_{B_\perp})\Gamma\!\int_0^{\infty}\!\dhd\alpha\!\int\!\dhd p_\perp
{\beta_Bs\over p_\perp^2[\alpha\beta_Bs-(p+p_B)_\perp^2+\ie]}e^{i{\alpha\over\sigma}}
\label{virtnalevo}
\end{eqnarray}
Similar to Eq. (\ref{virt}),  to get the last line we rotated the contour of integration over $\alpha$ in the upper half-plane of complex $\alpha$. Next, the contribution of diagrams in  Figs. \ref{fig:kvdiagrams}(d) and \ref{fig:kvdiagrams}(e) is
\begin{eqnarray}
&&\hspace{-1mm}
\langle\bsi(x^+,x_\perp,-\delta^-)[-\infty,y^+]_y\Gamma\rangle_{\Psi}^{\rm Fig. \ref{fig:kvdiagrams}d,e}~
\nonumber\\
&&\hspace{-1mm}
=~g^2c_F\!\int\! \dhd\beta'_B\dhd p_{B_\perp}e^{-i(p_B,x)}\Bsi(\beta_B,p_{B_\perp})\Gamma\!\int\!\dhd\alpha\dhd\beta\dhd p_\perp\Big[{1\over \alpha\beta s-p_\perp^2+\ie}
2\pi\delta(\alpha(\beta+\beta_B)s-(p+p_B)_\perp^2)
(\beta+\beta_B)s\theta(\alpha) 
\nonumber\\
&&\hspace{22mm}
+~{(\beta+\beta_B)s\over \alpha(\beta+\beta_B)s-(p+p_B)_\perp^2-\ie}
2\pi\delta( \alpha\beta s-p_\perp^2)\theta(\alpha)\Big]{e^{i\alpha\vro\delta^-}\over \beta-\ie}e^{-i\beta\Delta+i(p,\Delta)_\perp}
\nonumber\\
&&\hspace{-1mm}
=~g^2c_F\!\int\!\dhd\beta_B\dhd p_{B_\perp}\e^{-ip_Bx}\Gamma\Bsi(\beta_B,p_{B_\perp})
\!\int_0^\infty\!{\dhd\alpha\over\alpha}e^{i{\alpha\over\sigma}}\!\int\!\dhd p_\perp e^{i(p,\Delta)_\perp}
\bigg[{(\alpha\beta_Bs+p_\perp^2)
e^{-i{p_\perp^2\over\alpha s}\vro\Delta^+}
\over p_\perp^2[\alpha\beta_Bs-(p+p_B)_\perp^2+\ie]}
\nonumber\\
&&\hspace{22mm}
+~{(p+p_B)_\perp^2e^{i(\beta_B-{(p+p_B)_\perp^2\over\alpha s})\vro\Delta^+}
\over [\alpha\beta_B s-(p+p_B)_\perp^2+\ie]
[\alpha\beta_B s-(p+p_B)_\perp^2+p_\perp^2]}\bigg]
\label{realnalevo}
\end{eqnarray}
The sum of Eqs. (\ref{virtnalevo}) and (\ref{realnalevo}) can be rewritten as
\begin{eqnarray}
&&\hspace{-1mm}
\langle\bsi(x^+,x_\perp,-\delta^-)[-\infty,y^+]_y\Gamma\rangle_{\Psi}^{\rm Fig. \ref{fig:kvdiagrams}d-f}~
\nonumber\\
&&\hspace{-1mm}
=~g^2c_F\!\int\!\dhd\beta_B\dhd p_{B_\perp}\e^{-ip_Bx}\Gamma\Bsi(\beta_B,p_{B_\perp})
\!\int_0^\infty\!\dhd\alpha~e^{i{\alpha\over\sigma}}\!\int\!\dhd p_\perp e^{i(p,\Delta)_\perp}
\bigg[{\beta_Bs
\big(e^{-i{p_\perp^2\over\alpha s}\vro\Delta^+}-1\big)
\over p_\perp^2[\alpha\beta_Bs-(p+p_B)_\perp^2+\ie]}
\nonumber\\
&&\hspace{22mm}
+~{(p+p_B)_\perp^2\big[e^{i(\beta_B-{(p+p_B)_\perp^2\over\alpha s})\vro\Delta^+}
-e^{-i{p_\perp^2\over\alpha s}\vro\Delta^+}\big]
\over \alpha[\alpha\beta_B s-(p+p_B)_\perp^2+\ie]
[\alpha\beta_B s-(p+p_B)_\perp^2+p_\perp^2]}\bigg]
\label{nalevo}
\end{eqnarray}
Now, the integral
\begin{eqnarray}
&&\hspace{-1mm}
\!\int_0^\infty\!\dhd\alpha~e^{i{\alpha\over\sigma}}\!\int\!\dhd p_\perp e^{i(p,\Delta)_\perp}
\bigg[{\beta_Bs
\big(e^{-i{p_\perp^2\over\alpha s}\vro\Delta^+}-1\big)
\over p_\perp^2[\alpha\beta_Bs-(p+p_B)_\perp^2+\ie]}
+{(p+p_B)_\perp^2\big[e^{i(\beta_B-{(p+p_B)_\perp^2\over\alpha s})\vro\Delta^+}
-e^{-i{p_\perp^2\over\alpha s}\vro\Delta^+}\big]
\over \alpha[\alpha\beta_B s-(p+p_B)_\perp^2+\ie]
[\alpha\beta_B s-(p+p_B)_\perp^2+p_\perp^2]}\bigg]
\label{intnalevo}
\end{eqnarray}
differs from the corresponding integral in Eq. (\ref{napravo}) by complex conjugation and replacements $x\leftrightarrow y$,
$p_B\leftrightarrow -p_B$ so we get the result obtained from Eq. (\ref{napravotvet}) by the same manipulations
\begin{eqnarray}
&&\hspace{-0mm}
\sigma'{d\over d\sigma'}\langle\bsi(x^+,x_\perp,-\delta'^-)
\Gamma[x^+,-\infty]_x[-\infty,y^+]_y\rangle_{\Psi}^{\rm Fig. \ref{fig:kvdiagrams}d-f}
\nonumber\\
&&\hspace{2mm}
=~-{g^2\over 8\pi^2}c_F\!\int\!\dhd\beta_B\Bsi(\beta_B,x_\perp) \Gamma
e^{-ip_B\vro x^+}\ln\Big(-{i\over 4}(\beta_B+\ie)\sigma' s\Delta_\perp^2e^\gamma\Big)
~+~O\big({m_\perp^2\over\beta_B\sigma' s}\big)
\label{evolqminus}
\end{eqnarray}
Finally, let us discuss diagrams in Fig. \ref{fig:kvdiagrams}(g)-\ref{fig:kvdiagrams}(i). 
Since the separation between operators $\bsi(x)$ and $\psi(y)$ is spacelike, we can replace the
product of operators by the T-product and get
\begin{eqnarray}
&&\hspace{-0mm}
\langle {\rm T}\{\bsi(x^+,x_\perp,-{\delta'}^-)\Gamma\psi(y^+,y_\perp,-\delta^-)\}\rangle_{\Psi}^{\rm Fig. \ref{fig:kvdiagrams}g-i}~
\label{handbag}\\
&&\hspace{-0mm}
=~g^2c_F
\int\!\dhd \beta_B\beta'_B\dhd p_{B_\perp}\dhd p'_{B_\perp} e^{-ip'_Bx-ip_By}
\!\int\! \dhd p ~e^{i\vro\alpha(\delta'-\delta)_- +i\vro\beta\Delta_+ -i(p,\Delta)_\perp}{\Bsi(p_B)\gamma_\xi(\notp-\notp_B)\Gamma(\notp+\notp'_B)\gamma^\xi\Psi(p'_B)
\over (p^2+\ie)[(p-p_B)^2+\ie][(p+p'_B)^2+\ie]}
\nonumber
\end{eqnarray}
Here we introduced  two different point splittings $\delta^-$ and ${\delta'}^-$. 
This is a temporary auxiliary construction that simplifies 
solution of the differential evolution 
equations obtained below. In the final results we take ${\delta'}^-=\delta^-={1\over\rho\sigma}$ 
where $\sigma$ is our rapidity cutoff in $\alpha$.

Let us demonstrate that the integral (\ref{handbag}) does not depend on $\delta^-,{\delta'}^-$ in the region 
\begin{equation}
\Delta_\perp^2\gg \Delta^+\delta^-,\Delta^+{\delta'}^-~~\Leftrightarrow~~\sigma\beta_Bs,\sigma'\beta_Bs\gg m_\perp^2
\label{region}
\end{equation}
Consider
\begin{eqnarray}
&&\hspace{-11mm}
\!\int\! \dhd p~ e^{ip\cdot\tilde\Delta}
{(\notp-\notp_B)\Gamma(\notp+\notp'_B)
\over (p^2+\ie)[(p-p_B)^2+\ie][(p+p'_B)^2+\ie]}
\label{handbag1}\\
&&\hspace{-11mm}
=~{s\over 2}\!\int\! \dhd\alpha\dhd\beta\dhd p_\perp~
e^{i\alpha\vro(\delta'-\delta)^- +i\beta\vro\Delta^+}{e^{-i(p,\Delta)_\perp}[(\beta-\beta_B)\notp_2+(p-p_B)_\perp]\Gamma
[(\beta+\beta'_B)\notp_2+(p+p'_B)_\perp]
\over(\alpha\beta s-p_\perp^2+\ie)[\alpha(\beta-\beta_B)s-(p-p_B)_\perp^2+\ie]
[\alpha(\beta+\beta'_B)s-(p+p'_B)_\perp^2+\ie]}
\nonumber
\end{eqnarray}
where $\tilde{\Delta}=({\delta'}^{-}-\delta^-,\Delta^+,\Delta_\perp)$.
Since due to Eq. (\ref{region}) $\tilde{\Delta}^2=-\Delta_\perp^2+2\Delta^+(\delta-\delta')^-\neq 0$, there is no overall divergence and 
the integral in the left-hand-side (LHS) is UV convergent. 
Also, since $p_{B_\perp}, p'_{B_\perp}\neq 0$ the integral in the LHS of Eq. (\ref{handbag1}) is IR convergent.
Now, let us expand the RHS of this equation in powers of $(\delta-\delta')^-$. The first term of the expansion is the 
integral (\ref{handbag1}) with $\tilde\Delta$ replaced by $\Delta$ which is  also convergent since $\Delta^2=-\Delta_\perp^2<0$. Moreover, 
if one takes three residues over $\beta$ in the RHS 
[corresponding to three diagrams in Figs. \ref{fig:kvdiagrams}(g), \ref{fig:kvdiagrams}(h), and \ref{fig:kvdiagrams}(i)]
one gets integrals over $\alpha$ which converge at $\alpha\sim{m_\perp^2\over\beta_Bs}$.
Next, the integral over $\alpha$ in the second term of the expansion has an extra 
$\alpha$ but  is still convergent so $\alpha\vro(\delta'-\delta)^- \sim {m_\perp^2\over\sigma\beta_Bs}$ due to Eq. (\ref{region}).  Thus,  the  expansion of $e^{i\alpha\vro(\delta'-\delta)^-}$ gives the $\sigma,\sigma'$-independent 
term plus power corrections that we neglect. In other words,  the diagrams in Figs.  \ref{fig:kvdiagrams}(g)-\ref{fig:kvdiagrams}(i) 
do not contribute to the rapidity evolution in the Sudakov region.

Thus, the result of the calculation of diagrams in Fig. \ref{fig:kvdiagrams} reads
\begin{eqnarray}
&&\hspace{-0mm}
\Big(\sigma{d\over d\sigma}+\sigma'{d\over d\sigma'}\Big)
\langle\bsi^{\sigma'}\big(x^+,x_\perp\big)[x^+,-\infty]_x[-\infty,y^+]_y\Gamma\psi^\sigma\big(y^+,y_\perp\big)\rangle_{\Psi}
\nonumber\\
&&\hspace{2mm}
=~-{\alpha_s\over 2\pi}c_F\!\int\!\dhd\beta_B\dhd\beta'_B
\Bsi(\beta'_B,x_\perp)\Gamma\Psi(\beta_B,y_\perp) e^{-i\beta'_B\vro x^+ -i\beta_B\vro y^+}
\nonumber\\
&&\hspace{2mm}
\times~\Big[\ln\Big(-{i\over 4}(\beta_B+\ie)\sigma sb_\perp^2e^\gamma\Big)
+\ln\Big(-{i\over 4}(\beta'_B+\ie)\sigma' sb_\perp^2e^\gamma\Big)\Big]
~+~O\Big({m_\perp^2\over\beta_B\sigma s},{m_\perp^2\over\beta'_B\sigma' s}\Big)
\label{tmdevolminus}
\end{eqnarray}
where $b_\perp\equiv\Delta_\perp$ is a standard notation for the transverse separation of the TMD operator.

\subsection{Evolution equations for quark TMDs}

Promoting background fields in the RHS of Eq. 
(\ref{tmdevolminus}) to operators, one obtains the leading-order evolution equation of quark TMD operators in the form
\begin{eqnarray}
&&\hspace{-0mm}
\Big(\sigma{d\over d\sigma}+\sigma'{d\over d\sigma'}\Big)
\bsi^{\sigma'}(\beta'_B,x_\perp) \Gamma\psi^\sigma(\beta_B,y_\perp)
\nonumber\\
&&\hspace{2mm}
=~-{\alpha_s\over 2\pi}c_F
\bsi^{\sigma'}(\beta'_B,x_\perp) \Gamma\psi^\sigma(\beta_B,y_\perp)
\Big[\ln\Big(-{i\over 4}(\beta'_B+\ie)\sigma' s\Delta_\perp^2e^\gamma\Big)
+\ln\Big(-{i\over 4}(\beta_B+\ie)\sigma s\Delta_\perp^2e^\gamma\Big)\Big]
\label{loevoleq}
\end{eqnarray}
where standard TMD  gauge links are assumed
.
The solution of the evolution equation (\ref{loevoleq}) reads
\begin{eqnarray}
&&\hspace{-0mm}
\bsi^{\sigma'}(\beta'_B,x_\perp) \Gamma\psi^\sigma(\beta_B,y_\perp)
\label{evolo}\\
&&\hspace{0mm}
=~e^{ -{\alpha_sc_F\over 4\pi}\ln{\sigma'\over\sigma'_0}\big[\ln{\sigma'\sigma'_0}+2\ln \big(-{i\over 4}(\beta'_B+\ie)s\Delta_\perp^2e^\gamma\big)\big]}
\bsi^{\sigma'_0}(\beta'_B,x_\perp) \Gamma\psi^{\sigma_0}(\beta_B,y_\perp)
e^{ -{\alpha_sc_F\over 4\pi}\ln{\sigma\over\sigma_0}\big[\ln{\sigma\sigma_0}+2\ln \big(-{i\over 4}(\beta'_B+\ie)s\Delta_\perp^2e^\gamma\big)\big]}
\nonumber
\end{eqnarray}
It appears that two exponential factors in the RHS describe two independent evolutions of operators (\ref{Psidef}). Of course, this is not 
quite right since the left and right exponents come not only from ``virtual'' corrections of Fig. \ref{fig:kvdiagrams}(c) type but also 
from ``emission'' diagrams of Fig. \ref{fig:kvdiagrams}(a) and \ref{fig:kvdiagrams}(b) 
type which is reflected in the $\Delta_\perp$ dependence of these factors. Still, as we will see below, this
``factorized'' structure persists to quark-loop  corrections.
 \subsubsection{Leading-order evolution in the coordinate space and conformal invariance}

The evolution equation in the coordinate space is easily obtained by the Fourier transformation of Eq. (\ref{loevoleq})
\begin{eqnarray}
&&\hspace{-0mm}
\Big(\sigma{d\over d\sigma}+\sigma'{d\over d\sigma'}\Big)
\bsi^{\sigma'}(x^+,x_\perp)\Gamma\psi^{\sigma}(y^+,y_\perp)
\nonumber\\
&&\hspace{2mm}
=~{\alpha_s\over 4\pi^2}c_F\!\int\! dz^+~\Big\{
\Big[i{\ln\vro(-x^++z^++\ie)-\ln{\sigma b_\perp^2s\over 4}e^\gamma\over -x^++z^++\ie}+{\rm c.c.}\Big]
\bsi^{\sigma'}(z^+,x_\perp\big)\Gamma\psi^{\sigma}(y^+,y_\perp)
\nonumber\\
&&\hspace{2mm}
+~\Big[i{\ln\vro(-y^++w^++\ie)-\ln{\sigma'b_\perp^2s\over 4}e^\gamma\over -y^++w^++\ie}+{\rm c.c.}\Big]
\bsi^{\sigma'}(x^+,x_\perp)\Gamma\psi^{\sigma}(y^+,y_\perp)
\Big\}
\label{evoleqlo}
\end{eqnarray}
Note the ``causality'': $z^+\leq x^+$ and $w^+\leq y^+$: the evolved $\bsi$, $\psi$ operators  lag behind the original ones,
similar to the case of power corrections to TMD factorization  where the emission of additional projectile/target gluons 
also lags behind the original quark operators (see Refs. \cite{Balitsky:2017gis,Balitsky:2020jzt}).

The solution of this equation  has the form
\begin{eqnarray}
&&\hspace{-0mm}
\bsi^{\sigma'}(x^+,x_\perp)\Gamma\psi^{\sigma}(y^+,y_\perp)
~
=~e^{ -{\alpha_sc_F\over 4\pi}\big(\ln{\sigma'\over\sigma'_0}\ln{\sigma'\sigma'_0}+\ln{\sigma\over\sigma_0}\ln{\sigma\sigma_0}\big)}
\nonumber\\
&&\hspace{40mm}
\times~\int\!dz^+\bigg[{i\Gamma\big(1-{\alpha_sc_F\over 2\pi}\ln{\sigma'\over \sigma'_0}\big)
\over (z^+-x^++\ie)^{1-{\alpha_sc_F\over 2\pi}\ln{\sigma'\over \sigma'_0}}}
+{\rm c.c.}\bigg]
\int\!dw^+\bigg[{i\Gamma\big(1-{\alpha_sc_F\over 2\pi}\ln{\sigma\over \sigma_0}\big)
\over (w^+-y^++\ie)^{1-{\alpha_sc_F\over 2\pi}\ln{\sigma\over \sigma_0}}}
+{\rm c.c.}\bigg]
\nonumber\\
&&\hspace{40mm}
\times~{1\over 4\pi^2}\big(b_\perp^2 e^\gamma\sqrt{s/8}\big)^{ -{\alpha_sc_F\over 2\pi}\big(\ln{\sigma'\over\sigma'_0}+\ln{\sigma\over\sigma_0}\big)}
\bsi^{\sigma'_0}(z^+,x_\perp)\Gamma\psi^{\sigma_0}(w^+,y_\perp)
\label{tmdcoordevol}
\end{eqnarray}

In the leading order one does not take into account QCD running coupling  so one should expect 
some symmetries related to conformal invariance. Indeed,  
if we take $\sigma=\sigma'={\varsigma\sqrt{2}\over\vro|\Delta_\perp|}$ where $\varsigma$ is an evolution
parameter, the evolution (\ref{tmdcoordevol}) is invariant under a certain subgroup of conformal group
SO(2,4) (see the discussion  in Ref. \cite{Balitsky:2019ayf}).

\section{Quark loop correction \label{qloopq}}

\begin{figure}[htb]
\begin{center}
\includegraphics[width=121mm]{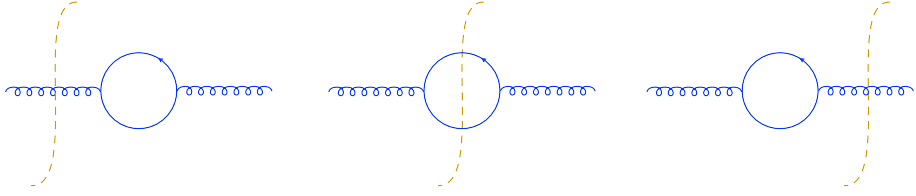}
\end{center}
\caption{Quark loop correction to cut gluon propagator.\label{fig:cutgluon1}}
\end{figure}

It is well-known that the argument of the coupling constant in the LO rapidity evolution equations 
[(\ref{loevoleq}) or (\ref{evolo})] cannot be determined. 
As we mentioned above, we will use the BLM method to fix the argument of the running coupling constant.
According to the BLM procedure, we need to calculate the contribution of the first quark loop  to our TMD evolution (\ref{tmdevolminus}) 
and promote $-{1\over 6\pi}n_f$ to full $b_0={11\over 12\pi}N_c-{1\over 6\pi}n_f$. Each gluon propagator
in diagrams in Fig. \ref{fig:kvdiagrams} should be replaced by a one-loop correction, i.e.
\begin{eqnarray}
{1\over p^2+\ie}~~&\rightarrow&~~{1\over p^2+\ie}\Big(1+b_0\alpha_s(\mu)\ln{\tmu^2\over -p^2-\ie}\Big)
\nonumber\\
{1\over p^2-\ie}~~&\rightarrow&~~{1\over p^2-\ie}\Big(1+b_0\alpha_s(\mu)\ln{\tmu^2\over -p^2+\ie}\Big)
\nonumber\\
2\pi\delta(p^2)\theta(p_0)~~&\rightarrow&~~{i\theta(p_0)\over p^2+\ie}\Big(1+b_0\alpha_s(\mu)\ln{\tmu^2\over -p^2-\ie}\Big)
-{i\theta(p_0)\over p^2-\ie}\Big(1+b_0\alpha_s(\mu)\ln{\tmu^2\over -p^2+\ie}\Big)
\label{luprop}
\end{eqnarray}
where $\tmu^2\equiv \bar\mu^2_{\rm MS}e^{5/3}$. The first two lines are trivial while the third line corresponds to the
sum of the diagrams shown in Fig. \ref{fig:cutgluon1}.
First, note that the convergence of integral (\ref{handbag}) representing diagrams in  Fig. \ref{fig:kvdiagrams}(h)-\ref{fig:kvdiagrams}(i)
is not affected by extra 
$\ln{\tmu^2\over -p^2-\ie}$. Repeating the arguments after Eq. (\ref{handbag}) we see that the contribution of diagrams in Fig. \ref{fig:kvdiagrams}(h)-\ref{fig:kvdiagrams}(i) with extra quark loops is still a power correction 
${m_\perp^2\over\sigma\beta_Bs}$ (multiplied by an extra log). Thus, we need to consider diagrams 
in Figs. \ref{fig:kvdiagrams}(a)-\ref{fig:kvdiagrams}(c) and Figs. \ref{fig:kvdiagrams}(d)-\ref{fig:kvdiagrams}(f).

It is convenient to start again with the diagram in Fig. \ref{fig:kvdiagrams}(c).  
Replacing Feynman gluon propagator ${1\over p^2+\ie}$ in Eq. (\ref{virt}) by the $\alpha_s$ correction
from the first line in Eq. (\ref{luprop}) we get
\begin{eqnarray}
&&\hspace{-1mm}
\langle{\rm T}\{[-\infty,y^+]_y\Gamma\psi(y^+,y_\perp,-\delta^-)\}\rangle_\Psi^{\rm loop~ 1c}
~=~-i4\pi b_0\alpha_s^2(\mu)c_F\!\int\! \dhd\beta_B\dhd p_{B_\perp}
\nonumber\\
&&\hspace{-1mm}
\times~e^{-ip_By}\!\int\!\dhd\alpha\dhd\beta\dhd p_\perp{1\over \beta+\ie}e^{-i\alpha\vro\delta^-}
{\ln{\tmu^2\over p_\perp^2-\alpha\beta s-\ie}\over \alpha\beta s-p_\perp^2+\ie}
{s(\beta-\beta_B)\over \alpha(\beta-\beta_B)s-(p-p_B)_\perp^2+\ie}\Gamma\Psi(\beta_B,p_{B_\perp})
\nonumber\\
&&\hspace{-1mm}
=~4\pi b_0\alpha_s^2(\mu)c_F\!\int\! \dhd\beta_B\dhd p_{B_\perp}e^{-ip_By}\!\int_{-\infty}^0\!\dhd\alpha\!\int\!\dhd p_\perp
{\beta_Bs\ln{\tmu^2\over p_\perp^2}\over p_\perp^2[\alpha\beta_Bs+(p-p_B)_\perp^2-\ie]}\Gamma\Psi(\beta_B,p_{B_\perp})e^{-i{\alpha\over\sigma}}
\nonumber\\
&&\hspace{-1mm}
=~-4\pi b_0\alpha_s^2(\mu)c_F\!\int\! \dhd\beta_B\dhd p_{B_\perp}e^{-ip_By}\!\int_0^{\infty}\!\dhd\alpha\!\int\!\dhd p_\perp
{\beta_Bs\ln{\tmu^2\over p_\perp^2}\over p_\perp^2[\alpha\beta_Bs+(p-p_B)_\perp^2+\ie]}\Gamma\Psi(\beta_B,p_{B_\perp})e^{-i{\alpha\over\sigma}}
\label{vloop}
\end{eqnarray}
where we made the same rotation of contour  over $\alpha$  on angle $\pi$ in the lower complex half-plane as in Eq. (\ref{virt}).
We get
\begin{eqnarray}
&&\hspace{-1mm}
\sigma{d\over d\sigma}\langle{\rm T}\{[-\infty,y^+]_y\Gamma\psi(y^+,y_\perp,-\delta^-)\}\rangle_\Psi^{\rm loop~1c}
\nonumber\\
&&\hspace{-1mm}
=~-i4\pi b_0\alpha_s^2(\mu)c_F{1\over\sigma}\!\int\! \dhd\beta_B\dhd p_{B_\perp}e^{-ip_By}\!\int_0^{\infty}\!\dhd\alpha\!\int\!\dhd p_\perp
{\alpha\beta_Bs\ln{\tmu^2\over p_\perp^2}\over p_\perp^2[\alpha\beta_Bs+(p-p_B)_\perp^2+\ie]}\Gamma\Psi(\beta_B,p_{B_\perp})e^{-i{\alpha\over\sigma}}
\label{vloopevol}
\end{eqnarray}

Next, consider diagrams in Figs. \ref{fig:kvdiagrams}(a) and \ref{fig:kvdiagrams}(b). 
Using Eqs. (\ref{luprop}) for various gluon propagators, we obtain the correction in the form
\begin{eqnarray}
&&\hspace{-0mm}
\langle[x^+,-\infty]_x\Gamma\psi(y^+,y_\perp,-\delta^-)\rangle_\Psi^{\rm loop~1a,b}
~=~4\pi b_0\alpha_s^2(\mu)c_F\!\int\! \dhd\beta'_B\dhd p_{B_\perp}e^{-i(p_B,y)}
\Gamma\Psi(\beta_B,p_{B_\perp})
\nonumber\\
&&\hspace{-0mm}
\times~\!\int\!\dhd\alpha\dhd\beta\dhd p_\perp\bigg[
\bigg({i\theta(\alpha)\over \alpha(\beta-\beta_B)s-(p-p_B)_\perp^2+\ie}-{i\theta(\alpha)\over \alpha(\beta-\beta_B)s-(p-p_B)_\perp^2-\ie}\bigg)
{(\beta-\beta_B)\ln{\tmu^2\over p_\perp^2-\alpha\beta s+\ie}\over \alpha\beta s-p_\perp^2-\ie}
\nonumber\\
&&\hspace{0mm}
+~{i(\beta-\beta_B)s\theta(\alpha)\over \alpha(\beta-\beta_B)s-(p-p_B)_\perp^2+\ie}
\bigg({\ln {\tmu^2\over p_\perp^2-\alpha\beta s-\ie}\over \alpha\beta s-p_\perp^2+\ie}
-{\ln {\tmu^2\over p_\perp^2-\alpha\beta s+\ie}\over \alpha\beta s-p_\perp^2-\ie}\bigg)
\bigg]{1\over\beta+\ie}e^{-i{\alpha\over\sigma}-i\beta\vro\Delta^++i(p,\Delta)_\perp}
\nonumber\\
&&\hspace{-1mm}
=~4\pi b_0\alpha_s^2(\mu)c_F\!\int\! \dhd\beta'_B\dhd p_{B_\perp}e^{-i(p_B,y)}
\Gamma\Psi(\beta_B,p_{B_\perp})
\!\int\!\dhd\alpha\dhd\beta\dhd p_\perp e^{-i{\alpha\over\sigma}-i\beta\vro\Delta^++i(p,\Delta)_\perp}~{1\over\beta+\ie}
\nonumber\\
&&\hspace{-0mm}
\times~\bigg[
{i(\beta-\beta_B)s\theta(\alpha)\ln {\tmu^2\over p_\perp^2-\alpha\beta s-\ie}\over [\alpha(\beta-\beta_B)s-(p-p_B)_\perp^2+\ie](\alpha\beta s-p_\perp^2+\ie)}
-{i(\beta-\beta_B)s\theta(\alpha)\ln {\tmu^2\over p_\perp^2-\alpha\beta s+\ie}\over [\alpha(\beta-\beta_B)s-(p-p_B)_\perp^2-\ie](\alpha\beta s-p_\perp^2-\ie)}
\bigg]\,,
\label{realoop}
\end{eqnarray}
where the second line comes from the Fig. \ref{fig:kvdiagrams}(b) diagram while the third line from Fig. \ref{fig:kvdiagrams}(a) diagram.
The  $\sigma$ evolution reads
\begin{eqnarray}
&&\hspace{-1mm}
\sigma{d\over d\sigma}\langle[-\infty,x^+]_x\Gamma\psi(y^+,y_\perp,\delta^-)\rangle_\Psi^{\rm loop~1a,b}
\nonumber\\
&&\hspace{-1mm}
=~i4\pi b_0\alpha_s^2(\mu)c_F{1\over\sigma}\!\int\! \dhd\beta_B\dhd p_{B_\perp}e^{-i(p_B,y)}
\Gamma\Psi(\beta_B,p_{B_\perp})
\!\int\!\dhd\alpha\dhd\beta\dhd p_\perp e^{-i{\alpha\over\sigma}-i\beta\vro\Delta^++i(p,\Delta)_\perp}~{1\over\beta+\ie}
\nonumber\\
&&\hspace{-0mm}
\times~\bigg[
{i\alpha(\beta-\beta_B)s\theta(\alpha)\ln {\tmu^2\over p_\perp^2-\alpha\beta s-\ie}\over [\alpha(\beta-\beta_B)s-(p-p_B)_\perp^2+\ie](\alpha\beta s-p_\perp^2+\ie)}
-{i\alpha(\beta-\beta_B)s\theta(\alpha)\ln {\tmu^2\over p_\perp^2-\alpha\beta s+\ie}\over [\alpha(\beta-\beta_B)s-(p-p_B)_\perp^2-\ie](\alpha\beta s-p_\perp^2-\ie)}
\bigg]
\label{realoopevol}
\end{eqnarray}

Now we shall prove that the dependence of RHS on $\Delta^+$ is a power correction. 
It is convenient to consider the derivative with respect to $x^+$
\begin{eqnarray}
&&\hspace{-1mm}
\Delta^+{d\over dx^+}\sigma{d\over d\sigma}\langle[-\infty,x^+]_x\Gamma\psi(y^+,y_\perp,-\delta^-)\rangle_\Psi^{\rm loop~1a,b}
\nonumber\\
&&\hspace{-1mm}
=~4\pi b_0\alpha_s^2(\mu)c_F\!\int\! \dhd\beta'_B\dhd p_{B_\perp}e^{-i(p_B,y)}
\Gamma\Psi(\beta_B,p_{B_\perp})
{\vro\Delta^+\over\sigma}\!\int\!\dhd\alpha\dhd\beta\dhd p_\perp e^{-i{\alpha\over\sigma}-i\beta\vro\Delta^++i(p,\Delta)_\perp}~
\nonumber\\
&&\hspace{-0mm}
\times\bigg[
{i\alpha(\beta-\beta_B)s\theta(\alpha)\ln {\tmu^2\over p_\perp^2-\alpha\beta s-\ie}\over [\alpha(\beta-\beta_B)s-(p-p_B)_\perp^2+\ie](\alpha\beta s-p_\perp^2+\ie)}
-{i\alpha(\beta-\beta_B)s\theta(\alpha)\ln {\tmu^2\over p_\perp^2-\alpha\beta s+\ie}\over [\alpha(\beta-\beta_B)s-(p-p_B)_\perp^2-\ie](\alpha\beta s-p_\perp^2-\ie)}
\bigg]
\label{realoopevold}
\end{eqnarray}
Let us consider case $\Delta^+ >0$. The second term in the square brackets in the RHS of Eq. (\ref{realoopevold}) 
vanishes while the first one can be rewritten as
\begin{eqnarray}
&&\hspace{-1mm}
{\vro\Delta^+\over\sigma}\!\int\!\dhd\alpha\dhd\beta\dhd p_\perp e^{-i{\alpha\over\sigma}-i\beta\vro\Delta^++i(p,\Delta)_\perp}~
{i\alpha(\beta-\beta_B)s\theta(\alpha)\ln {\tmu^2\over p_\perp^2-\alpha\beta s-\ie}\over [\alpha(\beta-\beta_B)s-(p-p_B)_\perp^2+\ie](\alpha\beta s-p_\perp^2+\ie)}
\label{tuterms}\\
&&\hspace{-0mm}
=~{\vro\Delta^+\over\sigma}\!\int\!\dhd\alpha\dhd\beta\dhd p_\perp e^{-i{\alpha\over\sigma}-i\beta\vro\Delta^++i(p,\Delta)_\perp}~
\theta(\alpha)
\bigg[{\ln {\tmu^2\over p_\perp^2-\alpha\beta s-\ie}\over \alpha\beta s-p_\perp^2+\ie}+
{i(p-p_B)_\perp^2\ln {\tmu^2\over p_\perp^2-\alpha\beta s-\ie}
\over [\alpha(\beta-\beta_B)s-(p-p_B)_\perp^2+\ie](\alpha\beta s-p_\perp^2+\ie)}\bigg]
\nonumber
\end{eqnarray}
The first term in the RHS can easily be calculated 
\begin{equation}
{2\vro\Delta^+\over\sigma s}\!\int\dhd^4p~{\ln {\tmu^2\over -p^2-\ie}\over p^2+\ie}e^{-ip\cdot\Telta}
~=~{2\beta_B\vro\Delta^+\over\sigma\beta_Bs}{\ln(-\Telta^2\mu^2+\ie)\over 4\pi^2(-\Telta^2+\ie)}
~\simeq~{2\beta_B\vro\Delta^+\over\sigma\beta_Bs}{\ln(\Delta_\perp^2\mu^2+\ie)\over 4\pi^2\Delta_\perp^2}
~\sim~\beta_B\vro\Delta^+\times O\Big({m_\perp^2\over\sigma\beta_Bs}\Big)
\end{equation}
where $\Telta^2=-2\Delta^+\delta^- -\Delta_\perp^2\simeq -\Delta_\perp^2$. Since, as discussed in the Introduction, we consider $\beta\vro\Delta^+\sim 1$, 
this term is a power correction $\sim{m_\perp^2\over\sigma\beta_Bs}$. As to the second term in the RHS 
of Eq. (\ref{tuterms}), it can be estimated as follows: 
since the integral over $p_\perp$ is convergent at $p_\perp\sim \Delta_\perp^{-1}\sim m_\perp$, 
one can replace $(p-p_B)_\perp^2$ in the numerator approximately by $m_\perp^2$ and get
\begin{equation}
\vro\Delta^+{2m_\perp^2\over\sigma s}\!\int\dhd^4p~{\ln {\tmu^2\over -p^2-\ie}\over (p^2+\ie)[(p-p_B)_\perp^2+\ie]}e^{-ip\cdot\Telta}
~\sim~\beta_B\vro\Delta^+O\Big({m_\perp^2\over\sigma\beta_Bs}\Big)
\end{equation}
Indeed, the integral over momenta in the RHS. can be represented as ($\baru\equiv 1-u$)
\begin{eqnarray}
&&\hspace{-1mm}
\!\int{\dhd^4p\over i}~{\ln {\tmu^2\over -p^2-\ie}\over (p^2+\ie)[(p-p_B)_\perp^2+\ie]}e^{-ip\cdot\Telta}
~=~\int_0^1\! du~e^{iu(p_B\cdot\Delta)_\perp}\!\int{\dhd^4p\over i}e^{-ip\cdot\Telta}{1+\ln\baru -\ln(p_{B_\perp}^2\baru u-p^2-\ie)/\tmu^2\over (p_{B_\perp}^2\baru u-p^2-\ie)^2}
\nonumber\\
&&\hspace{-0mm}
\simeq~{1\over 16\pi^2}\!\int_0^1\! du~e^{iu(p_B,\Delta)_\perp}\Big(\ln{\Delta_\perp^2\tmu^2\over 4}-\ln{p_{B_\perp}^2\over\mu^2}+2\gamma +\ln{\baru\over u}\Big)
K_0\big(\sqrt{p_{B_\perp}^2\Delta_\perp^2\baru u}\big)
\end{eqnarray}
where we used $\Telta^2\simeq -\Delta_\perp^2$. This integral is obviously $O(1)$ at $p_{B_\perp}^2\sim\Delta_\perp^{-2}\sim m_\perp^2$.

Summarizing, we proved that at $\Delta^+>0$ the RHS of integral (\ref{realoopevold}) is a power correction. Also, at $\Delta^+<0$ only the second term
in square brackets in  the RHS of integral (\ref{realoopevold}) contributes, and similar calculation shows that Eq. (\ref{realoopevold}) is a power correction. 
Thus,  with power accuracy $O\Big({m_\perp^2\over\sigma\beta_Bs}\Big)$ we can set $\Delta^+=0$.

Returning to Eq. (\ref{realoopevol}) and taking $\Delta^+=0$ we get
\begin{eqnarray}
&&\hspace{-1mm}
\sigma{d\over d\sigma}\langle[-\infty,x^+]_x\Gamma\psi(y^+,y_\perp,-\delta^-)\rangle_\Psi^{\rm loop~1a,b}
\nonumber\\
&&\hspace{-1mm}
=~i4\pi b_0\alpha_s^2(\mu)c_F{1\over\sigma}\!\int\! \dhd\beta'_B\dhd p_{B_\perp}e^{-i(p_B,y)}
\Gamma\Psi(\beta_B,p_{B_\perp})
\!\int\!\dhd\alpha\dhd\beta\dhd p_\perp e^{-i{\alpha\over\sigma}+i(p,\Delta)_\perp}~{1\over\beta+\ie}
\nonumber\\
&&\hspace{-0mm}
\times\bigg[
{i\alpha(\beta-\beta_B)s\theta(\alpha)\ln {\tmu^2\over p_\perp^2-\alpha\beta s-\ie}\over [\alpha(\beta-\beta_B)s-(p-p_B)_\perp^2+\ie](\alpha\beta s-p_\perp^2+\ie)}
-{i\alpha(\beta-\beta_B)s\theta(\alpha)\ln {\tmu^2\over p_\perp^2-\alpha\beta s+\ie}\over [\alpha(\beta-\beta_B)s-(p-p_B)_\perp^2-\ie](\alpha\beta s-p_\perp^2-\ie)}
\bigg]
\nonumber\\
&&\hspace{-1mm}
=~i4\pi b_0\alpha_s^2(\mu)c_F{1\over\sigma}\!\int\! \dhd\beta_B\dhd p_{B_\perp}e^{-ip_By}\Gamma\Psi(\beta_B,p_{B_\perp})\!\int_0^{\infty}\!\dhd\alpha\!\int\!\dhd p_\perp
{\alpha\beta_Bs\ln{\tmu^2\over p_\perp^2}\over p_\perp^2[\alpha\beta_Bs+(p-p_B)_\perp^2+\ie]}e^{-i{\alpha\over\sigma}}
\label{realloopevolotvet}
\end{eqnarray}
The total contribution of diagrams in Figs. \ref{fig:kvdiagrams}(a)-\ref{fig:kvdiagrams}(c)
is a sum of Eqs. (\ref{vloopevol}) and (\ref{realloopevolotvet})
\begin{eqnarray}
&&\hspace{-1mm}
\sigma{d\over d\sigma}\langle[-\infty,y^+]_y\Gamma\psi(y^+,y_\perp,-\delta^-)\rangle_\Psi^{\rm Fig1a-c~loop}
\nonumber\\
&&\hspace{-1mm}
=~i4\pi b_0\alpha_s^2(\mu)c_F\!\int\! \dhd\beta_B\dhd p_{B_\perp}e^{-ip_By}\Gamma\Psi(\beta_B,p_{B_\perp})
\!\int_0^{\infty}\!{\dhd\alpha\over\sigma}e^{-i{\alpha\over\sigma}}\!\int\!\dhd p_\perp
{\big(e^{i(p,\Delta)_\perp}-1\big)\alpha\beta_Bs\ln{\tmu^2\over p_\perp^2}\over p_\perp^2[\alpha\beta_Bs+(p-p_B)_\perp^2+\ie]}
\label{loopevoleqn1}
\end{eqnarray}

Next, similar to Eq. (\ref{napravo2}) one can demonstrate that $p_{B_\perp}$ dependence in the integral in the 
RHS can be omitted with power accuracy. Indeed,
\begin{eqnarray}
&&\hspace{-1mm}
\!\int_0^{\infty}\!{\dhd\alpha\over\sigma}e^{-i{\alpha\over\sigma}}\!\int\!\dhd p_\perp
\big(e^{i(p,\Delta)_\perp}-1\big)\alpha\beta_Bs\ln{\tmu^2\over p_\perp^2}
\Big[{1\over \alpha\beta_Bs+(p-p_B)_\perp^2+\ie}-{1\over \alpha\beta_Bs+p_\perp^2+\ie}\Big]
\label{estimate1}\\
&&\hspace{-1mm}
\simeq~\!\int\!\dhd p_\perp
\big(e^{i(p,\Delta)_\perp}-1\big)\ln{\tmu^2\over p_\perp^2}
\!\int_0^{\sigma}\!{\dhd\alpha\over\sigma}\Big[
{p_\perp^2\over \alpha\beta_Bs+p_\perp^2+\ie}-{(p-p_B)_\perp^2\over \alpha\beta_Bs+(p-p_B)_\perp^2+\ie}\Big]
\nonumber\\
&&\hspace{-1mm}
=~{1\over 2\pi}\!\int\!\dhd p_\perp
\big(1-e^{i(p,\Delta)_\perp}\big)\ln{\tmu^2\over p_\perp^2}
\Big[{(p-p_B)_\perp^2\over\sigma\beta_Bs}\ln{\sigma\beta_Bs+(p-p_B)_\perp^2+\ie\over (p-p_B)_\perp^2}
-(p_{B_\perp}\rightarrow 0)\Big]~\sim~O\Big({m_\perp^2\over\sigma\beta_Bs}\ln{\sigma\beta_Bs\over m_\perp^2}\Big)
\nonumber
\end{eqnarray}
Thus, we get
\begin{eqnarray}
&&\hspace{-1mm}
\sigma{d\over d\sigma}\langle[-\infty,y^+]_y\Gamma\psi(y^+,y_\perp,-\delta^-)\rangle_\Psi^{\rm Fig1a-c~loop}
\nonumber\\
&&\hspace{-1mm}
=~i4\pi b_0\alpha_s^2(\tmu)c_F\!\int\! \dhd\beta_B\dhd p_{B_\perp}e^{-ip_By}\Gamma\Psi(\beta_B,p_{B_\perp})
\!\int_0^{\infty}\!{\dhd\alpha\over\sigma}e^{-i{\alpha\over\sigma}}\!\int\!\dhd p_\perp
{\big(e^{i(p,\Delta)_\perp}-1\big)\alpha\beta_Bs\ln{\tmu^2\over p_\perp^2}\over p_\perp^2[\alpha\beta_Bs+p_\perp^2+\ie]}
\label{loopevoleqn1e}
\end{eqnarray}
The integral in the RHS of this equation is calculated in Appendix \ref{app:integrals}
[see Eq. (\ref{loopintegralotvet1})],  so we get the result with one-loop accuracy in the form
\begin{eqnarray}
&&\hspace{-0mm}
\sigma{d\over d\sigma}\langle[x^+,-\infty]_x[-\infty,y^+]_y\Gamma\psi(y^+,y_\perp,-\delta^-)\rangle_\Psi^{\rm Fig1a-c~loop}
\label{evolrunpart1}\\
&&\hspace{2mm}
=~-{\alpha_s(\tmu)\over 2\pi}c_F\!\int\!\dhd\beta_B\dhd p_{B_\perp}\Gamma\Psi(\beta_B,p_{B_\perp}) e^{-ip_By}
\Big\{\ln\big[-{i\over 4}(\beta_B+\ie)\sigma s\Delta_\perp^2e^\gamma\big]
\nonumber\\
&&\hspace{2mm}
+~{b_0\over 2}\alpha_s(\tmu)\Big[
\big(\ln{\Delta_\perp^2\tmu^4/4\over -i\sigma(\beta_B+\ie)s} +3\gamma\big)\ln\big[-{i\over 4}(\beta_B+\ie)\sigma s\Delta_\perp^2e^\gamma\big]-{\pi^2\over 2}\Big]\Big\}
~+~O\big({m_\perp^2\over\beta_B\sigma s}\big)
\nonumber\\
&&\hspace{2mm}
=~-{\alpha_s(\mu_\sigma)\over 2\pi}c_F\!\int\!\dhd\beta_B\Gamma\Psi(\beta_B,y_\perp) e^{-i\beta_B\vro y^+}
\big\{\ln\big[-{i\over 4}(\beta_B+\ie)\sigma s\Delta_\perp^2e^\gamma\big]
+O\big(\alpha_s(\mu_\sigma)\big)\big\}
~+~O\big({m_\perp^2\over\beta_B\sigma s}\big)
\nonumber
\end{eqnarray}
where $\mu^2_\sigma\equiv\sqrt{\sigma|\beta_B|s\over\Delta_\perp^2}$.  
Thus, the BLM scale for Sudakov evolution is halfway between the 
 transverse momentum and the energy  scale of TMD.
 
Performing a similar calculation of the loop contribution to diagrams in Figs. \ref{fig:kvdiagrams}(d)-\ref{fig:kvdiagrams}(f) we obtain 
\begin{eqnarray}
&&\hspace{-0mm}
\sigma'{d\over d\sigma'}\langle\bsi(x^+,x_\perp,-{\delta'}^-)\Gamma[x^+,-\infty]_x[-\infty,y^+]_y\rangle_\Psi^{\rm Fig1d-f}
\label{evolrunpart2}\\
&&\hspace{2mm}
=~-{\alpha_s(\mu'_{\sigma})\over 2\pi}c_F\!\int\!\dhd\beta'_B\Gamma\Bsi(\beta'_B,x_\perp) e^{-i\beta'_B\vro x^+}
\big\{\ln\big[-{i\over 4}(\beta'_B+\ie)\sigma' s\Delta_\perp^2e^\gamma\big]
+O\big((\alpha_s(\mu'_\sigma)\big)]\big\}
~+~O\big({m_\perp^2\over\beta_B\sigma' s}\big)
\nonumber
\end{eqnarray}
where
 $\mu^2_{\sigma'}\equiv\sqrt{\sigma'|\beta'_B|s\over\Delta_\perp^2}$.  
 
 Combining Eqs. (\ref{evolrunpart1}), (\ref{evolrunpart2}) and promoting background fields to operators we 
 obtain the evolution equation for the TMD operator in the form
\begin{eqnarray}
&&\hspace{-0mm}
\Big(\sigma{d\over d\sigma}+\sigma'{d\over d\sigma'}\Big)
\bsi^{\sigma'}(\beta'_B,x_\perp) \Gamma \psi^\sigma(\beta_B,y_\perp)
\label{evoleqrun}\\
&&\hspace{2mm}
=~-{c_F\over 2\pi}
\bsi^{\sigma'}(\beta'_B,x_\perp) \Gamma\psi^\sigma(\beta_B,y_\perp)
\Big[\alpha_s(\mu_{\sigma'})\ln\Big(-{i\over 4}(\beta'_B+\ie)\sigma' sb_\perp^2e^\gamma\Big)
+\alpha_s(\mu_\sigma)\ln\Big(-{i\over 4}(\beta_B+\ie)\sigma sb_\perp^2e^\gamma\Big)\Big]
\nonumber
\end{eqnarray}
where $b_\perp\equiv\Delta_\perp=(x-y)_\perp$
We see that in the Sudakov region we can define TMD operator (\ref{qtmdop}) with two independent ``left'' and ``right'' cutoffs
$\sigma$ and $\sigma'$ defined in Eqs. (\ref{Psidef}) and the evolutions with respect to those cutoffs are independent [except for 
$b_\perp= (x-y)_\perp$]. 
 
We can solve evolution equation (\ref{evoleqrun}) by replacing 
$\sigma{d\over d\sigma}=-{b_0\over 2}\alpha^2(\mu_{\sigma}){d\over d\alpha(\mu_{\sigma})}$
(and similarly for $\sigma'{d\over d\sigma'}$). We get then
\begin{eqnarray}
&&\hspace{-0mm}
\Big(\alpha^2(\mu_{\sigma}){d\over d\alpha(\mu_{\sigma})}+\alpha^2(\mu_{\sigma'}){d\over d\alpha(\mu_{\sigma'})}\Big)
\bsi^{\sigma'}(\beta'_B,x_\perp)\Gamma\psi^\sigma(\beta_B,y_\perp)
\label{evoleqrunsol}\\
&&\hspace{0mm}
=~{c_F\over \pi b_0}
\Big[\alpha_s(\mu_{\sigma'})\ln\Big(-{i\over 4}(\beta'_B+\ie)\sigma' sb_\perp^2e^\gamma\Big)
+\alpha_s(\mu_\sigma)\ln\Big(-{i\over 4}(\beta_B+\ie)\sigma sb_\perp^2e^\gamma\Big)\Big]
\bsi^{\sigma'}(\beta'_B,x_\perp) \Gamma\psi^\sigma(\beta_B,y_\perp)
\nonumber\\
&&\hspace{0mm}
=~ -{2c_F\over \pi b_0^2}\Big\{{\alpha_s(\mu_{\sigma'})\over\alpha_s(\tbe_\perp^{-1})}+
{\alpha_s(\mu_{\sigma})\over\alpha_s(\tbe_\perp^{-1})}-2
-{b_0\alpha_s(\mu_{\sigma'})\over 2}\ln[-i(\tau'_B+\ie)]-{b_0\alpha_s(\mu_{\sigma})\over 2}\ln[-i(\tau_B+\ie)]\Big\}
\bsi^{\sigma'}(\beta'_B,x_\perp) \Gamma\psi^\sigma(\beta_B,y_\perp)
\nonumber
\end{eqnarray}
where $\tbe_\perp^2={b_\perp^2\over 2}e^{\gamma/2}$ and 
$\tau_B={\beta_B\over |\beta_B|},\tau'_B={\beta'_B\over |\beta'_B|}$. Note that formally 
$\alpha_s\ln[-i(\tau_B+\ie)]$ exceeds our accuracy but we keep to ensure the correct causal structure in the coordinate space, 
similar to the leading order evolution (\ref{evoleqlo}). 

The solution of Eq. (\ref{evoleqrun}) has the form
\begin{eqnarray}
&&\hspace{-0mm}
\bsi^{\sigma'}(\beta'_B,x_\perp)\Gamma\psi^\sigma(\beta_B,y_\perp)~
=~e^{ -{2c_F\over \pi b_0^2}\big[
\big({1\over\alpha_s(\tbe_\perp^{-1})}-{b_0\over 2}\ln[-i(\tau'_B+\ie)]\big)\ln{\alpha_s(\mu_{\sigma'})\over\alpha_s(\mu_{\sigma'_0})}+{1\over \alpha_s(\mu_{\sigma'})}-{1\over \alpha_s(\mu_{\sigma'_0})}
\big]}
\nonumber\\
&&\hspace{0mm}
\times~e^{-{2c_F\over \pi b_0^2}\big[
\big({1\over\alpha_s(\tbe_\perp^{-1})}-{b_0\over 2}\ln[-i(\tau_B+\ie)]\big)\ln{\alpha_s(\mu_{\sigma})\over\alpha_s(\mu_{\sigma_0})}+{1\over \alpha_s(\mu_{\sigma})}-{1\over \alpha_s(\mu_{\sigma_0})}
\big]}
\bsi^{\sigma'_0}(\beta'_B,x_\perp)\Gamma\psi^{\sigma_0}(\beta_B,y_\perp)
\label{qevolution}
\end{eqnarray}
Using the expansion
\begin{eqnarray}
&&\hspace{-0mm}
\Big({1\over\alpha_s(\tbe_\perp^{-1})}-{b_0\over 2}\ln[-i(\tau'_B+\ie)]\Big)\ln{\alpha_s(\mu_{\sigma'})\over\alpha_s(\mu_{\sigma'_0})}
+{1\over \alpha_s(\mu_{\sigma'})}-{1\over \alpha_s(\mu_{\sigma'_0})}
\\
&&\hspace{-0mm}
=~{b_0^2\over 4}\alpha_s\ln{\sigma\over\sigma_0}
\Big[\ln\Big(-{i\over 4}(\beta_B+\ie)\sigma sb_\perp^2e^\gamma\Big)+\half\ln\sigma\sigma_0\Big]~+~O(\alpha_s^2),
\nonumber
\end{eqnarray}
it is easy to check that at the leading order we obtain the LO equation (\ref{evolo}).

Note that, as in the leading order, the structure of Sudakov evolution (\ref{qevolution})  looks like two independent 
exponential factors which describe two independent evolutions of operators (\ref{Psidef}) [see the discussion below Eq. (\ref{evolo})].
Of course, one should not expect this property beyond the Sudakov region.

Let us present the final result for the rapidity evolution of
quark TMDs with running coupling
\begin{eqnarray}
&&\hspace{-0mm}
\bsi^{\sigma}(\beta'_B,x_\perp)\Gamma\psi^\sigma(\beta_B,y_\perp)~
=~e^{-{2c_F\over \pi b_0^2}\big[
\big({1\over\alpha_s(\tbe_\perp^{-1})}-{b_0\over 2}\ln[-i(\tau'_B+\ie)]\big)\ln{\alpha_s(\mu_{\sigma'})\over\alpha_s(\mu_{\sigma'_0})}
+{1\over \alpha_s(\mu_{\sigma'})}-{1\over \alpha_s(\mu_{\sigma'_0})}
\big]}
\nonumber\\
&&\hspace{0mm}
\times~e^{-{2c_F\over \pi b_0^2}\big[
\big({1\over\alpha_s(\tbe_\perp^{-1})}-{b_0\over 2}\ln[-i(\tau_B+\ie)]\big)\ln{\alpha_s(\mu_{\sigma})\over\alpha_s(\mu_{\sigma_0})}+{1\over \alpha_s(\mu_{\sigma})}-{1\over \alpha_s(\mu_{\sigma_0})}
\big]}
\bsi^{\sigma_0}(\beta'_B,x_\perp)\Gamma\psi^{\sigma_0}(\beta_B,y_\perp)
\label{qevolutionfinal}
\end{eqnarray}
where $\mu^2_\sigma\equiv b_\perp^{-1}\sqrt{\sigma|\beta_B|s}$, 
$\mu^2_{\sigma}\equiv b_\perp^{-1}\sqrt{\sigma|\beta'_B|s}$, $\tbe_\perp^2={b_\perp^2\over 2}e^{\gamma/2}$ and 
$\tau_B={\beta_B\over |\beta_B|},\tau'_B={\beta'_B\over |\beta'_B|}$. 
Equation (\ref{qevolutionfinal}) is one of the main results of this paper, another being the similar Eq.  (\ref{gevolution}) for
 gluon TMDs.

 \subsection{Quark loop contribution from light-cone expansion \label{qloopqlico}}
 
\begin{figure}[htb]
\begin{center}
\includegraphics[width=3.7in]{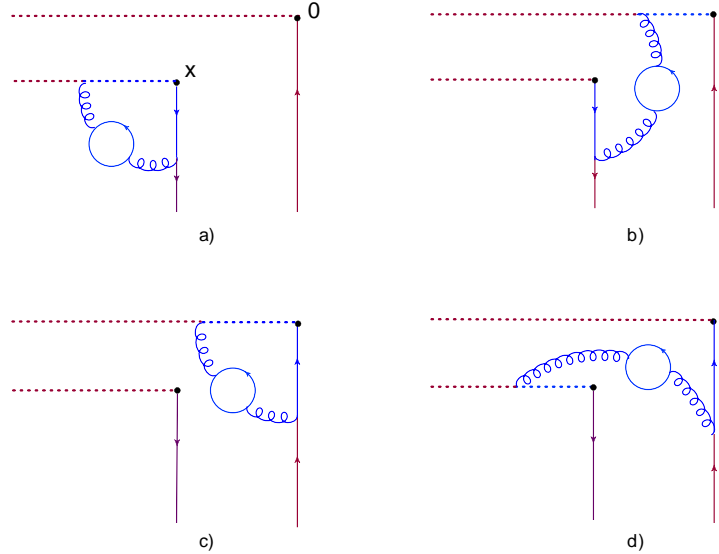}
\end{center}
\caption{Quark loop correction to quark TMD evolution.\label{fig:Tquark-running}}
\end{figure}

There is a simple way to check Eq. (\ref{evolrunpart1}). 
First, note that knowing the result  (\ref{evolrunpart1}), we can take $p_{B_\perp}=0$ from
the beginning. This means that our external field is on the mass shell so we can use 
the light-cone expansion (see e.g., Ref. \cite{Balitsky:1987bk}). 
\footnote{The reader may wonder why here we use the expansion at small $x_\perp^2$ while in other sections we say that $x_\perp^2$ 
is not small.  The reason is that the parameter of the near-light-cone expansion of Ref. \cite{Balitsky:1987bk} is $x^2 D^2$ and 
$D^2=0$ in our approximation so the first term of this light-cone expansion is sufficient for our purposes at any $x$, see Eq. (\ref{licoexp})
from Appendix \ref{sect:lico}.
 }
Second, as we demonstrated above, the $x^+$ dependence of  LHS. of Eq. (\ref{evolrunpart1}) 
is power suppressed so we can take $x^+=y^+$ from the beginning:
\begin{eqnarray}
&&\hspace{-0mm}
\sigma{d\over d\sigma}\langle[y^+,-\infty]_x[-\infty,y^+]_y\Gamma\psi(y^+,y_\perp,-\delta^-)\rangle_\Psi
~=~-\delta^-{d\over d\delta^-}\langle[y^+,-\infty]_x[-\infty,y^+]_y\Gamma\psi(y^+,y_\perp,-\delta^-)\rangle_\Psi^{\rm Fig1a-c~loop}
\end{eqnarray}
In this case, all relevant distances are spacelike so we can replace the product of operators 
in the matrix element in the LHS. by the T-product. Also, it is convenient to take $y_\perp=0$ and $y^+=0$. Thus, we need to calculate
\begin{eqnarray}
&&\hspace{-0mm}
{1\over b\alpha_s^2c_F}
\delta^-{d\over d\delta^-}\langle{\rm T}\{[0^+,-\infty]_x[-\infty,0^+]_0\Gamma\psi(0^+,0_\perp,-\delta^-)\}\rangle_\Psi^{\rm Fig1a-c~loop}
\nonumber\\
&&\hspace{-0mm}
=~2\delta^-{d\over d\delta^-}\Big[\!\int_{-\infty}^0\! dz^+(z^+,x_\perp|{\ln{\tmu^2\over -p^2}\over p^2}\Gamma\Psi{p^-\over p^2}|0^+,-\delta^-,0_\perp)~-~({x_\perp\rightarrow 0})\Big]
\label{klico1}
\end{eqnarray}
in the background field
$$
\Psi(z^+)~=~ \int\!\dhd\beta_B~e^{-i\vro\beta_B z^+}\Psi(\beta_B)
 $$ 
 where we used the BLM prescription (\ref{luprop}) for the Feynman gluon propagator. 
 Hereafter  we use Schwinger's notations defined as
 \begin{equation}
 (x|f(p)|y)=\int\!\dhd p~e^{-ip(x-y)}f(p), ~~~~~(x|\Psi|y)=\Psi(x)\delta(x-y)
\label{schwinger}
\end{equation}
and similarly $(x|A|y)=A(x)\delta(x-y)$ for future calculations in the gluon background. The relevant diagrams are shown 
in Fig. \ref{fig:Tquark-running}.

It is convenient to rewrite Eq. (\ref{klico1}) as follows
\begin{eqnarray}
&&\hspace{-0mm}
{\rm RHS~of~Eq.~(\ref{klico1})}~=
~-2i\delta^-{d\over d\delta^-}\Big[\!\int_{-\infty}^0\! dz^+(z^+,x_\perp|{\ln{\tmu^2\over -p^2}\over p^2}
\partial^-\Psi{1\over p^2}|0^+,-\delta^-,0_\perp)~-~({x_\perp\rightarrow 0})\Big]
\nonumber\\
&&\hspace{-0mm}
~+2i\delta^-{d\over d\delta^-}\Big[(0^+,x_\perp|{\ln{\tmu^2\over -p^2}\over p^2}\Psi{1\over p^2}|0^+,-\delta^-,0_\perp)~-~({x_\perp\rightarrow 0})\Big]
\label{fla65}
\end{eqnarray}
Since $\partial^2\Psi(x^+)=0$  we can  use light-cone expansion (\ref{licoexpanshen}) from Sec. \ref{sect:lico}.  First, note that
 the second term in the RHS of Eq. (\ref{fla65})  can be omitted since the light-cone expansion of the expression in the square brackets
depends only on $x_\perp^2$  and does not depend on $\delta^-$. Second, using Eq. (\ref{clicon1}) for the first line in Eq. (\ref{fla65}) we
get
\begin{eqnarray}
&&\hspace{-0mm}
{\rm RHS~of~Eq.~(\ref{klico1})}~=
~\delta^-{d\over d\delta^-}\Big[\!\int_{-\infty}^0\! dz^+
~{\Gamma\big(\ve)\over 8\pi^{d\over 2}(x_\perp^2-2z^+\delta^-)^\ve}
\label{fla66}\\
&&\hspace{-0mm}
\times~
\Big(\big[\ln{\tmu^2(x_\perp^2-2z^+\delta^-)\over 4}+{1\over \ve}-\psi(1+\ve)+\gamma \big]\!\int_0^1\! du~\partial^-\Psi(uz^+)
+\!\int_0^1\! du~\ln\baru~\partial^-\Psi(uz^+)\Big)
~-~({x_\perp\rightarrow 0})\Big]
\nonumber
\end{eqnarray}
Next, using formulas (\ref{glavormula}) and (\ref{glavormulas})
one obtains
\begin{eqnarray}
&&\hspace{-0mm}
{\rm RHS~of~Eq.~(\ref{klico1})}~=
~-{1\over 8\pi^{d\over 2}}\!\int_{-\infty}^0\! dz^+
~{\Gamma\big(\ve)\over (x_\perp^2-2z^+\delta^-)^\ve}
\Big[\ln{-\tmu^2(x_\perp^2-2z^+\delta^-)\over 4}+{1\over \ve}-\psi(1+\ve)-\psi(1)
\label{licokv1}\\
&&\hspace{-0mm}
-\!\int_0^1{dt\over 1-t}\ln{x_\perp^2-{2\over t}{x'}^+\delta^-\over x_\perp^2-2{x'}^+\delta^-}\Big]\partial^-\Psi(z^+)
~-~({x_\perp\rightarrow 0})
\nonumber\\
&&\hspace{-0mm}
=~{1\over 8\pi^{d\over 2}}\!\int_{-\infty}^0\! dz^+ ~{\partial\Psi(z^+)\over\partial z^+}\Big[\half\ln^2(x_\perp^2-2z^+\delta^-)+\ln(x_\perp^2-2z^+\delta^-)\big(\ln{\tmu^2\over 4}+2\gamma\big)-{\rm Li}_2\Big( {x_\perp^2\over x_\perp^2-2z^+\delta^-}\Big) \Big]
~-~({x_\perp\rightarrow 0})
\nonumber\\
&&\hspace{-0mm}
=~{1\over 16\pi^{d\over 2}}\!\int_{-\infty}^0\! dz^+ ~{\partial\Psi(z^+)\over\partial z^+}\Big[\ln{x_\perp^2-2z^+\delta^-\over -2z^+\delta^-}
\Big(\ln\tmu^2{x_\perp^2-2z^+\delta^-\over 4}+\ln\tmu^2{-2z^+\delta^-\over 4}+4\gamma\Big)-2{\rm Li}_2\Big( {x_\perp^2\over x_\perp^2-2z^+\delta^-}\Big) \Big]
\nonumber
\end{eqnarray}
To compare to Eq. (\ref{evolrunpart1}) we should take $\Psi(z^+)=\int\!\dhd\beta_Be^{-i\vro\beta_Bz^+}\Psi(\beta_B)$. After some algebra we get
\begin{eqnarray}
&&\hspace{-0mm}
-i\vro\beta_B\!\int_{-\infty}^0\! dz^+ ~e^{-i\vro\beta_Bz^+}
\Big[
\half\ln^2(-2z^+\delta^-)+\ln(-2z^+\delta^-)\big(\ln{\tmu^2\over 4}+2\gamma\big)\Big]
\nonumber\\
&&\hspace{-0mm}
=~
-\big[\ln {-i\beta_B+\epsilon\over 2 \delta^-}\vro+\gamma\big]\big(\ln{\tmu^2\over 4}+2\gamma\big)
+\half[\ln{(-i\beta_B+\epsilon)\vro\over 2 \delta^-}+\gamma]^2
+{\pi^2\over 12}
\label{furintegral1}
\end{eqnarray}
Also,  
\begin{eqnarray}
&&\hspace{-0mm}
-i\vro\beta_B\!\int_{-\infty}^0\! dz^+ ~e^{-i\vro\beta_Bz^+}\Big[
\half\ln^2(x_\perp^2-2z^+\delta^-)+\ln(x_\perp^2-2z^+\delta^-)\big(\ln{\tmu^2\over 4}+2\gamma\big)
-{\rm Li}_2\Big( {x_\perp^2\over x_\perp^2-2z^+\delta^-}\Big)\Big]
\nonumber\\
&&\hspace{-0mm}
=~
\half\ln^2x_\perp^2+\ln x_\perp^2\big(\ln{\tmu^2\over 4}+2\gamma\big)
-{\pi^2\over 6}~+~O\big({\delta^-\over \vro x_\perp^2\beta_B}\big)
\label{furintegral2}
\end{eqnarray}
 where we used the fact that  ${2z^+\delta^-\over x_\perp^2}\sim {1\over \sigma\beta_Bs}\sim {m_\perp^2\over\sigma\beta_Bs}\ll 1$.
 
Using these two integrals for Eq. (\ref{licokv1}) we get
\begin{eqnarray}
&&\hspace{-0mm}
\delta^-{d\over d\delta^-}\langle{\rm T}\{[0^+,-\infty]_x[-\infty,0^+]_0\Gamma\psi(0^+,0_\perp,-\delta^-)\}\rangle_\Psi^{\rm Fig1a-c~loop}
\nonumber\\
&&\hspace{-0mm}
=~{b\alpha_s^2c_F\over 16\pi^2}\!\int\!\dhd\beta_B\Psi(\beta_B)
\Big[\big(\ln{x_\perp^2\vro\over 2\delta^-}[-i\beta_B+\epsilon] +\gamma\big)
\big(\ln{x_\perp^2\tmu^4\delta^-\over 8(-i\beta_B+\epsilon)\vro} +3\gamma\big)-{\pi^2\over 2}
~+~O\big({\delta^-\over \vro x_\perp^2\beta_B}\big)\Big]
\label{licotvet1}
\end{eqnarray}
which agrees with Eq. (\ref{evolrunpart1}). We will use this method for calculation of quark loops in the gluon case below.

 \section{Evolution of quark TMDs with gauge links out to $+\infty$ \label{EvolutionPlusInf}}
 
The calculation of diagrams with Wilson lines going to $+\infty$ repeats that the 
of $-\infty$ case. For the diagrams Fig. \ref{fig:kvdiagrams}(a)-\ref{fig:kvdiagrams}(c) 
with gauge links out to $+\infty$, instead of Eq. (\ref{virt}), we get
\begin{eqnarray}
&&\hspace{-1mm}
\langle[\infty,y^+]_y\Gamma\psi(y^+,y_\perp,\delta^-)\rangle_{\Psi}^{\rm Fig1c~loop}
~=~-ig^2c_F\!\int\! \dhd\beta_B\dhd p_{B_\perp}
\nonumber\\
&&\hspace{-1mm}
\times~e^{-ip_By}\!\int\!\dhd\alpha\dhd\beta\dhd p_\perp{1\over \beta-\ie}{e^{i\alpha\vro\delta^-}\over \alpha\beta s-p_\perp^2+\ie}
{s(\beta-\beta_B)\over \alpha(\beta-\beta_B)s-(p-p_B)_\perp^2+\ie}\Gamma\Psi(\beta_B,p_{B_\perp})
\nonumber\\
&&\hspace{-1mm}
=~-g^2c_F\!\int\! \dhd\beta_B\dhd p_{B_\perp}e^{-ip_By}\!\int_0^\infty\!\dhd\alpha\!\int\!\dhd p_\perp
{\beta_Bse^{i{\alpha\over\sigma}}\over p_\perp^2[\alpha\beta_Bs+(p-p_B)_\perp^2-\ie]}\Gamma\Psi(\beta_B,p_{B_\perp})
\label{virtplus}
\end{eqnarray}
and in place of Eq. (\ref{reals}) or Eqs. (\ref{repart1}) and  (\ref{repart2})
\begin{eqnarray}
&&\hspace{-1mm}
\langle[x^+,\infty]_x\Gamma\psi(y^+,y_\perp,\delta^-)\rangle_{\Psi}^{\rm Fig1a,b~loop}~
\nonumber\\
&&\hspace{-1mm}
=~g^2c_F\!\int\! \dhd\beta_B\dhd p_{B_\perp}e^{-i(p_B,y)}\!\int\!\dhd\alpha\dhd\beta\dhd p_\perp\Big[2\pi\delta(\alpha(\beta-\beta_B)s-(p-p_B)_\perp^2)
(\beta-\beta_B)s\theta(\alpha) 
{1\over \alpha\beta s-p_\perp^2-\ie}
\nonumber\\
&&\hspace{22mm}
+~{(\beta-\beta_B)s\over \alpha(\beta-\beta_B)s-(p-p_B)_\perp^2+\ie}
2\pi\delta( \alpha\beta s-p_\perp^2)\theta(\alpha)\Big]{e^{i\alpha\vro\delta^-}\over \beta-\ie}e^{-i\beta\Delta+i(p,\Delta)_\perp}\Gamma\Psi(\beta_B,p_{B_\perp})
\nonumber\\
&&\hspace{-1mm}
=~g^2c_F\!\int\!\dhd\beta_B\dhd p_{B_\perp}\psi(\beta_B,p_{B_\perp}) e^{-ip_By}
\!\int_0^\infty\!\dhd\alpha~e^{i{\alpha\over\sigma}}\!\int\!\dhd p_\perp ~\bigg(
{\beta_Bs\big(e^{-i{p_\perp^2\over\alpha s}\vro\Delta^+ +i(p,\Delta)_\perp}-1\big)
\over  p_\perp^2[\alpha\beta_Bs+(p-p_B)_\perp^2-\ie]}
\nonumber\\
&&\hspace{44mm}
+~{(p-p_B)_\perp^2e^{i(p,\Delta)_\perp}\big[e^{-i(\beta_B+{(p-p_B)_\perp^2\over\alpha s})\vro\Delta^+}-e^{-i{p_\perp^2\over\alpha s}\vro\Delta^+}\big]
\over \alpha[\alpha\beta_B s+(p-p_B)_\perp^2+\ie][\alpha\beta_B s+(p-p_B)_\perp^2-p_\perp^2]}\bigg)\Gamma\Psi(\beta_B,p_{B_\perp})\,.
\label{realplus}
\end{eqnarray}
Note that for gauge links out to $+\infty$ the sign of cutoff $\delta^-$ does not matter: the IR cancellation occurs at whatever $\delta$. 
As discussed above, we  choose the sign of splitting in such a way that all relevant distances in the operators are spacelike. 
With this sign of point splitting in the ``$-$'' direction we can  use the complex conjugate versions of 
 integrals (\ref{eiknapravotvet})-(\ref{loopintegralotvet1}) in the Appendix \ref{app:integrals}.  

Similarly, one can easily demonstrate that the contribution of diagrams in Fig.\ref{fig:kvdiagrams}(d)-\ref{fig:kvdiagrams}(f)
with gauge links out to $+\infty$ differs from Eq. (\ref{evolqminus}) 
by replacements $\beta_B+\ie\rightarrow\beta_B-\ie$ and $\sigma\rightarrow -\sigma$.  Repeating the analysis of Sec. \ref {rapidityEvolqTMD}
and using (complex conjugate) integral (\ref{evolint}) from Appendix \ref{app:integrals} 
we get  the evolution equation for quark TMDs with gauge links out to $+\infty$ in the form
\begin{eqnarray}
&&\hspace{-0mm}
\Big(\sigma{d\over d\sigma}+\sigma'{d\over d\sigma'}\Big)
\bsi^{\sigma'}(\beta'_B,x_\perp)\Gamma\psi^\sigma(\beta_B,y_\perp)
\nonumber\\
&&\hspace{2mm}
=~-{g^2\over 8\pi^2}c_F\!\int\!\dhd\beta_B\dhd\beta'_B\dhd p_{B_\perp}\dhd p'_{B_\perp} 
\bsi^{\sigma'}(\beta'_B,x_\perp)\Gamma\psi^\sigma(\beta_B,y_\perp)e^{-i\beta'_B\vro x^+-i\beta_B\vro y^+}
\nonumber\\
&&\hspace{2mm}
\times~\Big[\ln\Big({i\over 4}(\beta_B-\ie)\sigma sb_\perp^2e^\gamma\Big)
+\ln\Big({i\over 4}(\beta'_B-\ie)\sigma' sb_\perp^2e^\gamma\Big)\Big]
~+~O\Big({m_\perp^2\over\beta_B\sigma s},{m_\perp^2\over\beta'_B\sigma' s}\Big)
\label{tmdevolplus}
\end{eqnarray}
and the solution is
\begin{eqnarray}
&&\hspace{-0mm}
\bsi^{\sigma'}(\beta'_B,x_\perp) \Gamma\psi^\sigma(\beta_B,y_\perp)
\label{evolsolplus}\\
&&\hspace{0mm}
=~e^{ -{\alpha_sc_F\over 4\pi}\ln{\sigma'\over\sigma'_0}\big[\ln{\sigma'\sigma'_0}+2\ln \big({i\over 4}(\beta'_B-\ie)sb_\perp^2e^\gamma\big)\big]}
\bsi^{\sigma'_0}(\beta'_B,x_\perp) \Gamma\psi^{\sigma_0}(\beta_B,y_\perp)
e^{- {\alpha_sc_F\over 4\pi}\ln{\sigma\over\sigma_0}\big[\ln{\sigma\sigma_0}
+2\ln \big({i\over 4}(\beta'_B-\ie)sb_\perp^2e^\gamma\big)\big]}
\nonumber
\end{eqnarray}
where $\bsi^{\sigma'}(\beta'_B,x_\perp)$ and $\psi^\sigma(\beta_B,y_\perp)$ are given by 
formulas (\ref{Psidef}) with gauge links out to $+\infty$:
\begin{eqnarray}
&&\hspace{-0mm}
\bsi^\sigma(x^+,x_\perp) ~\equiv~\bsi\big(x^+,x_\perp,-{1\over\rho\sigma}\big)[x^+,-\infty]_x,~~~~
\psi^\sigma(y^+,y_\perp) ~\equiv~[-\infty,y^+]_y\psi\big(y^+,y_\perp,-{1\over\rho\sigma}\big),
\label{Psidefplus}
\end{eqnarray}

In the coordinate space Eq. (\ref{evolsolplus}) corresponds to
\begin{eqnarray}
&&\hspace{-0mm}
\bsi^{\sigma'}(x^+,x_\perp) \Gamma\psi^\sigma(y^+,y_\perp) 
~=~{1\over 4\pi^2}e^{- {\alpha_sc_F\over 4\pi}\big(\ln{\sigma'\over\sigma'_0}\ln{\sigma'\sigma'_0}+\ln{\sigma\over\sigma_0}\ln{\sigma\sigma_0}\big)}
\label{tmdcoordevolplus}\\
&&\hspace{0mm}
\times ~\int\!dz^+
\bigg[{i\Gamma\big(1-{\alpha_sc_F\over 2\pi}\ln{\sigma'\over \sigma'_0}\big)
\over (x^+-z^++\ie)^{1-{\alpha_sc_F\over 2\pi}\ln{\sigma'\over \sigma'_0}}}
+{\rm c.c.}\bigg]
\int\!dw^+\bigg[{i\Gamma\big(1-{\alpha_sc_F\over 2\pi}\ln{\sigma\over \sigma_0}\big)
\over (y^+-w^++\ie)^{1-{\alpha_sc_F\over 2\pi}\ln{\sigma\over \sigma_0}}}
+{\rm c.c.}\bigg]
\bsi^{\sigma'_0}(x^+,x_\perp) \Gamma\psi^{\sigma_0}(y^+,y_\perp) 
\nonumber
\end{eqnarray}
Now we have $z^+\geq x^+$ and $w^+\geq y^+$ so the evolution goes out to $+\infty$.

For completeness, let us present the final result for the evolution with running coupling which is obtained from Eq. (\ref{qevolutionfinal}) by replacement $-i\tau_B+\epsilon$ to $i\tau_B+\epsilon$
\begin{eqnarray}
&&\hspace{-0mm}
\bsi^{\sigma}(\beta'_B,x_\perp)\Gamma\psi^\sigma(\beta_B,y_\perp)~
=~e^{-{2c_F\over \pi b_0^2}\big[
\big({1\over\alpha_s(\tbe_\perp^{-1})}-{b_0\over 2}\ln[i(\tau'_B-\ie)]\big)\ln{\alpha_s(\mu_{\sigma'})\over\alpha_s(\mu_{\sigma'_0})}
+{1\over \alpha_s(\mu_{\sigma'})}-{1\over \alpha_s(\mu_{\sigma'_0})}
\big]}
\nonumber\\
&&\hspace{0mm}
\times~e^{-{2c_F\over \pi b_0^2}\big[
\big({1\over\alpha_s(\tbe_\perp^{-1})}-{b_0\over 2}\ln[i(\tau_B-\ie)]\big)\ln{\alpha_s(\mu_{\sigma})\over\alpha_s(\mu_{\sigma_0})}
+{1\over \alpha_s(\mu_{\sigma})}-{1\over \alpha_s(\mu_{\sigma_0})}
\big]}
\bsi^{\sigma_0}(\beta'_B,x_\perp)\Gamma\psi^{\sigma_0}(\beta_B,y_\perp)
\label{qevolution2}
\end{eqnarray}
%

\section{Rapidity evolution of gluon TMDs \label{rapiditygTMD}}
  \subsection{Leading-order contribution}

 The gluon TMD is defined by the operator (\ref{gtmdop})
\begin{equation}
\hspace{-1mm}
F^{-i}(x^+,x_\perp)[x,x\pm\infty n]~[x_\perp,y_\perp]_{\pm\infty n}[\pm\infty n+y,y]F^{-j}(y^+,y_\perp)
\label{gtmdope}
\end{equation}
The typical process determined by gluon TMD (with gauge links  out to $-\infty$) is 
Higgs production by gluon-gluon fusion in the Sudakov region. If one approximates $t$-quark loop by a 
 point vertex, the differential cross section is determined by  the ``hadronic tensor''  given by the formula similar to Eq. (\ref{TMDfakt}) with gluon TMDs \cite{Mulders:2000sh}
\begin{eqnarray}
&&\hspace{-0mm}
G(x_B,z_\perp,\eta)~=~{-x_B^{-1}\over 2\pi p^-}\!\int\! dz^+ ~e^{-ix_Bp^-z^+}
\langle P|F^{-i,a}(z^+,z_\perp)([z,z \pm\infty n]~[z_\perp,0_\perp]_{\pm\infty n}[\pm\infty n,0])^{ab}F^{-,b}_{~~i}(0)|P\rangle
\label{TMDg}
\end{eqnarray}
 in place of quark ones (see the discussion in Ref. \cite{Balitsky:2017flc}). 
 \footnote{It should also be noted that  at small $x$ the Sudakov double logs for Higgs production in $pA$ collisions were studied  in Refs. \cite{Mueller:2012uf,Mueller:2013wwa} using the $k_T$ factorization approach.}
 
The leading-order rapidity evolution was found in Ref.  \cite{Balitsky:2019ayf}.  Here we first repeat the 
LO derivation and then obtain the running-coupling correction by the BLM prescription. Similar to the quark case,
we define rapidity-regularized operators by
\begin{equation}
\hspace{-0mm}
\tilde\scrf^{\sigma,a}_i(x_\perp,x^+)~=~(F^{-}_{~~i})^b(x^+,x_\perp, -\delta^-)[x^+,-\infty]_x^{ba},~~~~~
\scrf^{\sigma,a}_i(y_\perp,y^+)~=~[-\infty,y^+]_y^{ab}(F^{-}_{~~i})^b(x^+,x_\perp, -\delta'^-)
\label{Fdef}
\end{equation}
where $\delta^-={1\over\vro\sigma}$. Let us emphasize again that we use the point-splitting operators in the RHSs 
 only for the perturbative calculations.
 
Similar to the quark case considered above, to find the LO evolution equation we calculate diagrams in  the 
background field $A^{\rm ext}_\mu(x_+,x_\perp)$ and use point splitting for regularization of rapidity divergences,
\begin{equation}
\hspace{-1mm}
\Big(\sigma{d\over d\sigma}+\sigma'{d\over d\sigma'}\Big)
\langle F^{-i,a}(x^+,x_\perp, -\delta'^-)[x^+,-\infty]_x^{ab} ~[x_\perp,y_\perp]_{-\infty^+}^{bc}
[-\infty,y^+]_y^{cd}F^{-,d}_{~~i}(y^+,y_\perp,-\delta^-)\rangle_\cala\,.
\label{gtmdael}
\end{equation}
Also, we use the $A^-_{\rm ext}=0$ gauge for the background field  
and background-Feynman (bF) gauge for quantum gluons. As we mentioned above, in such a gauge the contribution
of gauge link $[x_\perp,y_\perp]_{-\infty^+}$ can be neglected. Moreover, in the bF gauge the product
$gA^{\rm ext}_\mu$ is renorm invariant so there is no need to consider self-energy diagrams,  and the one-loop evolution 
of the operator (\ref{gtmdop})
looks the same as in Fig. \ref{fig:kvdiagrams} but with gluons instead of quarks. We will also use the notation $\cala_\mu\equiv gA^{\rm ext}_\mu$
and $\calf_{\mu\nu}=\partial_\mu\cala_\nu-\partial_\nu\cala_\mu-i[\cala_\mu,\cala_\nu]$.
\begin{figure}[htb]
\begin{center}
\includegraphics[width=131mm]{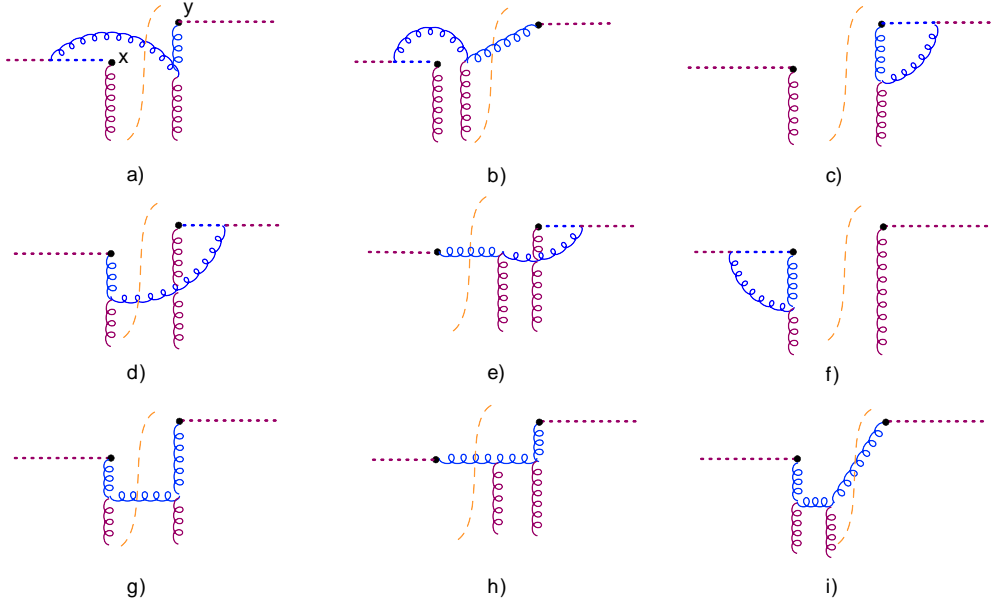}
\end{center}
\caption{One-loop diagrams for gluon TMD operator (\ref{gtmdop}) in the background gluon field.\label{fig:gldiagrams}}
\end{figure}
Gluon propagators in the bF gauge are
\begin{eqnarray}
&&\hspace{-0mm} 
\langle {\rm T}\{A^a_\mu(x)A^b_\nu(y)\}\rangle~=~(x|{-i\over \calp^2g_{\mu\nu}+2i\calf_{\mu\nu}+\ie}|y)^{ab}~=~
(x|{-ig_{\mu\nu}g^{ab}\over p^2+\ie}|y) -2(x|{1\over p^2+\ie}\calf^{ab}_{\mu\nu}{1\over p^2+\ie}|y)~+~O(\calf^2)
\nonumber\\
&&\hspace{-0mm} 
\langle \tilde{\rm T}\{A^a_\mu(x)A^b_\nu(y)\}\rangle~=~(x|{i\over \calp^2g_{\mu\nu}+2i\calf_{\mu\nu}-\ie}|y)^{ab}~=~
(x|{ig_{\mu\nu}g^{ab}\over p^2-\ie}|y) +2(x|{1\over p^2-\ie}\calf^{ab}_{\mu\nu}{1\over p^2-\ie}|y)~+~O(\calf^2)
\nonumber\\
&&\hspace{-0mm} 
\langle A^a_\mu(x)A^b_\nu(y)\rangle
~=~-(x|{1\over \calp^2g_{\mu\xi}+2i\calf_{\mu\xi}-\ie}p^2\dbar(p^2)\theta(p_0)p^2{1\over \calp^2g_{\xi\nu}+2i\calf_{\xi\nu}+\ie}|y)^{ab}
\label{gluprops}\\
&&\hspace{-0mm} 
=~
 -g_{\mu\nu}g^{ab}(x|2\pi\delta(p^2)\theta(p_0)|y)+4\pi i(x|{1\over p^2-\ie}\big[\calf_{\mu\nu}^{ab}\delta(p^2)\theta(p_0)
+\delta(p^2)\theta(p_0)\calf^{ab}_{\mu\nu}\big]{1\over p^2+\ie}|y)
~+~O(\calf^2)
\nonumber
\end{eqnarray}
Here $a$ and $b$ are adjoint indices and 
$$(x|f(p)|y)=\int\!\dhd p~e^{-p(x-y)}f(p), ~~~~~(x|f(\cala)|y)=f(\cala(x))\delta(x-y)$$
are Schwinger's notations for propagators in background fields. 
 
Let us start with diagrams in Fig.  \ref{fig:gldiagrams}(a)-\ref{fig:gldiagrams}(c). 
The contribution of the virtual diagram in Fig.  \ref{fig:gldiagrams}(c) is
\begin{eqnarray}
&&\hspace{-0mm} 
\langle \scrf^{\sigma,a}(y^+,y_\perp)\rangle_\cala^{\rm Fig.~\ref{fig:gldiagrams}c}
~=~-ig^2\!\int_{-\infty}^{y^+}\! dy'^+ \langle A^{-,ab}(y'^+,y_\perp)(D^- A^{j,b}-D^j A^{-,b})(y^+,y_\perp,-\delta^-)\rangle_\cala
\label{VFvirtua}\\
&&\hspace{-0mm} 
=~-2g^2N_c\!\int_{-\infty}^{y^+}\!\! dy'^+\vro~(y'^+,y_\perp|{1\over p^2+\ie}\calf^{-j,a}{1\over p^2+\ie}p^-|y^+,y_\perp,-\delta^-)
\nonumber\\
&&\hspace{-0mm}
=-2ig^2N_c(y|{1\over\beta+\ie}{1\over p^2+\ie}\calf^{-j,a}{\beta\over p^2+\ie}|y-\delta^-)
\nonumber\\
&&\hspace{-0mm}
=~-isg^2N_c\!\int\!\dhd\beta_B\dhd p_{B_\perp}\calf^{-j,a}(\beta_B,p_{B_\perp}) e^{-ip_By}\!\int\!\dhd\alpha{\dhd\beta\over\beta+\ie}~
\!\int\!\dhd p_\perp{(\beta-\beta_B)e^{-i\alpha\vro \delta^-}\over(\alpha\beta s-p_\perp^2+\ie)(\alpha(\beta-\beta_B) s-(p-p_B)_\perp^2+\ie)}
\nonumber
\end{eqnarray}
where $\calf^{-i}(\beta_B,p_{B_\perp})$ is the Fourier transform of the background field:
\begin{eqnarray}
&&\hspace{-0mm}
\calf^{-i}(\beta_B,p_{B_\perp})~=~\!\int\! dz^+dz_\perp~\calf^{-i}(z^+,z_\perp)~e^{i\beta_B\vro z^+-i(p_B,z)_\perp}.~~~~~~~
\label{normi}
\end{eqnarray}
It is worth noting that similar to the quark case, in a general gauge one should replace background fields by
\begin{eqnarray}
&&\hspace{-0mm}
\calf^a_{-i}(z^+,z_\perp)~\rightarrow~[z^+,\pm\infty^+]^{ab}\calf^b_{-i}(z^+,z_\perp)
\end{eqnarray}
where the direction of Wilson lines corresponds to the choice of $\pm\infty$ in Eq. (\ref{gtmdope}).

The integral over momenta in Eq. (\ref{VFvirtua}) is the same as in Eq. (\ref{virt}) so we get
\begin{eqnarray}
&&\hspace{-0mm} 
\langle \scrf^{\sigma,a}_j(y_\perp,y^\perp)\rangle^{\rm Fig.~\ref{fig:gldiagrams}c}_\cala
~=~-ig^2\!\int_{-\infty}^{y^+}\! dy'^+ ~A^{-,ab}(y'^+,y_\perp, -\delta^-)(D^- A^{j,b}-D^j A^{-,b})(y^+,y_\perp,-\delta^-)
\label{VFvirtual}\\
&&\hspace{-0mm} 
=~-g^2N_c\!\int\! \dhd\beta_B\dhd p_{B_\perp}e^{-ip_By}\calf^{-j,a}(\beta_B,p_{B_\perp})\!\int_0^{\infty}\!\dhd\alpha\!\int\!\dhd p_\perp
{\beta_Bs\over p_\perp^2[\alpha\beta_Bs+(p-p_B)_\perp^2+\ie]}e^{-i{\alpha\over\sigma}}
\end{eqnarray}

Next, consider diagrams in Fig. \ref{fig:gldiagrams}(a) and \ref{fig:gldiagrams}(b)
\begin{eqnarray}
&&\hspace{-0mm} 
\langle [x^+,-\infty]_x^{ba}F^{-j,a}(y^+,y_\perp,-\delta^-)\rangle_\cala^{\rm Fig.~\ref{fig:gldiagrams}a,b}
~=~ig^2\!\int_{-\infty}^{x^+}\! dx'^+ ~A^{-,ba}(x'^+,x_\perp)(D^- A^{j,a}-D^j A^{-,a})(y^+,y_\perp,-\delta^-)
\nonumber\\
&&\hspace{-0mm}  
=~-2ig^2N_c\!\int_{-\infty}^{x^+}\!\! dx'^+(x'|\big({1\over p^2-\ie}\calf^{-j,b}2\pi\delta(p^2)\theta(p_0)+2\pi\delta(p^2)\theta(p_0)\calf^{-j,b}{1\over p^2+\ie}\big)p^-|y) 
\nonumber\\
&&\hspace{-0mm}
=~2g^2N_c(x|{1\over\beta+\ie}\big({1\over p^2-\ie}\calf^{-j,b}\dbar(p^2)\theta(p_0)\beta+\dbar(p^2)\theta(p_0)\calf^{-j,b}{\beta\over p^2+\ie}\big)|y)
\nonumber\\
&&\hspace{-0mm}
=~sg^2N_c\!\!\int\!\!\dhd\beta_B\dhd p_{B_\perp}\calf^{-j,b}(\beta_B,p_{B_\perp}\!) e^{-ip_By}\!\!\int\!\!\dhd\alpha{\dhd\beta\over\beta+\ie}
~e^{-i\beta\vro\Delta^+}
\nonumber\\
&&\hspace{-0mm}
\times~\int\!\dhd p_\perp~e^{i(p,\Delta)_\perp}\Big[{(\beta-\beta_B)\theta(\alpha)\over\alpha\beta s-p_\perp^2-\ie}
2\pi\delta[(\beta-\beta_B)\alpha s-(p-p_B)_\perp^2]
+{(\beta-\beta_B)\theta(\alpha)2\pi\delta(\alpha\beta s-p_\perp^2)\over\alpha(\beta-\beta_B) s-(p-p_B)_\perp^2+\ie}\Big]
\label{VFreal}
\end{eqnarray}
The integral over momenta is the same as in the quark case [see Eq. (\ref{reals})] so similar to Eqs. (\ref{repart1}) and (\ref{repart2}) we get
\begin{eqnarray}
&&\hspace{-0mm} 
\langle [x^+,-\infty]_x^{ba}F^{-j,a}(y^+,y_\perp,-\delta^-)\rangle_\cala
\nonumber\\
&&\hspace{-0mm}
=~g^2N_c\!\!\int\!\!\dhd\beta_B\dhd p_{B_\perp}\calf^{-j,b}(\beta_B,p_{B_\perp}\!) e^{-ip_By}
\!\int_0^\infty\!\dhd\alpha\!\int\!\dhd p_\perp \Big\{
{\beta_Bse^{-i{p_\perp^2\over\alpha s}\vro\Delta^+ +i(p,\Delta)}
\over p_\perp^2[\alpha\beta_Bs+(p-p_B)_\perp^2+\ie[}e^{-i{\alpha\over\sigma}}
\nonumber\\
&&\hspace{-0mm}
+~
{(p-p_B)_\perp^2e^{i(p,\Delta)_\perp}\big[e^{-i(\beta_B+{(p-p_B)_\perp^2\over\alpha s})\vro\Delta^+}-e^{-i{p_\perp^2\over\alpha s}\vro\Delta^+}\big]
\over [\alpha\beta_B s+(p-p_B)_\perp^2+\ie][\alpha\beta_B s+(p-p_B)_\perp^2-p_\perp^2]}e^{-i{\alpha\over\sigma}}\Big\}
\label{realglu}
\end{eqnarray}
and therefore
\begin{eqnarray}
&&\hspace{-0mm} 
\langle [x^+,-\infty]_x^{ab}[-\infty,y^+]_y^{bc}F^{-j,c}(y^+,y_\perp,-\delta^-)\rangle_\cala^{\rm Fig.~\ref{fig:gldiagrams}a-c}
\label{glnapravo}\\
&&\hspace{22mm}
=~g^2N_c\!\!\int\!\!\dhd\beta_B\dhd p_{B_\perp}\calf^{-j,b}(\beta_B,p_{B_\perp}\!) e^{-ip_By}
\!\int_0^\infty\!\dhd\alpha~e^{-i{\alpha\over\sigma}}\!\int\!\dhd p_\perp ~\bigg(
{\beta_Bs\big(e^{-i{p_\perp^2\over\alpha s}\vro\Delta^+ +i(p,\Delta)_\perp}-1\big)
\over  p_\perp^2[\alpha\beta_Bs+(p-p_B)_\perp^2+\ie]}
\nonumber\\
&&\hspace{44mm}
+~{(p-p_B)_\perp^2e^{i(p,\Delta)_\perp}\big[e^{-i(\beta_B+{(p-p_B)_\perp^2\over\alpha s})\vro\Delta^+}-e^{-i{p_\perp^2\over\alpha s}\vro\Delta^+}\big]
\over \alpha[\alpha\beta_B s+(p-p_B)_\perp^2+\ie][\alpha\beta_B s+(p-p_B)_\perp^2-p_\perp^2]}\bigg)
\nonumber\\
\end{eqnarray}
The integral is the same as in Eq. (\ref{napravo}) for the quark case, so similar to Eq. (\ref{napravotvet}) 
we get the contribution of diagrams in Fig.  \ref{fig:gldiagrams}(a)-\ref{fig:gldiagrams}(c) in the form
\begin{eqnarray}
&&\hspace{-0mm}
\sigma{d\over d\sigma}\langle [x^+,-\infty]_x^{ab}[-\infty,y^+]_y^{bc}gF^{-j,c}\big(y^+,y_\perp,-{1\over\vro\sigma}\big)
\rangle_\cala^{\rm Fig. ~ \ref{fig:gldiagrams}a-c}
\nonumber\\
&&\hspace{2mm}
=~-{g^2\over 8\pi^2}N_c\!\int\!\dhd\beta_B~\calf^{-j,b}(\beta_B,y_\perp) e^{-i\beta_B\vro y^+}\ln\Big(-{i\over 4}(\beta_B+\ie)\sigma s\Delta_\perp^2e^\gamma\Big)
~+~O\big({m_\perp^2\over\beta_B\sigma s}\big)
\label{napravotvetg}
\end{eqnarray}
where
\begin{eqnarray}
&&\hspace{-0mm}
\calf^{-i}(\beta_B,z_\perp)~=~\!\int\! dz^+~\calf^{-i}(z^+,z_\perp)e^{i\beta_B\vro z^+}~~~~~~~
\label{normir}
\end{eqnarray}
in accordance with Eq. (\ref{normi}).

A similar calculation of diagrams in Fig. \ref{fig:gldiagrams}(d)-\ref{fig:gldiagrams}(f) yields
\begin{eqnarray}
&&\hspace{-0mm} 
\langle F^{-i,a}(x^+,x_\perp,-\delta'^-)[x^+,-\infty]_x^{ab}[-\infty,y^+]_y^{bc}\rangle_\cala
\nonumber\\
&&\hspace{-0mm}
=~\langle (D^- A^{i,a}-D^i A^{-,a})(x^+,x_\perp,-\delta'^-)ig^2
\bigg[\!\int_{-\infty}^{x^+}\! dx'^+A^{-,ac}(x^+,x_\perp)
-\!\int_{-\infty}^{y^+}\! dy'^+A^{-,ac}(y^+,y_\perp)\bigg]\rangle_\cala
\label{FV}\\
&&\hspace{11mm} 
=~g^2N_c\!\int\!\dhd\beta_B\dhd p_{B_\perp}
e^{-ip_Bx}g\calf^{-i,c}(\beta_B,p_{B_\perp})\!\int\!\dhd p_\perp
\!\int_0^{\infty}\!\dhd\alpha~e^{i{\alpha\over \sigma'}}\bigg(
{\beta_B s\big(e^{i(p,\Delta)_\perp-i{p_\perp^2\over\alpha s}\vro\Delta^+}-1\big)\over \alpha\beta_B s-(p+p_B)_\perp^2+\ie}
\nonumber\\
&&\hspace{22mm}
+~{(p+p_B)_\perp^2
\big[e^{i(p,\Delta)_\perp}e^{-i{(p+p_B)_\perp^2\over\alpha s}\vro\Delta^+
+i\beta_B\vro\Delta^+}-e^{-i{p_\perp^2\over\alpha s}\vro\Delta^+}\big]
\over\alpha[\alpha\beta_B s+p_\perp^2-(p+p_B)_\perp^2+\ie][\alpha\beta_B s-(p+p_B)_\perp^2+\ie]}
\bigg)
\end{eqnarray}
and therefore [see Eq. (\ref{evolqminus})] we get
\begin{eqnarray}
&&\hspace{-0mm}
\sigma'{d\over d\sigma'}\langle gF^{-i,a}\big(x^+,x_\perp,-{1\over\vro\sigma'}\big)
[x^+,-\infty]_x[-\infty,y^+]_y\rangle_\Psi^{\rm Fig. ~\ref{fig:gldiagrams}d-f}
\nonumber\\
&&\hspace{2mm}
=~-{g^2\over 8\pi^2}N_c\!\int\!\dhd\beta_B\calf^{-i,c}(\beta_B,x_\perp)
e^{-i\beta_B\vro x^+}\ln\Big(-{i\over 4}(\beta_B+\ie)\sigma' s\Delta_\perp^2e^\gamma\Big)
~+~O\big({m_\perp^2\over\beta_B\sigma' s}\big)
\label{evolgminus}
\end{eqnarray}

Finally, let us discuss ``handbag'' diagrams in Figs. \ref{fig:gldiagrams}(g)-\ref{fig:gldiagrams}(i). 
Similar to the quark case, since the separation
between $x$ and $y$ is spacelike we can replace the product of operators in Eq. (\ref{gtmdael}) by the T-product and get
\begin{eqnarray}
&&\hspace{-0mm}
\langle {\rm T}\{F^{-i,a}(x^+,x_\perp, -\delta'^-)
F^{-,a}_{~~i}(y^+,y_\perp,-\delta^-)\}\rangle_\cala
~=-i(x|(\calp^-\delta^i_\xi-\calp^i\delta^-_\xi){1\over \calp^2g_{\xi\eta}+2ig\calf_{\xi\eta}+\ie}(\calp^-g_{i\eta}-\calp_i\delta^-_\eta)|y)^{aa}_{\rm Fig. \ref{fig:gldiagrams}g-i} 
\nonumber\\
&&\hspace{-0mm} 
=~2g^2(x|\calp^i{1\over\calp^2+\ie}\calf^{-i}{1\over\calp^2+\ie}\calf^-_{~~i}{1\over\calp^2+\ie}|y)^{aa}
=~
-4g^2N_c\!\int\!\dhd\beta_B\dhd p_{B_\perp}\dhd\beta'_B\dhd {p'}_{B_\perp}
e^{-ip'_Bx-ip_By}\calf^{-i,a}(\beta'_B,p'_{B_\perp})
\nonumber\\
&&\hspace{0mm}
\times~\calf^{-,a}_{~~i}(\beta_B,p_{B_\perp})
(x|{(p+p'_B)^i\over (p+p_B)^2+\ie}{1\over p^2+\ie}{(p-p_B)_i\over (p-p_B)^2+\ie}|y)
\nonumber\\
&&\hspace{-0mm}
=~
-2sg^2N_c\!\int\!\dhd\beta_B\dhd p_{B_\perp}\dhd\beta'_B\dhd {p'}_{B_\perp}
e^{-ip'_Bx-ip_By}\calf^{-i,a}(\beta'_B,p'_{B_\perp})\calf^{-,a}_{~~i}(\beta_B,p_{B_\perp})
\nonumber\\
&&\hspace{0mm}
\times~
\!\int\! \dhd\alpha\dhd\beta\dhd p_\perp~
e^{i\alpha\vro(\delta'-\delta)^- +i\beta\vro\Delta^+}{e^{-i(p,\Delta)_\perp}
(p+p'_B)^i(p-p_B)_i
\over(\alpha\beta s-p_\perp^2+\ie)[\alpha(\beta-\beta_B)s-(p-p_B)_\perp^2+\ie]
[\alpha(\beta+\beta'_B)s-(p+p'_B)_\perp^2+\ie]}
\nonumber
\end{eqnarray}
The integral in the RHS is the same as for  one of the terms in Eq. (\ref{handbag1}), namely the $\sim\gamma_i\Gamma\gamma_j$ term.
As discussed below Eq. (\ref{handbag1}), it is a sum of contributions independent of $\delta,\delta'$ and power corrections so it can be neglected 
for the evolution with respect to $\sigma$ and $\sigma'$.

Thus, similar to the quark case (\ref{loevoleq}), we get the leading-order evolution equation for gluon TMD in the form 
\begin{eqnarray}
&&\hspace{-0mm}
\Big(\sigma{d\over d\sigma}+\sigma'{d\over d\sigma'}\Big)
\scrf^{i,a;\sigma'}(\beta'_B,x_\perp)\scrf^{a:\sigma}_{~~i}(\beta_B,y_\perp)
\nonumber\\
&&\hspace{2mm}
=~-{\alpha_s\over 2\pi}N_c
\scrf^{i,a;\sigma'}(\beta'_B,x_\perp)\scrf^{a:\sigma}_{~~i}(\beta_B,y_\perp)
\Big[\ln\Big(-{i\over 4}(\beta'_B+\ie)\sigma' sb_\perp^2e^\gamma\Big)
+\ln\Big(-{i\over 4}(\beta_B+\ie)\sigma sb_\perp^2e^\gamma\Big)\Big]
\label{gloevoleq}
\end{eqnarray}
and the solution is
\begin{eqnarray}
&&\hspace{-0mm}
\scrf^{i,a;\sigma'}(\beta'_B,x_\perp)\scrf^{a:\sigma}_{~~i}(\beta_B,y_\perp)
\label{evolog}\\
&&\hspace{0mm}
=~e^{ -{\alpha_sc_F\over 4\pi}\ln{\sigma'\over\sigma'_0}\big[\ln{\sigma'\sigma'_0}+2\ln \big(-{i\over 4}(\beta'_B+\ie)s\Delta_\perp^2e^\gamma\big)\big]}
\scrf^{i,a;\sigma'_0}(\beta'_B,x_\perp)\scrf^{a:\sigma_0}_{~~i}(\beta_B,y_\perp)e^{ -{\alpha_sc_F\over 4\pi}\ln{\sigma\over\sigma_0}\big[\ln{\sigma\sigma_0}+2\ln \big(-{i\over 4}(\beta'_B+\ie)s\Delta_\perp^2e^\gamma\big)\big]}
\nonumber
\end{eqnarray}
where again $b_\perp\equiv \Delta_\perp$.
The only difference between the evolution of quark and gluon TMDs at the leading order is the replacement $c_F\leftrightarrow N_c$. Also, the leading-order  evolution of gluon TMDs in the coordinate space has the same conformal form 
as Eq. (\ref{tmdcoordevol}) with the $c_F\rightarrow N_c$ replacement (see the discussion in Ref. \cite{Balitsky:2019ayf}).

 \subsection{Quark loop contribution from light-cone expansion \label{sec:qloop}}
 
\begin{figure}[htb]
\begin{center}
\includegraphics[width=141mm]{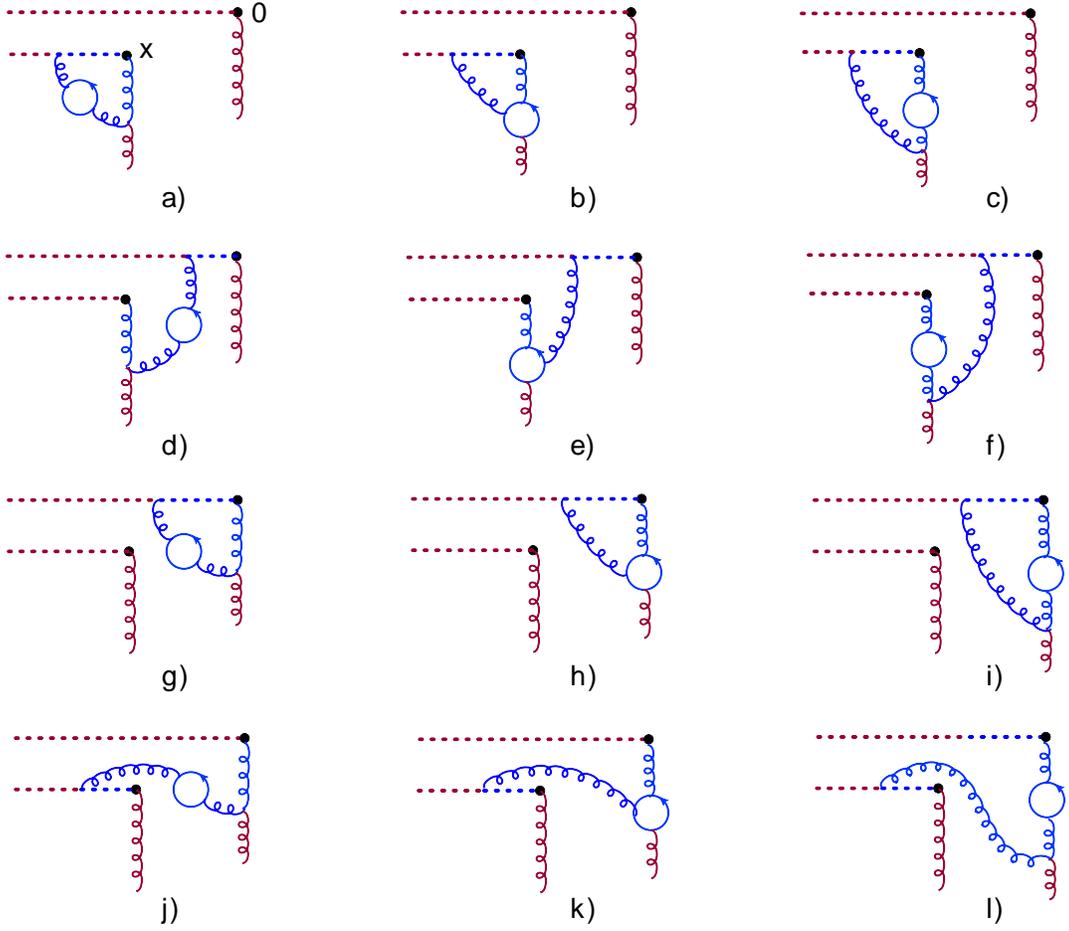}
\end{center}
\caption{Quark loop correction to gluon TMD evolution.\label{fig:Tgluon-running}}
\end{figure}

\begin{figure}[htb]
\begin{center}
\includegraphics[width=131mm]{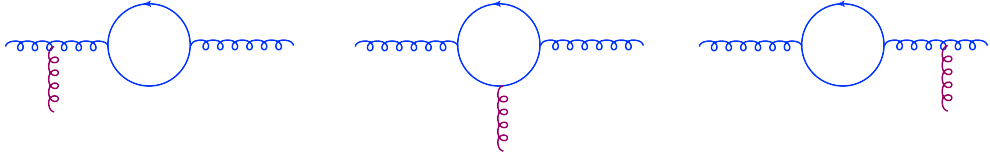}
\end{center}
\caption{Quark loop correction to the gluon propagator in the background field.\label{fig:1loopgprop}}
\end{figure}

As we saw in Sec. \ref{qloopq}, while the diagrams for quark TMDs in the external field depend 
on virtualities of background-field gluons, the rapidity 
evolution of these diagrams does not. It is natural to assume that the same will happen for  gluon TMDs. In this section we will calculate 
the quark-loop contribution to the rapidity evolution of  gluon TMDs using light-cone expansion of quark and gluon propagators. Similar to 
the calculations in Sec. \ref{qloopqlico}, we assume that the background-field gluons are on the mass shell and that $x^+=y^+$. As
we discussed, at $x^+=y^+$ all relevant operators are at 
spacelike separations so we can calculate ordinary Feynman diagrams 
(instead of cut diagrams depicted in Fig. \ref{fig:gldiagrams}); see Fig. \ref{fig:Tgluon-running}.
fffff
The quark-loop contribution to the gluon propagator in the bF gauge has the form
\begin{eqnarray}
&&\hspace{-1mm}
\langle {\rm T}\{A^a_\mu(x) A^b_\nu(y)\}\rangle_{\rm quark~loop}
\nonumber\\
&&\hspace{-0mm}=~\int\!dz_1dz_2
(x|{1\over P^2g_{\mu\alpha}+2i\calf_{\mu\alpha}}|z_1)^{am}
{\rm Tr}~ t^m\gamma_\alpha(z_1|{1\over\slashed{P}}|z_2)t^n\gamma_\beta(z_2|{1\over\slashed{P}}|z_1)
(z_2|{1\over P^2g_{\beta\nu}+2i\calf_{\beta\nu}}|y)^{nb}
\label{gluprop1}
\end{eqnarray}
which we need to calculate near the light cone $(x-y)^2=0$ in the background field with the only component
\begin{equation}
\calf^{-i}(x^+)
\label{bfield}
\end{equation}
with one-$\calf$ accuracy.  The relevant diagrams for the gluon propagator are shown in Fig. \ref{fig:1loopgprop}.

We start from the calculation of the light-cone expansion of the quark loop. Using light-cone expansion of a quark propagator \cite{Balitsky:1987bk}
\begin{eqnarray}
&&\hspace{-0mm}
(z_1|{1\over\hat{P}}|z_2)
\nonumber\\
&&\hspace{-0mm}=~{\slashed{z}_{12}\Gamma\big({d\over 2}\big)[z_1,z_2]\over 2\pi^2(-\Delta^2)^{d\over 2}}
 +{g\Gamma\big({d\over 2}-1\big)\over  16\pi^2(-z_{12}^2)^{{d\over 2}-1}}
\!\int_0^1\! du~[z_1,z_u]
\big(\baru\slashed{z}_{12}\sigma \calf(z_u)+u\sigma \calf(z_u)\slashed{z}_{12}\big)[z_u,z_2]~+~O(D^\mu \calf_{\mu\nu},\calf^2) 
\label{kvprop1}
\end{eqnarray}
where $\baru\equiv 1-u$ and $z_u\equiv uz_1+\baru z_2$, we get
\begin{eqnarray}
&&\hspace{-0mm}
{\rm Tr}~t^a\gamma_\alpha(z_1|{1\over\hat{P}}|z_2)t^b\gamma_\beta(z_2|{1\over\hat{P}}|z_1)~
\nonumber\\
&&\hspace{-0mm}=~
i{B\big({d\over 2},{d\over 2}\big)\over 4\pi^{d\over 2}4^{-\ve}}[z_1,z_2]^{ab}
(z_1|(p^2 g_{\alpha\beta}-p_\alpha p_\beta){\Gamma(-\ve)\over(-p^2)^{-\ve}}|z_2)~
\nonumber\\
&&\hspace{-0mm}
+~{B\big({d\over 2},{d\over 2}-1\big)\over 8\pi^d}{g\Gamma(d-1)\over  (-z_{12}^2)^{d-1}}
\!\int_0^1\! du~\Big([z_1,z_u]z_{12}^\xi z_{12}^\eta\big(2i\baru g_{\alpha\xi} \calf_{\beta\eta}(z_u)-2iu g_{\beta\xi} \calf_{\alpha\eta}(z_u)
+iz_{12}^2 \calf_{\alpha\beta}(z_u)\big)[z_u,z_2]\Big)^{ab}
\label{polope}
\end{eqnarray}
where $B(a,b)=\Gamma(a)\Gamma(b)/\Gamma(a+b)$ and $\ve\equiv{d\over 2}-2$.  
To perform integration over $z_1$ and $z_2$ in Eq. (\ref{gluprop1}),
it is convenient to use 
Eq. (\ref{firsterm}) from the Appendix \ref{sect:lico} (at $a=-\ve$)
and represent Eq. (\ref{polope}) as follows
\begin{eqnarray}
&&\hspace{-0mm}
{\rm Tr} ~t^a\gamma_\alpha(x|{1\over\hat{P}}|0)t^b\gamma_\beta(0|{1\over\hat{P}}|x)~=~
\nonumber\\
&&\hspace{-0mm}
=~i{B\big({d\over 2},{d\over 2}\big)\over \pi^{d\over 2}4^{1+\ve}}
\Big\{(z_1|(P^2g_{\alpha\beta}{\Gamma(-\ve)\over(-P^2)^{-\ve}}
-P_\alpha{\Gamma\big(-\ve)\over(-P^2)^{-\ve}}P_\beta)|z_2)~
\nonumber\\
&&\hspace{-0mm}
+~ig(z_1|{\Gamma(-\ve)\over(-p^2)^{-\ve}}|0)\!\int_0^1\! du ~\calf_{\alpha\beta}(z_u)\Big[d-2+{d-4\over 2(d-2)}\Big]
+{gd\over d-2}(z_1|{p^\xi\Gamma(1-\ve)\over(-p^2)^{1-\ve}}|z_2)
\!\int_0^1\! du~\baru u(D_\alpha \calf_{\beta\xi}(z_u)+\alpha\leftrightarrow\beta)
\nonumber\\
&&\hspace{-0mm}
+~{2ig\over d-2}\!\int_0^1\! du ~
\Big[ -up^\xi\Big(\calf_{\alpha\xi}(z_u)(z_1|{\Gamma(1-\ve)\over(-p^2)^{1+\ve}} |z_2) \Big)\stackrel{\leftarrow}p_\beta
+\baru p_\alpha\Big(\calf_{\beta\xi}(z_u)(z_1|{\Gamma(1-\ve)p^\xi\over(-p^2)^{1-\ve}} |z_2) \Big)\Big]\Big\}~+~O(D\calf,\calf^2)
 \label{polope4}
 \end{eqnarray}
 Subtracting the counterterm
$$
\half\delta Z_3^FA^{a\mu}\big(D^2g_{\mu\nu}-2i\calf_{\mu\nu}-D_\mu D_\nu\big)^{ab}A^b_\nu
$$
where $\delta Z_3^F={g^2\over 24\pi^2\ve}$, we get
\begin{eqnarray}
&&\hspace{-0mm}
g^2{\rm Tr}~t^a\gamma_\alpha(z_1|{1\over\hat{P}}|z_2)t^b\gamma_\beta(z_2|{1\over\hat{P}}|z_1)~-~{\rm counterterm}
\label{kvloop}\\
&&\hspace{-0mm}
=~{ig^2\over 4\pi^2}\Big\{g_{\alpha\beta}(z_1|P^2\ln {\tmu^2\over -P^2}|z_2)
-(z_1|P_\alpha\ln{\tmu^2\over -P^2}P_\beta|z_2)
+ig\!\int_0^1\! du ~\Big[u\Big(\calf_{\alpha \xi}(z_u)(z_1|{p^\xi\over p^2}|z_2)\Big)\stackrel{\leftarrow}P_\beta
\nonumber\\
&&\hspace{2mm}
-P_\alpha\Big(\baru \calf_{\beta \xi}(z_u)(z_1|{p^\xi\over p^2}|z_2)\Big)
+~(z_1|2\ln{\tmu^2\over -p^2}-{5\over 2}|z_2)\calf_{\alpha\beta}(z_u)
 +2i\baru u(z_1|{p^\xi\over p^2}|z_2)(D_\alpha \calf_{\beta \xi}(z_u)+\alpha\leftrightarrow\beta)\Big]
\Big\}
 \nonumber
\end{eqnarray}
(recall that $\tmu^2\equiv \bar\mu^2_{\rm MS}e^{5/3}$). Substituting this expression to Eq. (\ref{gluprop1}), we obtain
\begin{eqnarray}
&&\hspace{-0mm}
\langle {\rm T}\{A^a_\mu(x) A^b_\nu(y)\}\rangle_{\rm quark~loop}~=~{g^2\over 24\pi^2}\Big\{ig_{\mu\nu}(x|{\ln -\tmu^2/ P^2\over P^2}|y)
-i(x|P_\mu{\ln -\tmu^2/ P^2\over P^4}P_\nu|y)
+2g(x|{1\over p^2}\big\{\calf_{\mu\nu},\ln{\tmu^2\over -p^2}\big\}{1\over p^2}|y)
\nonumber\\
&&\hspace{5mm}
+~g\!\int\! dz_1dz_2\!\int_0^1\! du ~\Big[-u(x|{1\over p^2}]z_1)\calf_{\mu \xi}(z_u)(z_1|{p^\xi\over p^2}|z_2)(z_2|{p_\nu\over p^2}|y)
+\baru (x|{p_\mu\over p^2}]z_1)\calf_{\nu \xi}(z_u)(z_1|{p^\xi\over p^2}|z_2)(z_2|{1\over p^2}|y)
\nonumber\\
&&\hspace{10mm}
-~(x|{1\over p^2}|z_1)\Big(
 2i\baru u(z_1|{p^\xi\over p^2}|z_2)(D_\mu \calf_{\nu \xi}(z_u)+\mu\leftrightarrow\nu)
 +(z_1|2\ln{\tmu^2\over -p^2}-{5\over 2}|z_2)\calf_{\mu\nu}(z_u)\Big)(z_2|{1\over p^2}|y)\Big]
\Big\}
\label{gluproplicoe}
\end{eqnarray}
for one flavor of massless quarks. We will need also
\begin{eqnarray}
&&\hspace{-0mm}
\langle {\rm T}\{ A^a_\alpha(x) F^b_{\mu\nu}(y)\}\rangle_{\rm quark~loop}~=~{g^2\over 24\pi^2}\Big\{
-g_{\alpha\nu}(x|{\ln {\tmu^2\over -P^2}\over P^2}P_\mu|y)
+2ig(x|{1\over p^2}\big\{\calf_{\alpha\nu},\ln{\tmu^2\over -p^2}\big\}{p_\mu\over p^2}|y)
\nonumber\\
&&\hspace{5mm}
+~{ig\over 2}(x|p_\alpha{\ln{\tmu^2\over -p^2}\over p^4}\calf_{\mu\nu}|y)+~\!\int\! dz_1dz_2\!\int_0^1\! du ~\Big[
i\baru (x|{p_\alpha\over p^2}]z_1)\calf_{\nu \xi}(z_u)(z_1|{p^\xi\over p^2}|z_2)(z_2|{p_\mu\over p^2}|y)
\label{gluproplicog}\\
&&\hspace{10mm}
-~ig(x|{1\over p^2}|z_1)\Big(
 2i\baru u(z_1|{p^\xi\over p^2}|z_2)(D_\alpha \calf_{\nu \xi}(z_u)+\alpha\leftrightarrow\nu)
 +(z_1|2\ln{\tmu^2\over -p^2}-{5\over 2}|z_2)\calf_{\alpha\nu}(z_u)\Big)(z_2|{p_\mu\over p^2}|y)\Big]
\Big\}-\mu\leftrightarrow\nu
\nonumber
\end{eqnarray}

Let us calculate now the quark-loop contribution to Eq. (\ref{gtmdael}). As we discussed above, we can put $x^+=y^+=0$ and
 $y_\perp=0$,
\begin{equation}
\hspace{-1mm}
\delta^-{d\over d\delta^-}\langle{\rm T}\{ F^{-i,a}(0^+,x_\perp, -\delta'^-)[x^+,-\infty^+]_x^{ac} 
[-\infty^+,0^+]^{cd}F^{-,d}_{~~i}(0^+,0_\perp,-\delta^-)\}\rangle_\cala 
\label{gtmdael1e}
\end{equation}
We start with the term coming from the diagram in Fig. \ref{fig:Tgluon-running}(j)-\ref{fig:Tgluon-running}(l);
\begin{eqnarray}
&&\hspace{-1mm}
\delta^-{d\over d\delta^-}\langle{\rm T}\{[0^+,-\infty^+]_x^{ac} ~
[-\infty^+,0^+]^{cd}F^{-,d}_{~~i}(0^+,0_\perp,-\delta^-)\}\rangle_\cala
\nonumber\\
&&\hspace{5mm}
=~-gf^{abc} \delta^-{d\over d\delta^-}
\!\int_{-\infty}^0\! dz^+\langle A^{-b}(z^+,0^-,x_\perp)F^{-,c}_{~~i}(0^+,0_\perp,-\delta^-)
~-~(x_\perp\rightarrow 0)\rangle_\cala
\label{gtmdael1}
\end{eqnarray}
First, let us demonstrate that the terms in the second line in Eq. (\ref{gluproplicog}) do not contribute to Eq. (\ref{gtmdael1}). 
Indeed, consider, for example, the first term
\begin{eqnarray}
&&\hspace{-1mm}
\delta^-{d\over d\delta^-}\!\int_{-\infty}^0\! dz^+(0^-,z^+,x^\perp|p^-{\ln{\tmu^2\over -p^2}\over p^4}|0^+,y_\perp,-\delta^-)\calf^-_{~i}(0^+,y_\perp)
-(x_\perp\rightarrow 0)
\nonumber\\
&&\hspace{5mm}
=~\delta^-{d\over d\delta^-}(0^-,0^+,x^\perp|{\ln{\tmu^2\over -p^2}\over p^4}|0^+,0_\perp,-\delta^-)\calf^-_{~i}(0^+,0_\perp)
-(x_\perp\rightarrow 0)
~=~0
\label{vanishes1}
\end{eqnarray}
because $(0,0^+,x^\perp|{\ln{\tmu^2\over -p^2}\over p^4}|0^+,0_\perp,-\delta^-)$ does not depend on $\delta^-$. Similarly,
for the second term in the second line of Eq. (\ref{gluproplicog}) one gets
\begin{eqnarray}
&&\hspace{-1mm}
\delta^-{d\over d\delta^-}\!\int_{-\infty}^0\! dz^+\!\int_0^1\! du ~
\baru \Big[(0^-,z^+,x^\perp|{p^-\over p^2}]z_1)\calf_{\nu \xi}(z_u)(z_1|{p^\xi\over p^2}|z_2)(z_2|{p_\mu\over p^2}|0^+,0_\perp,-\delta^-)
~-~(\mu\leftrightarrow\nu)-(x_\perp\rightarrow 0)\Big]
\nonumber\\
&&\hspace{0mm}
=~\delta^-{d\over d\delta^-}\!\int_0^1\! du~\baru \Big[(0^-,0^+,x^\perp|{1\over p^2}]z_1)\calf_{\nu \xi}(z_u)(z_1|{p^\xi\over p^2}|z_2)(z_2|{p_\mu\over p^2}|0^+,0_\perp,-\delta^-)~-~(\mu\leftrightarrow\nu)~-~(x_\perp\rightarrow 0)\Big]
\nonumber\\
&&\hspace{0mm}
=~\delta^-{d\over d\delta^-}
\Big[{i(x_\perp+\delta^-)_\mu(x_\perp+\delta^-)^\xi\over 16\pi^2x_\perp^2}\!\int_0^1\! du~u\ln u~ \calf_{\nu \xi}(0^+)
~-~(\mu\leftrightarrow\nu)
+{i\over 32\pi^{d\over 2}}{\Gamma(\ve)\over (x_\perp^2)^{\ve}}\!\int_0^1\! du ~ (\ln u+\baru u)~\calf_{\mu\nu}(0)
\nonumber\\
&&\hspace{0mm}
~-~(x_\perp\rightarrow 0)\Big]~=~~-{1\over 64\pi^2} \calf_{\nu j}(0)\delta^-{d\over d\delta^-}
\Big[{i(x_\perp+\delta^-)_\mu x_\perp^j\over x_\perp^2}
~-~(\mu\leftrightarrow\nu)\Big]
\label{termvanishes}
\end{eqnarray}
where we used Eq. (\ref{vanishterm}) from the Appendix \ref{formulas-lighcone} 
and the fact that $D^\xi \calf_{\nu\xi}=0$ for our background field. 
Now, we have either ($\mu=i,\nu=-$) or {\it vice versa}. In the first case nothing in square brackets depends on $\delta^-$ while
in the second case $\calf_{ij}=0$ for our background field (\ref{bfield}).
\footnote{The $(x_\perp\rightarrow 0)$  term is singular as $x_\perp\rightarrow 0$ so one should regularize this divergency, for example, 
taking small gluon mass $m$, and then $x_\mu x^\xi{\Gamma(1+\epsilon)\over (x_\perp^2)^{1+\epsilon}}-(x_\perp\rightarrow 0)\calf_{\nu\xi}$
should be replaced by $\big[x_\mu x^\xi\big({m\over x_\perp}\big)^{1+\epsilon}K_{1+\epsilon}(mx_\perp)-\delta^\mu\delta^\xi m^{2+2\epsilon}\Gamma(-1-\epsilon)\big]\calf_{\nu\xi}$. The second term here vanishes 
for our background field (\ref{bfield}) whereas the first term gives  the expression in square brackets in the RHS of Eq. (\ref{termvanishes})}

Thus, the second line in Eq. (\ref{gluproplicog}) can be ignored and we get
\begin{eqnarray}
&&\hspace{-1mm}
\delta^-{d\over d\delta^-}g\langle{\rm T}\{ [x^+,-\infty]_x^{ac} 
[-\infty,y^+]_y^{cd}F^{-i,d}(0^+,0_\perp,-\delta^-)\}\rangle_\cala 
 \nonumber\\
&&\hspace{0mm}
=~{g^2\over 24\pi^2}iN_c\delta^-{d\over d\delta^-}\!\int_{-\infty}^0\! dz^+\Big\{
2i(z^+,x_\perp,0^-|{1\over p^2}\big\{\calf^{-i,a},\ln{\tmu^2\over -p^2}\big\}{p^-\over p^2}|y_\delta)
\nonumber\\
&&\hspace{0mm}
-~\!\int\! dz_1dz_2\!\int_0^1\! du ~(z^+,x_\perp,0^-|{1\over p^2}|z_1)\Big[i(z_1|2\ln{\tmu^2\over -p^2}-{5\over 2}|z_2)\calf^{-i,a}(z_u)
(z_2|{p^-\over p^2}|y_\delta)
\nonumber\\
&&\hspace{0mm}
+~
2\baru u(z_1|{p^+\over p^2}|z_2)D^- \calf^{-,i}(z_u)
(z_2|{p^-\over p^2}|y_\delta)
+~4\baru u(z_1|{p_j\over p^2}|z_2)D^- \calf^{- j,a}(z_u)(z_2|{p^i\over p^2}|y_\delta)\Big]~-~(x_\perp\rightarrow 0)
\label{gluproplico2}
\end{eqnarray}
where $|y_\delta)\equiv|0^+,0_\perp,-\delta^-) $.

The first contribution to RHS of Eq. (\ref{gluproplico2}) is proportional to 
\begin{eqnarray}
&&\hspace{-0mm}
\delta^-{d\over d\delta^-}\!\int_{-\infty}^0\! dz^+\Big(2i
(z^+,x_\perp,0^-|{1\over p^2}\big\{\calf^{-,i},\ln{\tmu^2\over -p^2}\big\}{p^-\over p^2}|0^+,0_\perp,-\delta^-)
\label{firstem1}\\
&&\hspace{-0mm}
-~i\!\int\! dz_1dz_2\!\int_0^1\! du ~(z^+,x_\perp,0^-|{1\over p^2}|z_1)(z_1|2\ln{\tmu^2\over -p^2}-{5\over 2}|z_2)\calf^{-,i}(z_u)
(z_2|{p^-\over p^2}|0^+,0_\perp,-\delta^-)~-~(x_\perp\rightarrow 0)\Big)
\nonumber\\
&&\hspace{-0mm}
~=~\delta^-{d\over d\delta^-}\Big[-2(0^+,x_\perp,0^-|{1\over p^2}\big\{\calf^{-,i},\ln{\tmu^2\over -p^2}\big\}{1\over p^2}|0^+,0_\perp,-\delta^-)
\nonumber\\
&&\hspace{-0mm}
+~\!\int\! dz_1dz_2\!\int_0^1\! du ~(0^+,x_\perp,0^-|{1\over p^2}|z_1)(z_1|2\ln{\tmu^2\over -p^2}-{5\over 2}|z_2)\calf^{-,i}(z_u)
(z_2|{1\over p^2}|0^+,0_\perp,-\delta^-)~-~(x_\perp\rightarrow 0)\Big]
\nonumber\\
&&\hspace{0mm}
+~\delta^-{d\over d\delta^-}\!\int_{-\infty}^0\! dz^+\Big(
2(z^+,x_\perp,0^-|{1\over p^2}\big\{D^-\calf^{-,i},\ln{\tmu^2\over -p^2}\big\}{1\over p^2}|0^+,0_\perp,-\delta^-)
\nonumber\\
&&\hspace{-0mm}
-~\!\int\! dz_1dz_2\!\int_0^1\! du ~(z^+,x_\perp,0^-|{1\over p^2}|z_1)(z_1|2\ln{\tmu^2\over -p^2}-{5\over 2}|z_2)D^-\calf^{-,i}(z_u)
(z_2|{1\over p^2}|0^+,0_\perp,-\delta^-)~-~(x_\perp\rightarrow 0)\Big)\,,
\nonumber
\end{eqnarray}
where we used Eq. (\ref{speminusom}).  Similar to Eq. (\ref{vanishes1}), the first term in the 
RHS vanishes since the expression in the square brackets 
does not actually depend on $\delta^-$. Using   Eqs. (\ref{clico1}), (\ref{fla85}), and (\ref{glavormula}), the second term can be rewritten as
\begin{eqnarray}
&&\hspace{0mm}
{\rm RHS~of~Eq. ~(\ref{firstem1})}~=~\delta^-{d\over d\delta^-}\!\int_{-\infty}^0\! dz^+\!\int_0^1\! du
\Big[(z^+,x_\perp,0^-|{2\ln{\tmu^2\over -p^2}+{9\over 2}\over p^4}|0^+,0_\perp,-\delta^-)~D^-\calf^{-,i}(uz^+)
~-~(x_\perp\rightarrow 0)\Big]
\nonumber\\
&&\hspace{-0mm}
=~-\!\int_{-\infty}^0\! dz^+D^-\calf^{-,i}(z^+)
\Big[(z^+,x_\perp,0^-|{2\ln{\tmu^2\over -p^2}+{9\over 2}\over p^4}|0^+,0_\perp,-\delta^-)~-~(x_\perp\rightarrow 0)\Big]
\nonumber\\
&&\hspace{-0mm}
=~-{i\over 8\pi^2}\!\int_{-\infty}^0\! dz^+D^-\calf^{-,i}(z^+)\Big[{\Gamma(\ve)\over (x_\perp^2-2z^+\delta^-)^\ve}
\Big({1\over\ve}+\ln{\tmu^2 (x_\perp^2-2z^+\delta^-)\over 4}~-~\psi(1+\ve)+\gamma +{5\over 4}\Big)~-~(x_\perp\rightarrow 0)\Big]
\nonumber\\
&&\hspace{-0mm}
=~{i\over 16\pi^2}\!\int_{-\infty}^0\! dz^+D^-\calf^{-,i}(z^+)
\ln{x_\perp^2-2z^+\delta^-\over -2z^+\delta^-}\Big[\ln\tmu^2{x_\perp^2-2z^+\delta^-\over 4}
+\ln\tmu^2{-2z^+\delta^-\over 4}+4\gamma +{5\over 2}   
\Big]
\end{eqnarray}
We get
\begin{eqnarray}
&&\hspace{0mm}
{g^2\over 24\pi^2}iN_c\delta^-{d\over d\delta^-}\!\int_{-\infty}^0\! dz^+\Big[
2i(z^+,x_\perp,0^-|{1\over p^2}\big\{\calf^{-,i},\ln{\tmu^2\over -p^2}\big\}{p^-\over p^2}|y_\delta)
\nonumber\\
&&\hspace{0mm}
-~i\!\int\! dz_1dz_2\!\int_0^1\! du ~(z^+,x_\perp,0^-|{1\over p^2}|z_1)(z_1|2\ln{\tmu^2\over -p^2}-{5\over 2}|z_2)\calf^{-,i}(z_u)
(z_2|{p^-\over p^2}|y_\delta)~-~(x_\perp\rightarrow 0)\Big]
\nonumber\\
&&\hspace{0mm}
=~-{g^2N_c\over 384\pi^2}\!\int_{-\infty}^0\! dz^+D^- \calf^{- i}(z^+)\ln{x_\perp^2-2z^+\delta^-\over -2z^+\delta^-}\Big[\ln\tmu^2{x_\perp^2-2z^+\delta^-\over 4}
+\ln\tmu^2{-2z^+\delta^-\over 4}+4\gamma +{5\over 2}\Big]
\label{glavterm}
\end{eqnarray}

Let us now consider the two remaining terms in  the RHS of Eq. (\ref{gluproplico2}).  With our accuracy the last term
in  the RHS of Eq. (\ref{gluproplico2}) reduces to
\begin{eqnarray}
&&\hspace{-0mm}
\delta^-{d\over d\delta^-}\!\int_{-\infty}^0\! dz^+\!\int\! dz_1dz_2\!\int_0^1\! du ~\baru u
(z^+,x_\perp,0^-|{1\over p^2}|z_1)(z_1|{p^j\over p^2}|z_2)D^- \calf_{- j}(z_u)(z_2|{p^i\over p^2}|0^+,0_\perp,-\delta^-)
~-~(x_\perp\rightarrow 0)
\nonumber\\
&&\hspace{0mm}
=~-i{\partial\over\partial y_i}\delta^-{d\over d\delta^-}\!\int_{-\infty}^0\! dz^+
\!\int\! dz_1dz_2\!\int_0^1\! du ~\baru u
(z^+,x_\perp,0^-|{1\over p^2}|z_1)(z_1|{p_j\over p^2}|z_2)D^- \calf^{- j}(z_u)(z_2|{1\over p^2}|0^+,y_\perp,-\delta^-)\Big|_{y_\perp=0}
\nonumber\\
&&\hspace{0mm}
~-~(x_\perp\rightarrow 0)
\nonumber\\
&&\hspace{0mm}
=~i{1\over 64\pi^{d\over 2}}
\delta^-{d\over d\delta^-}\!\int_{-\infty}^0\! dz^+\!\int_0^1\! du ~\baru u
\Big[{g_{ij}\Gamma(\ve)\over (x_\perp^2-2z^+\delta^-)^\ve} 
+{2x_i x_j\over x_\perp^2-2z^+\delta^-}\Big]D^- \calf^{- j}(uz^+)~-~(x_\perp\rightarrow 0)
\nonumber\\
&&\hspace{0mm}
=~i{1\over 64\pi^2}\!\int_{-\infty}^0\! dz^+
\!\int_0^1\! du~\Big[{g_{ij}\baru ux_\perp^2\over ux_\perp^2-2z^+\delta^-} 
+2u(1-2u){x_i x_j\over ux_\perp^2-2z^+\delta^-} \Big]
D^- \calf^{- j}(z^+)
\end{eqnarray}
where we used Eq. (\ref{flasdg}) to get the fourth line and Eq. (\ref{nado3}) to get the last line. 
As we discussed above, the characteristic $z^+$ are $z^+_{\rm char}\sim {1\over\beta_B\vro}$
so $x_\perp^2\gg 2z^+_{\rm char}\delta^-$, and we get
\begin{eqnarray}
&&\hspace{-1mm}
{g^2\over 24\pi^2}iN_c\delta^-{d\over d\delta^-}\!\int_{-\infty}^0\! dz^+\!\int\! dz_1dz_2\!\int_0^1\! du ~\Big[
4\baru u(z_1|{p_j\over p^2}|z_2)(D^- \calf^{- j}(z_u)(z_2|{p^i\over p^2}|y_\delta)~-~(x_\perp\rightarrow 0)\Big]
\nonumber\\
&&\hspace{0mm}
=~-{g^2N_c\over 768\pi^2}\!\int_{-\infty}^0\! dz^+D^- \calf^{- i}(z^+)~+~O\Big({z^+_{\rm char}\delta^-\over x_\perp^2}\Big)
=~-{g^2N_c\over 768\pi^2}\calf^{- i}(0^+)~+~O\Big({m_\perp^2\over\sigma\beta_Bs}\Big)
\label{term3}
\end{eqnarray}
Finally, from Eq. (\ref{flagen}) we get
\begin{eqnarray}
&&\hspace{-1mm}
\delta^-{d\over d\delta^-}\!\int_{-\infty}^0\! dz^+\!\int\! dz_1dz_2\!\int_0^1\! du ~\baru u
(z^+,x_\perp,0^-|{1\over p^2}|z_1)
(z_1|{p^+\over p^2}|z_2)D^- \calf^{-,i}(z_u)(z_2|{p^-\over p^2}|0^+,0_\perp,-\delta^-)
~-~(x_\perp\rightarrow 0)
\nonumber\\
&&\hspace{0mm}
=~-i{\partial\over\partial y^+}\delta^-{d\over d\delta^-}\!\int_{-\infty}^0\! dz^+
\!\int\! dz_1dz_2\!\int_0^1\! du ~\baru u
(z^+,x_\perp,0^-|{1\over p^2}|z_1)(z_1|{p^+\over p^2}|z_2)D^- \calf^{-,i}(z_u)
\nonumber\\
&&\hspace{22mm}
\times~(z_2|{1\over p^2}|y^+,0_\perp,-\delta^-)\Big|_{y^+=0}
~-~(x_\perp\rightarrow 0)
\nonumber\\
&&\hspace{0mm}
=~-i{\partial\over\partial y^+}\delta^-{d\over d\delta^-}\!\int_{-\infty}^0\! dz^+
\!\int_0^1\! du ~\baru u
~(z^+,x_\perp,0^-|{p^+\over p^6}|y^+,0_\perp,-\delta^-)D^- \calf^{-,i}(uz^++\baru y^+)
\Big|_{y^+=0}
~-~(x_\perp\rightarrow 0)
\nonumber\\
&&\hspace{0mm}
=~i\delta^-{d\over d\delta^-}\!\int_{-\infty}^0\! dz^+{\partial\over\partial z^+}
\!\int_0^1\! du ~\baru u
~(z^+,x_\perp,0^-|{p^+\over p^6}|0^+,0_\perp,-\delta^-)D^- \calf^{-,i}(uz^+)
~-~(x_\perp\rightarrow 0)
\nonumber\\
&&\hspace{0mm}
-~i\delta^-{d\over d\delta^-}\!\int_{-\infty}^0\! dz^+
\!\int_0^1\! du ~\baru u
~(z^+,x_\perp,0^-|{p^+\over p^6}|0^+,0_\perp,-\delta^-)(D^-)^2 \calf^{-,i}(uz^+)
~-~(x_\perp\rightarrow 0)
\nonumber\\
&&\hspace{0mm}
=~
-~i\delta^-{d\over d\delta^-}\!\int_{-\infty}^0\! dz^+
\!\int_0^1\! du ~\baru u
~(z^+,x_\perp,0^-|{p^+\over p^6}|0^+,0_\perp,-\delta^-)(D^-)^2 \calf^{-,i}(uz^+)
~-~(x_\perp\rightarrow 0)
\label{poldela}
\end{eqnarray}
because the term in the fifth line vanishes similar to Eq. (\ref{vanishes1}).
Using the first of Eqs.  (\ref{nado3}) with $\calo(uz^+)=uz^+(D^-)^2 \calf^{-,i}(uz^+)$, we obtain
\begin{eqnarray}
&&\hspace{0mm}
{\rm RHS~of~Eq. ~(\ref{poldela})}~
=~-i{1\over 64\pi^2}
\!\int_{-\infty}^0\! dz^+(D^-)^2 \calf^{-,i}(z^+)\!\int_0^1\! du ~
z_+\ln{ux_\perp^2-2z^+\delta^-\over -2z^+\delta^-}
\nonumber\\
&&\hspace{0mm}
=~{i\over 64\pi^2}
\!\int_{-\infty}^0\! dz^+D^-\calf^{-,i}(z^+)\!\int_0^1\! du ~
\Big[\ln{ux_\perp^2-2z^+\delta^-\over -2z^+\delta^-}-{ux_\perp^2\over ux_\perp^2-2z^+\delta^-}\Big]
\nonumber\\
&&\hspace{0mm}
=~{i\over 64\pi^2}
\!\int_{-\infty}^0\! dz^+D^-\calf^{-,i}(z^+)
\Big[\ln{x_\perp^2\over -2z^+\delta^-}-2\Big]~+~O\Big({m_\perp^2\over\sigma\beta_Bs}\Big)
\end{eqnarray}
because the characteristic $z^+$ are $\sim {1\over\beta_B\vro}$. Thus, the first term in the last line in Eq. (\ref{gluproplico2}) is
\begin{eqnarray}
&&\hspace{-1mm}
{g^2\over 24\pi^2}iN_c\delta^-{d\over d\delta^-}\!\int_{-\infty}^0\! dz^+\!\int\! dz_1dz_2\!\int_0^1\! du ~2\baru u(z_1|{p^+\over p^2}|z_2)(D^- \calf^{-,i}(z_u)(z_2|{p^-\over p^2}|0^+,0_\perp,-\delta^-)~-~(x_\perp\rightarrow 0)
\nonumber\\
&&\hspace{0mm}
=~-{g^2N_c\over 768\pi^4}\!\int_{-\infty}^0\! dz^+D^- \calf^{- i}(z^+)\Big[\ln{x_\perp^2\over -2z^+\delta^-}-2\Big]
~+~O\Big({m_\perp^2\over\sigma\beta_Bs}\Big)
\label{term2}
\end{eqnarray}
Let us now assemble the result for the contribution (\ref{gtmdael1}) given by a sum of Eqs. (\ref{glavterm}), (\ref{term2}),
and (\ref{term3})
\begin{eqnarray}
&&\hspace{-0mm}
\delta^-{d\over d\delta^-}\langle{\rm T}\{[0^+,-\infty]_x^{ac} ~
[-\infty,0^+]_0^{cd}F^{-,d}_{~~i}(0^+,0_\perp,-\delta^-)\}\rangle_\cala
\label{gluotvet1}\\
&&\hspace{0mm}
=-{g^2N_c\over 384\pi^4}\!\int_{-\infty}^0\!\! dz^+D^- \calf^{- i,a}(z^+)\Big\{
\ln{x_\perp^2-2z^+\delta^-\over -2z^+\delta^-}\Big[\ln\tmu^2{x_\perp^2-2z^+\delta^-\over 4}
+\ln\tmu^2{-2z^+\delta^-\over 4}+4\gamma+3\Big]-\half\Big\}~+~O\Big({m_\perp^2\over\sigma\beta_Bs}\Big)
\nonumber
\end{eqnarray}
Note that double-log terms are the same as in the quark case, see Eq. (\ref{licokv1}).

Performing Fourier transformation using Eqs. (\ref{furintegral1}) and (\ref{furintegral2})  we get
\begin{eqnarray}
&&\hspace{-0mm}
\delta^-{d\over d\delta^-}\langle{\rm T}\{ [x^+,-\infty]_x^{ab}[-\infty,0^+]_0^{bc}F^{-j,c}\big(0^+,0_\perp,-{1\over\vro\sigma}\big)\}
\rangle_\cala^{\rm Fig. ~ \ref{fig:gldiagrams}a-c~loop}
~=~ -{g^2N_c\over 384\pi^4}\!\int\!\dhd\beta_B~ e^{-i\beta_B\vro y^+}
\label{gnapravotvet}\\
&&\hspace{2mm}
\times~\Big\{\Big(\ln{x_\perp^2\vro\over 2\delta^-}[-i\beta_B+\epsilon] +\gamma\Big)
\Big(\ln{x_\perp^2\tmu^4\delta^-\over 8\vro(-i\beta_B+\epsilon)} +3\gamma+3\Big)-\half-{\pi^2\over 6}~+~O\Big({m_\perp^2\over\beta_B\sigma s}\Big)\Big\}\calf^{-i,a}(\beta_B,0_\perp)
\nonumber
\end{eqnarray}
Recall that we calculated the contribution due to one quark flavor, so for $n_f$ flavors we should multiply Eq. (\ref{gnapravotvet})
by $n_f$, and to use the BLM prescription we must replace $-{1\over 6\pi}n_f$ by $b_0={11\over 12\pi}N_c-{1\over 6\pi}n_f$. We obtain then
\begin{eqnarray}
&&\hspace{-0mm}
\sigma^-{d\over d\sigma^-}\langle {\rm T}\{[x^+,-\infty]_x^{ab}[-\infty,0^+]_0^{bc}F^{-j,c}\big(0^+,0_\perp,-{1\over\vro\sigma}\big)\}
\rangle_\cala^{\rm Fig. ~ \ref{fig:gldiagrams}a-c~loop}
~=~ -{g^2N_c b_0\over 64\pi^3}\!\int\!\dhd\beta_B~ e^{-i\beta_B\vro y^+}
\label{gnapravotvete}\\
&&\hspace{2mm}
\times~\Big\{\Big(\ln{x_\perp^2\sigma s\over 4}[-i\beta_B+\epsilon] +\gamma\Big)
\Big(\ln{x_\perp^2\tmu^4\over 4\sigma s(-i\beta_B+\epsilon)} +3\gamma+3\Big)-\half
-{\pi^2\over 6}~+~O\Big({m_\perp^2\over\beta_B\sigma s}\Big)\Big\}\calf^{-i,a}(\beta_B,0_\perp)
\nonumber
\end{eqnarray}
Adding the leading-order term and restoring $y^+,y_\perp$ we get
\begin{eqnarray}
&&\hspace{-0mm}
\sigma{d\over d\sigma}\langle {\rm T}\{[x^+,-\infty]_x^{ab}[-\infty,y^+]_y^{bc}F^{-j,c}\big(y^+,y_\perp,-{1\over\vro\sigma}\big)\}
\rangle_\cala^{\rm Fig. ~ \ref{fig:gldiagrams}a-c~+~loop}
\\
&&\hspace{2mm}
=~-{\alpha_s(\tmu)\over 2\pi}N_c\!\int\!\dhd\beta_B~\calf^{-j,b}(\beta_B,y_\perp) e^{-i\beta_B\vro y^+}
\Big\{\ln\Big({\Delta_\perp^2\over 4}(-i\beta_B+\epsilon)\sigma se^\gamma\Big)
~+~O\big({m_\perp^2\over\beta_B\sigma s}\big)
\nonumber\\
&&\hspace{0mm}
+~{b\alpha_s(\tmu)\over 8\pi}\Big\{\Big(\ln{\Delta_\perp^2\sigma s\over 4}[-i\beta_B+\epsilon] +\gamma\Big)
\Big(\ln{\Delta_\perp^2\tmu^4\over 4\sigma s(-i\beta_B+\epsilon)} +3\gamma+3\Big)-\half-{\pi^2\over 6}~+~O\Big({m_\perp^2\over\beta_B\sigma s}\Big)\Big\}
\nonumber\\
&&\hspace{2mm}
=~-{\alpha_s(\mu_\sigma)\over 2\pi}N_c\!\int\!\dhd\beta_B\calf^{-j,b}(\beta_B,y_\perp) e^{-i\beta_B\vro y^+}
\big\{\ln\big[-{i\over 4}(\beta_B+\ie)\sigma s\Delta_\perp^2e^\gamma\big]
+O\big(\alpha_s(\mu_\sigma)\big)\big\}
~+~O\big({m_\perp^2\over\beta_B\sigma s}\big)
\nonumber
\end{eqnarray}
where
 $\mu^2_\sigma\equiv\sqrt{\sigma|\beta_B|s\over\Delta_\perp^2}$. 
 
We see that  the result is the same as Eq. (\ref{evolrunpart1}) for tquark TMD up to the replacement $c_F\rightarrow N_c$ and 
 $O\big(\alpha_s(\mu_\sigma)\big)$ corrections.  Because of that, we can just recycle all formulas for the evolution from
 the quark TMD case replacing $c_F\rightarrow N_c$ when appropriate.
 The evolution equation for gluon TMDs will be 
\begin{eqnarray}
&&\hspace{-0mm}
\Big(\sigma{d\over d\sigma}+\sigma'{d\over d\sigma'}\Big)
\scrf^{i,a;\sigma'}(\beta'_B,x_\perp)\scrf^{a;\sigma}_{~~i}(\beta_B,y_\perp)
\label{evolegrun}\\
&&\hspace{2mm}
=~-{N_c\over 2\pi}
\scrf^{i,a;\sigma'}(\beta'_B,x_\perp)\scrf^{a;\sigma}_{~~i}(\beta_B,y_\perp)
\Big[\alpha_s(\mu_{\sigma'})\ln\Big(-{i\over 4}(\beta'_B+\ie)\sigma' sb_\perp^2e^\gamma\Big)
+\alpha_s(\mu_\sigma)\ln\Big(-{i\over 4}(\beta_B+\ie)\sigma sb_\perp^2e^\gamma\Big)\Big]
\nonumber
\end{eqnarray}
where $b_\perp\equiv\Delta_\perp$ as usual. 
The solution of this equation is the same as (\ref{qevolution}) with $c_F\rightarrow N_c$ replacement
\begin{eqnarray}
&&\hspace{-0mm}
\scrf^{i,a;\sigma'}(\beta'_B,x_\perp)\scrf^{a;\sigma}_{~~i}(\beta_B,y_\perp)~
=~e^{-{2N_c\over \pi b_0^2}\big[\ln{\alpha_s(\mu_{\sigma'})\over\alpha_s(\mu_{\sigma'_0})}
\big({1\over\alpha_s(\tbe_\perp^{-1})}+\ln[-i(\tau'_B+\ie)]\big)+{1\over \alpha_s(\mu_{\sigma'})}-{1\over \alpha_s(\mu_{\sigma'_0})}
\big]}
\nonumber\\
&&\hspace{0mm}
\times~e^{-{2N_c\over \pi b_0^2}\big[\ln{\alpha_s(\mu_{\sigma})\over\alpha_s(\mu_{\sigma_0})}
\big({1\over\alpha_s(\tbe_\perp^{-1})}+\ln[-i(\tau_B+\ie)]\big)+{1\over \alpha_s(\mu_{\sigma})}-{1\over \alpha_s(\mu_{\sigma_0})}
\big]}
\scrf^{i,a;\sigma'_0}(\beta'_B,x_\perp)\scrf^{a;\sigma_0}_{~~i}(\beta_B,y_\perp)
\label{gluonevolution}
\end{eqnarray}

Let us now set $\sigma'=\sigma$ and present the final form of the evolution with the rapidity cutoff (cf. Eq. (\ref{qevolutionfinal}))
\begin{eqnarray}
&&\hspace{-0mm}
\scrf^{i,a;\sigma}(\beta'_B,x_\perp)\scrf^{a;\sigma}_{~~i}(\beta_B,y_\perp)~
=~e^{-{2N_c\over \pi b_0^2}\big[\ln{\alpha_s(\mu_{\sigma'})\over\alpha_s(\mu_{\sigma'_0})}
\big({1\over\alpha_s(\tbe_\perp^{-1})}+\ln[-i(\tau'_B+\ie)]\big)+{1\over \alpha_s(\mu_{\sigma'})}-{1\over \alpha_s(\mu_{\sigma'_0})}
\big]}
\nonumber\\
&&\hspace{0mm}
\times~e^{-{2N_c\over \pi b_0^2}\big[\ln{\alpha_s(\mu_{\sigma})\over\alpha_s(\mu_{\sigma_0})}
\big({1\over\alpha_s(\tbe_\perp^{-1})}+\ln[-i(\tau_B+\ie)]\big)+{1\over \alpha_s(\mu_{\sigma})}-{1\over \alpha_s(\mu_{\sigma_0})}
\big]}
\scrf^{i,a;\sigma'_0}(\beta'_B,x_\perp)\scrf^{a;\sigma_0}_{~~i}(\beta_B,y_\perp)
\label{gevolution}
\end{eqnarray}

It should also be mentioned that  the result for the evolution of gluon TMDs with gauge links out to $+\infty$ is Eq. 
(\ref{gevolution}) with the replacement  $-i(\tau_B+\ie)\rightarrow i(\tau_B-\ie)$, the same as in Eq. (\ref{qevolution2})
for quark TMDs.

\section{Conclusions \label{Conclusions}}
This paper was devoted to the study of the rapidity evolution of quark and gluon TMDs using the small-$x$ methods. 
As customary for studies of small-$x$ amplitudes, we used a rapidity-only cutoff for longitudinal divergences due to infinite gauge links. 
With such cutoff for TMDs, there is only one evolution parameter--this rapidity cutoff. However, as we mentioned in the Introduction,
the argument the of coupling constant in such an evolution is undetermined at the leading order. 
To fix it, one needs to go beyond the leading order and employ some additional  BLM/renormalon considerations, 
as was done for NLO BK evolution in Refs. \cite{Balitsky:2006wa,Kovchegov:2006vj}. 
In this paper, we have done such BLM analysis  for both quark and gluon TMDs, 
and the result is very simple: the effective BLM scale for Sudakov evolution is halfway
(in the logarithmical scale) between transverse momentum and longitudinal ``energy'' of TMD. 

Let us present the final form of the
running-coupling evolution for the  cutoff $\varsigma$ such that $\sigma=\sigma'={\varsigma\sqrt{2}\over\vro|\Delta_\perp|}$.  As we mentioned above, in the leading order the evolution with such a cutoff is conformally invariant, 
see Eq. (\ref{tmdcoordevol}). With the running coupling, 
the evolution equation for quark TMDs reads ($b_\perp\equiv x_\perp-y_\perp$)
\begin{eqnarray}
&&\hspace{-0mm}
\varsigma{d\over d\varsigma}
\bsi^{\varsigma}(\beta'_B,x_\perp) \Gamma\psi^\varsigma(\beta_B,y_\perp)
\label{ouresult1}\\
&&\hspace{2mm}
=~-{c_F\over 2\pi}
\bsi^{\varsigma}(\beta'_B,x_\perp) \Gamma\psi^\varsigma(\beta_B,y_\perp)
\Big[\alpha_s(\mu'_{\varsigma})\ln\Big(-{i\over \sqrt{2}}(\beta'_B+\ie)\varsigma\vro b_\perp e^\gamma\Big)
+\alpha_s(\mu_{\varsigma})\ln\Big(-{i\over \sqrt{2}}(\beta_B+\ie)\varsigma\vro b_\perp e^\gamma\Big)\Big]
\nonumber
\end{eqnarray}
where $\mu_\varsigma=b_\perp^{-1}\big(|\beta_B|\vro\varsigma\sqrt{2}\big)^{1/4},~\mu'_\varsigma=b_\perp^{-1}\big(|\beta'_B|\vro\varsigma\sqrt{2}\big)^{1/4}$ and the solution has the form
\begin{eqnarray}
&&\hspace{-0mm}
\bsi^{\varsigma}(\beta'_B,x_\perp)\Gamma\psi^\varsigma(\beta_B,y_\perp)~
=~e^{-{2c_F\over \pi b_0^2}\big[\ln{\alpha_s(\mu'_{\varsigma})\over\alpha_s(\mu'_{\varsigma_0})}
\big({1\over\alpha_s(\tbe_\perp^{-1})}+\ln[-i\tau'_B+\epsilon]\big)
+{1\over \alpha_s(\mu'_{\varsigma})}-{1\over \alpha_s(\mu'_{\varsigma_0})}
\big]}
\nonumber\\
&&\hspace{0mm}
\times~e^{-{2c_F\over \pi b_0^2}\big[\ln{\alpha_s(\mu_{\varsigma})\over\alpha_s(\mu_{\varsigma_0})}
\big({1\over\alpha_s(\tbe_\perp^{-1})}+\ln[-i\tau_B+\epsilon]\big)
+{1\over \alpha_s(\mu_{\varsigma})}-{1\over \alpha_s(\mu_{\varsigma_0})}
\big]}
\bsi^{\varsigma_0}(\beta'_B,x_\perp)\Gamma\psi^{\varsigma_0}(\beta_B,y_\perp)
\label{ouresult2}
\end{eqnarray}
where $\tbe_\perp^2={b_\perp^2\over 2}e^{\gamma/2}$ and $ \tau_B={\beta_B\over |\beta_B|}, \tau'_B={\beta'_B\over |\beta'_B|}$. 
As we mentioned above, although formally $\alpha_s\ln[-i\tau_B+\epsilon]$ exceeds our accuracy, it determines the direction of evolution
of operators in the coordinate space: $+$ positions of operators move to the left as a result of evolution, see the discussion after Eq. (\ref{evoleqlo}). Consequently, the evolution of quark TMDs with gauge links out to $+\infty$ has the same form (\ref{ouresult2}) 
but with $\ln[i\tau_B+\epsilon]$ [see Eq. (\ref{tmdcoordevolplus})], and $+$ positions of operators move to the right.
 
Another result of our paper is that with BLM scale setting and the rapidity evolution of gluon TMDs has the same form as the one for 
quark TMDs with trivial replacement $c_F\rightarrow N_c$ [see e.g., Eq. (\ref{gevolution})].

It should be noted that, although we used the small-$x$ methods (rapidity-only factorization, etc.), our results (\ref{ouresult1}) and
(\ref{ouresult2}) are correct at any $x_B\equiv\beta_B$ as long as $\sigma x_Bs\gg  \sim b_\perp^{-2}$. 
\footnote{The 
usual requirement of pQCD applicability means that $\alpha_s(b_\perp)$ should be a valid small parameter.}
The difference
between moderate and small $x$ comes at the end point of evolution.  As discussed in Ref. \cite{Balitsky:2019ayf}, the 
double-log logarithmical evolution  ((\ref{qevolution}) or (\ref{gevolution})) can be used until $\sigma x_Bs\sim  b_\perp^{-2}$.  
At this point, if $x_B\sim 1$, 
the situation is similar to Deep Inelastic Scattering (DIS) at moderate $x$ so one should use  single-log DGLAP evolution plus some
phenomenological models for TMDs based on relations to ordinary PDFs \cite{Collins:1984kg,Collins:2017oxh}
 If, however, $x_B\ll 1$, the situation is more like 
DIS at small $x$ so the BFKL/Balitsky-Kovchegov (BK) evolution should be  applicable. A plausible scenario of matching these
evolutions is discussed in Appendix \ref{sec:beyond}.

Also, we saw that one should be very careful with rapidity cutoff in order not
to spoil analytic properties of Feynman diagrams which may bring out the noncancellation of IR divergences.
While the ``rigid cutoff'' $\sigma\gg \alpha$ did not cause any IR problems in the analysis of dipole evolution, we saw 
that in such an analysis of TMD evolution it is not applicable and one should use ``smooth cutoff'' $e^{\pm i\alpha/\sigma}$ 
to avoid IR divergence. 
\footnote{We checked that the use of a smooth cutoff instead of a rigid one does not lead to any change in NLO BK calculations
in Refs. \cite{Balitsky:2006wa,Balitsky:2007feb,Balitsky:2009xg}}

Finally, an obvious outlook is to study the TMD factorization with rapidity-only cutoffs and find the cross section  of the Higgs
production or the Drell-Yan process at  $q_\perp\sim$ few GeV in the one-loop approximation using Eq. (\ref{gevolution}) and the would-be result 
for the one-loop ``coefficient factor''.  In addition, at that point it would be possible to compare our result with the 
two-loop results obtained by CSS method \cite{Gutierrez-Reyes:2017glx,Li:2016axz,Gehrmann:2014yya,Gutierrez-Reyes:2018iod}.
The study is in progress.

\begin{acknowledgments}
The authors are grateful to V. Braun, A. Prokudin, and A. Vladimirov for valuable discussions. 
The work of I.B.  was supported by DOE Contract No. DE-AC05-06OR23177 and by Grant No. DE-FG02-97ER41028.
\end{acknowledgments}

\section{Appendix}
\subsection{Rapidity cutoff and causality \label{app:cutoff}}

In this appendix we discuss the effects of rapidity cutoff on general properties of Feynman diagrams. As we saw in Sect. \ref{sect:diagsac},
the rigid cutoff does not ensure cancellation between real and virtual gluon emissions while point-splitting cutoff preserves this cancellation.
Thus, one should be very careful imposing cutoffs on Feynman diagrams since one may violate properties of causality and unitarity build-in into Feynman diagrams. 

It is a textbook subject that perturbative series in a quantum field theory preserve causality so if one calculates diagrams for some
commutator at spacelike distances one should get zero as a result. (Some caution must be applied in a gauge theory where
this property is correct for gauge-invariant operators.)  Similarly, one should expect the same property in a quantum theory in the background field,
namely diagrams in the background field for the commutator at spacelike distances should sum up to zero. 
Let us check this causality property for our typical commutators and discuss whether this property survives our rapidity cutoff for 
Feynman diagrams.

To avoid the above-mentioned specific complications in gauge theories, 
we consider a massless scalar theory in the background field described by the Lagrangian
$$
L~=~{1\over 2}\partial^\mu\vfi\partial_\mu\vfi+{\lambda\over 2}\vfi^2\bfi
$$
where $\bfi(x)~=~\bfi(x^+,x_\perp)$ is a background scalar field which does not depend on $x^-$.  Let us calculate the expectation value
of the commutator $[\phi(x_\perp,x^+),\phi(y_\perp, y^+)]$ in this theory. 
A simple calculation yields
\begin{eqnarray}
&&\hspace{-0mm}
 \theta(x^+-y^+)\langle[ \phi(x^+,x_\perp),\vfi(y^+,y_\perp)]\rangle_\bfi~=~
\theta(x^+-y^+)\Big(\langle \vfi(x^+,x_\perp)\vfi(y^+,y_\perp)\rangle_\bfi-\langle \tilde{\rm T}\{\vfi(x^+,x_\perp)\vfi(y^+,y_\perp)\rangle
\Big)
\label{kommut1}\\
&&\hspace{-0mm}
=~\int\! dz\big[-i\langle\tilde{\rm T}\{ \vfi(x)\vfi(z)\}\rangle\bfi(z)\langle\vfi(z)\vfi(y)\rangle
+i\langle \vfi(x)\vfi(z)\rangle\bfi(z)\langle{\rm T}\{\varphi(z)\varphi(y)\}\rangle
\nonumber\\
&&\hspace{11mm}
-~i\langle\vfi(x)\vfi(z)\rangle\bfi(z)\langle\vfi(y)\vfi(z)\rangle
+i\langle\tilde{\rm T}\{\vfi(x)\vfi(z)\rangle\phi(z)\}\langle\tilde{\rm T}\{\vfi(z)\vfi(y)\}\rangle\big]
\nonumber\\
&&\hspace{-0mm}
=~{\lambda s^2\over 4}\!\int\! dz\!\int\!\dhd\alpha\dhd\beta\dhd p_\perp~
\!\int\!\dhd\alpha'\dhd\beta'\dhd p'_\perp~e^{-i\alpha\vro(z-z')^- -i\beta'\vro(x-z)^+ +i(p',x-z)_\perp}
e^{-i\beta\vro(z-y)^+ +i(p,z-y)_\perp}
\nonumber\\
&&\hspace{22mm}
\times~\bigg[-{1\over \alpha'\beta' s-{p'_\perp}^2-\ie}
\dbar(\alpha\beta s-p_\perp^2)\theta(\alpha)
-~\dbar(\alpha'\beta' s-{p'_\perp}^2){1\over \alpha\beta s-p_\perp^2+\ie}\theta(\alpha')
\nonumber\\
&&\hspace{44mm}
-~i{1\over \alpha'\beta' s-{p'_\perp}^2-\ie}{1\over \alpha\beta s-p_\perp^2-\ie}
\bigg]\bfi(z^+,z_\perp)
\nonumber\\
&&\hspace{11mm}
=~{-i\lambda\over 4\vro}\!\int\! dz^+ d^2z_\perp~\bfi(z^+,z_\perp)\!\int\!{\dhd\alpha\over\alpha^2}\!\int\!\dhd p_\perp \dhd p'_\perp~
~e^{-i{{p'}_\perp^2\over\alpha s}\vro(x-z)^+ +i(p',x-z)_\perp-i{p_\perp^2\over\alpha s}\vro(z-y)^++i(p,z-y)_\perp}
\nonumber\\
&&\hspace{29mm}
\times~
\big[\theta(\alpha)\theta(z-x)^+-\theta(\alpha)\theta(z-y)^+
-\theta(-\alpha)\theta(x-z)^+\theta(z-y)^+\big]
\nonumber\\
&&\hspace{44mm}
=~{i\lambda\over 32\pi^2s\vro}
\!\int^{x^+}_{y^+}\! dz^+\!\int\! d^2z_\perp~{\bfi(z^+,z_\perp)\over (x-z)^+(z-y)^+}\!\int\!\dhd\alpha
~e^{i\alpha\vro\big[{(x-z)_\perp^2\over 2(x-z)^+}+{(y-z)_\perp^2\over 2(z-y)^+}\big]}
~=~0~
\nonumber
\end{eqnarray}
because the expression is square brackets in the exponent is strictly positive. 

Similarly,
\begin{eqnarray}
&&\hspace{-0mm}
 \theta(y^+-x^+)\langle[ \phi(x^+,x_\perp),\vfi(y^+,y_\perp)]\rangle_\bfi~=~\theta(y^+-x^+)\big(\langle \vfi(x^+,x_\perp)\vfi(y^+,y_\perp)\rangle_\bfi-\langle {\rm T}\{\vfi(x^+,x_\perp)\vfi(y^+,y_\perp)\rangle_\bfi\big)~=~
\nonumber\\
&&\hspace{11mm}
=~{-i\lambda\over 4\vro}\!\int\! dz^+ d^2z_\perp~\bfi(z^+,z_\perp)\!\int\!{\dhd\alpha\over\alpha^2}\!\int\!\dhd p_\perp \dhd p'_\perp~
~e^{-i{{p'}_\perp^2\over\alpha s}\vro(x-z)^+ +i(p',x-z)_\perp-i{p_\perp^2\over\alpha s}\vro(z-y)^++i(p,z-y)_\perp}
\nonumber\\
&&\hspace{29mm}
\times~
\big[\theta(\alpha)\theta(z-x)^+-\theta(\alpha)\theta(z-y)^+
+\theta(-\alpha)\theta(y-z)^+\theta(z-x)^+\big]
\nonumber\\
&&\hspace{44mm}
=~{-i\lambda\over 32\pi^2s}
\!\int_{x^+}^{y^+}\! dz_\ast\!\int\! d^2z_\perp~{\bfi(z^+,z_\perp)\over (y-z)_\ast(z-x)_\ast}\!\int\!\dhd\alpha
~e^{-i\alpha\vro\big[{(z-x)_\perp^2\over 2(z-x)^+}+{(y-z)_\perp^2\over 2(y-z)^+}\big]}
~=~0
\label{kommut2}
\end{eqnarray}
We see that without rapidity cutoff we have causality. However, if we adopt a rigid cutoff $\sigma>|\alpha|$,  we get an integral
\begin{eqnarray}
&&\hspace{-11mm}
{1\over\sigma}\!\int_{\sigma}^\sigma\!\dhd\alpha
~e^{i\alpha\vro\big[{(x-z)_\perp^2\over 2(x-z)^+}+{(y-z)_\perp^2\over 2(z-y)^+}\big]}
~=~{2\over\vro\sigma}\Big[{(x-z)_\perp^2\over 2(x-z)^+}+{(y-z)_\perp^2\over 2(z-y)^+}\Big]^{-1}
\sin\vro\sigma\Big[{(x-z)_\perp^2\over 2(x-z)^+}+{(y-z)_\perp^2\over 2(z-y)^+}\Big]
\end{eqnarray}
which does not vanish. Thus, rigid cutoff violates analytical properties of Feynman diagrams and hence there is no surprise that
there is no cancellation between ``real and virtual emissions'' 
represented by the fifth and the sixth lines in Eq. (\ref{kommut1}), respectively.

Let us now introduce a ``point-splitting cutoff''  $\delta^-$ such that the separation between $x$ and $y^\delta=y-\delta^-$ is spacelike. 
We get 
\begin{eqnarray}
&&\hspace{-0mm}
\theta(x^+-y^+)\langle[ \phi(x^+,x_\perp),\vfi(y^+,y_\perp,\delta^-)]\rangle_\bfi
\nonumber\\
&&\hspace{-0mm}
=~{i\lambda\over 32\pi^2s\vro}\theta(x^+-y^+)
\!\int^{x^+}_{y^+}\! dz^+\!\int\! d^2z_\perp~{\bfi(z^+,z_\perp)\over (x-z)^+(z-y)^+}\!\int\!\dhd\alpha
~e^{i\alpha\vro\big[{(x-z)_\perp^2\over 2(x-z)^+}+{(y-z)_\perp^2\over 2(z-y)^+}+\delta^-\big]}
\label{withpropercutoff1}
\end{eqnarray}
and
\begin{eqnarray}
&&\hspace{-0mm}
\theta(y^+-x^+)\langle[ \phi(x^+,x_\perp),\vfi(y^+,y_\perp,-\delta^-)]\rangle_\bfi
\label{kommut3}\\
&&\hspace{-0mm}
=~
-~{i\lambda\over 32\pi^2s}\theta(y^+-x^+)
\!\int_{x^+}^{y^+}\! dz^+\!\int\! d^2z_\perp~{\bfi(z^+,z_\perp)\over (y-z)^+(z-x)^+}\!\int\!\dhd\alpha
~e^{-i\alpha\vro\big[{(z-x)_\perp^2\over 2(z-x)^+}+{(y-z)_\perp^2\over 2(y-z)^+}+\delta^-\big]}
~=~0
\nonumber
\end{eqnarray}
We see that the sign of $\delta^-$ matters and should be chosen in such a way that $(x-y^\delta)^2=(x^+-y^+)(\pm\delta^-)-(x-y)_\perp^2<0$.  In this paper we use such a point-splitting cutoff for perturbative calculations [see the discussion
after Eq. (\ref{psmael})].

\subsection{Gauge invariance of rapidity-only evolution equations \label{app:lorentz}}
The proof of gauge invariance of evolution equations follows from Ward identities for propagators in the background field and 
for Wilson lines. In this appendix we will demonstrate that the use of background-Lorenz gauge for gluon propagators 
leads to the same evolution equation.

Let us start with the diagrams for leading-order evolution of quark TMDs. As we discussed above, with our
point-splitting cutoff (\ref{Psidef}), all relevant distances are spacelike so we can replace the 
product of operators  in the matrix element in the LHS. 
by the T-product. The gluon propagator in the Lorenz gauge has the form:
\begin{eqnarray}
&&\hspace{-11mm}
i\langle {\rm T}A_\mu(x)A_\nu(y)\rangle
~=~(x|\big(g_{\mu\alpha}-P_\mu{1\over P^2} P_\alpha\big){1\over P^2g_{\alpha\beta}+2iF_{\alpha\beta}}
\big(g_{\beta\nu}-P_\beta {1\over P^2} P_\nu\big)|y)^{ab}
\end{eqnarray}
where all singularities are of the form ${1\over P^2+\ie}$. For calculation of logarithmic part of evolution of
quark TMD we can neglect extra $F_{\alpha\beta}$ and use
\begin{eqnarray}
&&\hspace{-11mm}
i\langle {\rm T}A_\mu(x)A_\nu(y)\rangle
~=~(x|{g_{\mu\nu}\over P^2} -P_\mu{1\over P^2} P_\nu|y)^{ab}
\end{eqnarray}
We will demonstrate that the contribution of the second term 
\begin{eqnarray}
&&\hspace{-1mm}
i\langle {\rm T}A_\mu(x)A_\nu(y)\rangle^{\rm gauge}
~\equiv~(x|P_\mu{-1\over P^2} P_\nu|y)^{ab}
\label{gaugetermprop}
\end{eqnarray}
to
\begin{equation}
\langle\rmT\bsi\big(x^+,x_\perp,-{\delta'}^-)[x^+,-\infty]_x[-\infty,y^+]_y\Gamma\psi\big(y^+,y_\perp,-\delta^-\big)\rangle
\end{equation}
leads to power corrections $\sim {q_\perp^2\over \alpha_A\sigma ts}$ to the evolution equation. 

\begin{figure}[htb]
\begin{center}
\includegraphics[width=99mm]{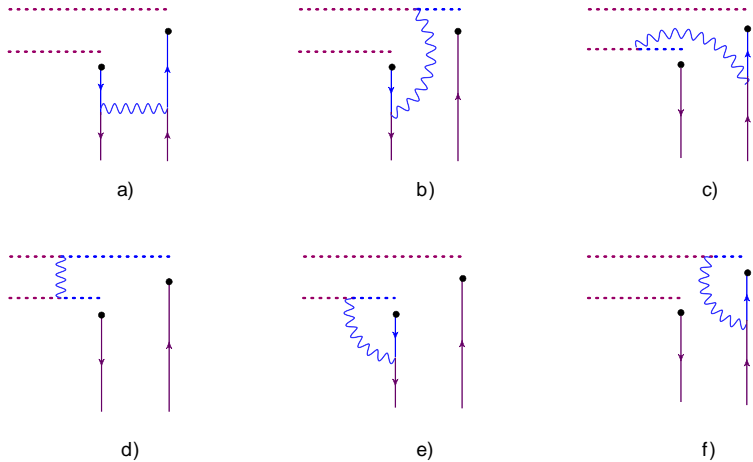}
\end{center}
\caption{Diagrams with the gauge-dependent part of gluon propagator (denoted by wavy line) \label{fig:gpart}}
\end{figure}

The relevant diagrams are shown in Fig. \ref{fig:gpart}
where the wavy line denotes the gauge contribution to the gluon propagator (\ref{gaugetermprop}).
Let us start with the ``handbag'' diagram in Fig. \ref{fig:gpart}(a). 
Using standard Ward identities and the equation of motion for background fields
$\Bsi\stackrel{\leftarrow}{\slashed{P}}=\slashed{P}\Psi=0$, we get 
\begin{equation}
(y|{1\over \slashed{P}}|z)\gamma^\mu t^b\Psi(z)\big(\!-i\overleftarrow{D}^z_\mu\big)^{ba}~=~\delta^{(4)}(z-y)t^a\Psi(y),~~~~~
(iD_\mu^z)^{ab}\Bsi(z)t^b\gamma^\mu(z|{1\over \slashed{P}}|x)~=~\delta^{(4)}(z-x)\Bsi(x)t^a
\label{wardentity1}
\end{equation}
and therefore the contribution of gauge part of gluon propagator (\ref{gaugetermprop}) takes the form
\begin{eqnarray}
&&\hspace{-1mm}
\langle \bsi(x-{\delta'}^-)\Gamma\psi(y-\delta^-)\rangle^{\rm gauge}
~=~-ig^2\!\int\! dz_1dz_2\Gamma(y-\delta^-|{1\over \slashed{P}}|z_1)\gamma^\mu t^a\Psi(z_1)
(z_1|P^\mu{1\over P^4}P^\nu|z_2)^{ab}\Bsi(z_2)\gamma_\nu t^b(z_2|{1\over \slashed{P}}|x-\delta^-)
\nonumber\\
&&\hspace{-1mm}
=~ig^2\Bsi(x-\delta^-)t^a(x|{1\over P^4}|y-\delta^-)t^a\Gamma\Psi(y-\delta^-)
~\simeq~ig^2c_F\Bsi(x-\delta^-)(x-\delta^-|{1\over p^4}|y-\delta^-)\Gamma\Psi(y-\delta^-)
\end{eqnarray}
where $x=x_\perp+x^+$, $y=y_\perp+y^+$, and superscript ``gauge'' means the contribution of the 
gauge part of the gluon propagator (\ref{gaugetermprop}). Next, let us consider diagram in Fig. \ref{fig:gpart}(b). 
Using Eq. (\ref{wardentity1}) and a similar formula for Wilson lines
\begin{eqnarray}
&&\hspace{-1mm}
\!\int_{-\infty}^{y^+}\! d{y'}^+
(z|P_\mu{1\over P^4}P^-|{y'})^{ab}t~=~
-i\!\int_{-\infty}^{y^+}\! 
\!d{y'}^+{d\over d{y'}^+}(z|P_\mu{1\over P^4}|{y'}^+ +{y'}_\perp)^{ab}
=~-i(z|P_\mu{1\over P^4}|y^++y_\perp)^{ab}
\label{wardentity2}
\end{eqnarray}
we obtain
\begin{eqnarray}
&&\hspace{-1mm}
\bsi(x-\delta^-)[-\infty^+,y^+]_y\Gamma\Psi(y-\delta^-)
~\stackrel{\rm gauge}=~-g^2\!\int\! dz~\Bsi(z)\gamma^\mu t^a(z|{1\over\slP}|x-\delta^-)
(z|P_\mu{1\over P^4}P^-|y'+y_\perp)^{ab}t^b\Gamma\Psi(y-\delta^-)
\nonumber\\
&&\hspace{-1mm}
=~-ig^2\Bsi(x-\delta^-)t^at^b(x-\delta^-|{1\over P^4}|y'+y_\perp)^{ab}\Gamma\Psi(y-\delta^-)~
\simeq~-ig^2c_F\Bsi(x-\delta^-)\Gamma\Psi(y-\delta^-)(x-\delta^-|{1\over p^4}|y'+y_\perp)
\nonumber\\
\label{figab}
\end{eqnarray}
(recall that we use $A^+=0$ gauge for background fields). 
Similarly, for the diagram in Fig. \ref{fig:gpart}(c) we get
\begin{eqnarray}
&&\hspace{-1mm}
\Bsi(x-\delta^-)\Gamma[x^+,-\infty^+]_x\psi(y-\delta^-)
~\stackrel{\rm gauge}=~g^2\!\int_{-\infty}^{x^+}\!d{x'}^+\Bsi(x-\delta^-)\Gamma t^a
(y-\delta^-|{1\over\slP}|z)t^b\gamma^\mu\Psi(z)
({x'}^++x_\perp|P^-{1\over P^4}P_\mu|z)^{ab}
\nonumber\\
&&\hspace{-1mm}
=~-ig^2\Bsi(x-\delta^-)\Gamma t^at^b\Psi(y-\delta^-)(x^+ +x_\perp|{1\over P^4}|y-\delta^-)^{ab}
~=~-ig^2c_F\Bsi(x-\delta^-)\Gamma \Psi(y-\delta^-)(x^+ +x_\perp|{1\over p^4}|y-\delta^-)
\nonumber\\
\label{figac}
\end{eqnarray}
where we used  Eq. (\ref{wardentity1}) and the formula
\begin{eqnarray}
&&\hspace{-1mm}
\!\int_{-\infty}^{x^+}\! d{x'}^+
({x'}^+ +x_\perp|P^-{1\over P^4}P_\mu|z)^{ab}
~=~
i\!\int_{-\infty}^{x^+}\! 
\!d{x'}^+{d\over d{x'}^+}({x'}^+ +x_\perp|{1\over P^4}P_\mu|z)^{ab}
=~i({x}^+ +x_\perp|{1\over P^4}P_\mu|z)^{ab}
\label{wardentity3}
\end{eqnarray}
Next, the contribution of the diagram in Fig. \ref{fig:gpart}(d) can be obtained using Eqs. (\ref{wardentity2}) and (\ref{wardentity3}):
\begin{eqnarray}
&&\hspace{-1mm}
\Bsi(x-\delta^-)\Gamma[x^+,-\infty^+]_x[-\infty^+,y^+]_y\Psi(y-\delta^-)
\\
&&\hspace{-1mm}
\stackrel{\rm gauge}=~ig^2\Bsi(x-\delta^-)\Gamma t^at^b\Psi(y-\delta^-)
\!\int_{-\infty}^{x^+}\! d{x'}^+\!\int_{-\infty}^{y^+}\!d{y'}^+ ({x'}^++x_\perp|P^-{1\over P^4}P^-|{y'}^++y_\perp)^{ab}
\nonumber\\
&&\hspace{-1mm}
=~ig^2c_F\Bsi(x-\delta^-)\Gamma \Psi(y-\delta^-)({x}^++x_\perp|{1\over p^4}|{y}^++y_\perp)
\end{eqnarray}
Finally, let us consider diagrams in Fig. \ref{fig:gpart}(e) and \ref{fig:gpart}(f) . 
The result for the diagram  in Fig. \ref{fig:gpart}(e) can be obtained by taking $x=y$ in Eq. (\ref{figab}):
\begin{eqnarray}
&&\hspace{-1mm}
\bsi(x-\delta^-)[x^+,-\infty^+]_x\Gamma\Psi(y-\delta^-)
~\stackrel{\rm gauge}=~ig^2c_F\Bsi(x-\delta^-)\Gamma\Psi(y-\delta^-)(x^++x_\perp-\delta^-|{1\over p^4}|x^++x_\perp)
\label{figabe}
\end{eqnarray}
The integral in the RHS is a pure divergence which does not depend on $\delta^-$ and should be set to $0$ in the dimensional
regularization framework. Similarly, the contribution of the diagram in Fig. \ref{fig:gpart}(f) vanishes.

Thus, the sum of diagrams in Fig. \ref{fig:gpart} takes the form
\begin{eqnarray}
&&\hspace{-1mm}
\bsi(x-\delta^-)\Gamma[x^+,-\infty^+]_x[-\infty^+,y^+]_y\psi(y-\delta^-)
\label{gaugecorrection}\\
&&\hspace{-1mm}
=~ig^2c_F\Bsi(x-\delta^-)\Gamma \Psi(y-\delta^-)
\bigg[({x}^++x_\perp|{1\over p^4}|{y}^++y_\perp)-(x^+ +x_\perp|{1\over p^4}|y^++y_\perp-\delta^-)
\nonumber\\
&&\hspace{-1mm}
-~(x^++x_\perp-\delta^-|{1\over p^4}|y'+y_\perp)+(x^+ +x_\perp-\delta^-|{1\over p^4}|y^++y_\perp-\delta^-)\bigg]
\nonumber\\
&&\hspace{-1mm}
=~{g^2c_F\over 16\pi^2}\Bsi(x-\delta^-)\Gamma \Psi(y-\delta^-)\Big[\ln\Delta_\perp^2-\ln\big(\Delta_\perp^2-\Delta^+\delta^-\big)
-\ln\big(\Delta_\perp^2+\Delta^+{\delta'}^-\big)+\ln\big(\Delta_\perp^2+\Delta^+({\delta'}^- -\delta^-)\big)\Big]
\nonumber\\
&&\hspace{-1mm}
=~{g^2c_F\over 16\pi^2}\Bsi(x-\delta^-)\Gamma \Psi(y-\delta^-)\Big[-\ln\Big(1-{\Delta^+\delta^-\over \Delta_\perp^2}\Big)
-\ln\Big(1+{\Delta^+{\delta'}^-\over \Delta_\perp^2}\Big)+\ln\Big(1+{\Delta^+({\delta'}^- -\delta^-)\over \Delta_\perp^2}\Big)\Big]
\nonumber
\end{eqnarray}
Since ${\Delta^+\delta^-\over \Delta_\perp^2}\sim {q_\perp^2\over \beta_B\sigma s}\ll 1$, the sum (\ref{gaugecorrection}) is a power correction
so the leading-order evolution equation (\ref{loevoleq}) is gauge invariant.

Let us discuss now the invariance of the one-loop quark correction. Since the effect of the one-loop correction reduces to replacement
${1\over p^2}\rightarrow -{b\alpha_s\over 4\pi}{1\over p^2}\ln(-p^2/\tmu^2)$, the corresponding contribution of ``gauge correction
diagrams'' in Fig. \ref{fig:gpart} with the extra quark loop is
\begin{eqnarray}
&&\hspace{-1mm}
-ig^2c_F{b\alpha_s\over 4\pi}\Bsi(x-\delta^-)\Gamma \Psi(y-\delta^-)
\bigg[({x}^++x_\perp|{\ln{-p^2\over \tmu^2}\over p^4}|{y}^++y_\perp)
-(x^+ +x_\perp|{\ln{-p^2\over \tmu^2}\over p^4}|y^++y_\perp-\delta^-)
\nonumber\\
&&\hspace{-1mm}
-~(x^++x_\perp-\delta^-|{\ln{-p^2\over \tmu^2}\over p^4}|y'+y_\perp)
+(x^+ +x_\perp-\delta^-|{\ln{-p^2\over \tmu^2}\over p^4}|y^++y_\perp-\delta^-)\bigg]
\nonumber\\
&&\hspace{-1mm}
=~{g^2c_F\over 16\pi^2}\Bsi(x-\delta^-)\Gamma \Psi(y-\delta^-)\Big[\ln^2\Delta_\perp^2\tmu^2
-\ln^2\big(\Delta_\perp^2-\Delta^+\delta^-\big)\tmu^2
\nonumber\\
&&\hspace{-1mm}
-~\ln^2\big(\Delta_\perp^2+\Delta^+{\delta'}^-\big)\tmu^2+\ln^2\big(\Delta_\perp^2+\Delta^+({\delta'}^- -\delta^-)\big)\tmu^2\Big]
\end{eqnarray}
which is again a power correction due to ${\Delta^+\delta^-\over \Delta_\perp^2}\ll 1$. Consequently, the running-coupling evolution equation (\ref{evoleqrun}) is gauge invariant. In a similar way one can prove gauge invariance of the evolution equation of gluon TMD operators.

\subsection{Necessary integrals \label{app:integrals}}

In this appendix we calculate some integrals used in the main text.
Let us start with the integral
\begin{eqnarray}
&&\hspace{-11mm}
16\pi^2\!\int\!{\dhd p_\perp\over p_\perp^2}
\!\int_0^{\infty}\!\dhd\alpha~e^{-i{\alpha\over \sigma}}
{\big(1-e^{i(p,\Delta)_\perp}\big)\beta_B s
\over \alpha\beta_B s+(p-p_B)_\perp^2+\ie}\,.
\end{eqnarray}

At $\beta_B>0$ we get
\begin{eqnarray}
&&\hspace{-11mm}
16\pi^2\!\int\!{\dhd p_\perp\over p_\perp^2}
\!\int_0^{\infty}\!\dhd\alpha~e^{-i{\alpha\over \sigma}}
{\big(1-e^{i(p,\Delta)_\perp}\big)\beta_B s\over \alpha\beta_B s+p_\perp^2}
~\stackrel{\Lambda\equiv\sigma\beta_Bs}=~8\pi\!\int\!{\dhd p_\perp\over p_\perp^2}
\!\int_0^{\infty}\!dt~e^{-i{t\over \Lambda}}
{\big(1-e^{i(p,\Delta)_\perp}\big)\over t+p_\perp^2}
~=~
\nonumber\\
&&\hspace{-11mm}
=~2\!\int_0^\infty\! {dt\over t}e^{-i{t\over\Lambda}}
\Big[\ln{t\Delta_\perp^2\over 4}+2\gamma +2K_0\big(\sqrt{t\Delta_\perp^2}\big)\Big]
~=~\ln^2\Big(-{i\over 4}\sigma\beta_Bs\Delta_\perp^2e^\gamma\Big)+{\pi^2\over 2}
~+~O\big({\Delta_\perp^{-2}\over\sigma\beta_Bs}\big)
\label{betabeplus}
\end{eqnarray}
while at $\beta_B<0$  we can rotate the contour of integration over $\alpha$ in the lower half-plane of complex $\alpha$ and get
\begin{eqnarray}
&&\hspace{-11mm}
16\pi^2\!\int\!{\dhd p_\perp\over p_\perp^2}
\!\int_0^{\infty}\!\dhd\alpha~e^{-i{\alpha\over \sigma}}
{\big(1-e^{i(p,\Delta)_\perp}\big)\beta_B s\over \alpha\beta_B s+p_\perp^2+\ie}
~=~16\pi^2\!\int\!{\dhd p_\perp\over p_\perp^2}
\!\int_0^{\infty}\!\dhd\alpha~e^{i{\alpha\over \sigma}}
{\big(1-e^{i(p,\Delta)_\perp}\big)|\beta_B| s\over \alpha|\beta_B| s+p_\perp^2}
\nonumber\\
&&\hspace{-11mm}
=~\ln^2\Big({i\over 4}\sigma|\beta_B|s\Delta_\perp^2e^\gamma\Big)+{\pi^2\over 2}
~+~O\big({\Delta_\perp^{-2}\over\sigma|\beta_B|s}\big)\,.
\label{betabeminus}
\end{eqnarray}
The combination of Eqs. (\ref{betabeplus}) and  (\ref{betabeminus}) can be written as
\begin{eqnarray}
&&\hspace{-11mm}
16\pi^2\!\int\!{\dhd p_\perp\over p_\perp^2}
\!\int_0^{\infty}\!\dhd\alpha~e^{-i{\alpha\over \sigma}}
{\big(1-e^{i(p,\Delta)_\perp}\big)\beta_B s
\over \alpha\beta_B s+p_\perp^2+\ie}
~=~\ln^2\Big(-{i\over 4}(\beta_B+\ie)\sigma s\Delta_\perp^2e^\gamma\Big)+{\pi^2\over 2}
~+~O\big({m_\perp^2\over\beta_B\sigma s}\big)
\label{eiknapravotvet}
\end{eqnarray}
which reflects the ``causal'' structure discussed after Eq. (\ref{evoleqlo}). 
From Eq. (\ref{napravo2}) we get
\begin{eqnarray}
&&\hspace{-11mm}
16\pi^2\!\int\!{\dhd p_\perp\over p_\perp^2}
\!\int_0^{\infty}\!\dhd\alpha~e^{-i{\alpha\over \sigma}}
{\big(1-e^{i(p,\Delta)_\perp}\big)\beta_B s
\over \alpha\beta_B s+(p-p_B)_\perp^2+\ie}
\nonumber\\
&&\hspace{-11mm}
=~\ln^2\Big(-{i\over 4}(\beta_B+\ie)\sigma s\Delta_\perp^2e^\gamma\Big)+{\pi^2\over 2}
-8\pi\!\int\!{\dhd p_\perp\over p_\perp^2}\big(1-e^{i(p,\Delta)_\perp}\big)
\ln{(p-p_B)_\perp^2\over p_\perp^2}~+~O\big({m_\perp^2\over\beta_B\sigma s}\big)
\label{eiknapravotvete}
\end{eqnarray}
so
\begin{eqnarray}
\sigma{d\over d\sigma}\!\int_0^\infty\!\dhd\alpha~e^{-i{\alpha\over\sigma}}\!\int\!{\dhd p_\perp\over  p_\perp^2}
{\beta_Bs\big(e^{i(p,\Delta)_\perp}-1\big)\over  \alpha(\beta_B+\ie)s+p_\perp^2}
~=~{\alpha_s\over 2\pi}\ln\Big(-{i\over 4}(\beta_B+\ie)\sigma s\Delta_\perp^2e^\gamma\Big)
\label{evolint}
\end{eqnarray}

Now let us consider the integral with an extra $\ln{\tmu^2\over p_\perp^2}$ in Eq. (\ref{loopevoleqn1}),
\begin{eqnarray}
&&\hspace{-1mm}
\!\int_0^{\infty}\!{\dhd\alpha\over\sigma}e^{-i{\alpha\over\sigma}}\!\int\!\dhd p_\perp
{\big(e^{i(p,\Delta)_\perp}-1\big)\alpha\beta_Bs\ln{\tmu^2\over p_\perp^2}\over p_\perp^2[\alpha\beta_Bs+p_\perp^2+\ie]}
~
=~\!\int_0^{\infty}\!{\dhd\alpha\over\sigma}e^{-i{\alpha\over\sigma}}\!\int\!\dhd p_\perp
\big(e^{i(p,\Delta)_\perp}-1\big)
\Big[{1\over p_\perp^2}-{1\over \alpha\beta_Bs+p_\perp^2+\ie}\Big]\ln{\tmu^2\over p_\perp^2}
\nonumber\\
&&\hspace{-1mm}
=~\!\int_0^{\infty}\!{\dhd\alpha\over\sigma}e^{-i{\alpha\over\sigma}}\!\int\!\dhd p_\perp
\Big[{e^{i(p,\Delta)_\perp}-1\over p_\perp^2}+{1\over \alpha\beta_Bs+p_\perp^2+\ie}\Big]\ln{\tmu^2\over p_\perp^2}
-\!\int_0^{\infty}\!{\dhd\alpha\over\sigma}e^{-i{\alpha\over\sigma}}\!\int\!\dhd p_\perp
{e^{i(p,\Delta)_\perp}\over \alpha\beta_Bs+p_\perp^2+\ie}\ln{\tmu^2\over p_\perp^2}
\label{loopintegral1}
\end{eqnarray}
It is easy to see that the last term in the RHS is actually a power correction:
\begin{eqnarray}
&&\hspace{-1mm}
\!\int_0^{\infty}\!{\dhd\alpha\over\sigma}e^{-i{\alpha\over\sigma}}\!\int\!\dhd p_\perp
{e^{i(p,\Delta)_\perp}\over \alpha\beta_Bs+p_\perp^2+\ie}\ln{\tmu^2\over p_\perp^2}
~=~\!\int\!\dhd p_\perp ~e^{i(p,\Delta)_\perp}\ln{\tmu^2\over p_\perp^2}
\!\int_0^{\infty}\!{\dhd\alpha\over\sigma\beta_Bs}e^{-i{\alpha\over\sigma}}
{d\over d\alpha}\ln{\alpha\beta_Bs+ p_\perp^2+\ie\over \tmu^2}
\nonumber\\
&&\hspace{-1mm}
\simeq~{1\over \sigma\beta_Bs}\!\int\!\dhd p_\perp ~e^{i(p,\Delta)_\perp}\ln{p_\perp^2\over\tmu^2}\Big[{1\over 2\pi}\ln{p_\perp^2\over\tmu^2}
+i\!\int_0^{\sigma}\!{\dhd\alpha\over\sigma}\ln{\alpha\beta_Bs+ p_\perp^2+\ie\over \tmu^2}\Big]~\sim~O\Big({m_\perp^2\over\sigma\beta_Bs}\Big)
\end{eqnarray}
As to the first term in the RHS of Eq. (\ref{loopintegral1}), it is easily calculated using 
standard trick $\big({p^2\over\tmu^2}\big)^\delta=1+\delta\ln {p^2\over\tmu^2}+O(\delta^2)$
\begin{eqnarray}
&&\hspace{-1mm}
\!\int_0^{\infty}\!{\dhd\alpha\over\sigma}e^{-i{\alpha\over\sigma}}\!\int\!\dhd p_\perp
\Big[{e^{i(p,\Delta)_\perp}-1\over p_\perp^2}+{1\over \alpha\beta_Bs+p_\perp^2+\ie}\Big]\ln{\tmu^2\over p_\perp^2}
\nonumber\\
&&\hspace{-1mm}
=~{i\over 16\pi^2}\Big[\big(\ln{\Delta_\perp^2\tmu^2\over 4} +2\gamma\big)^2
-\big(\ln{-i\sigma(\beta_B+\ie)s\over \tmu^2} -\gamma\big)^2-{\pi^2\over 2}\Big]
\nonumber\\
&&\hspace{-1mm}
=~{i\over 16\pi^2}\Big[\big(\ln{\Delta_\perp^2\over 4}[-i\sigma(\beta_B+\ie)s] +\gamma\big)
\big(\ln{\Delta_\perp^2\tmu^4/4\over -i\sigma(\beta_B+\ie)s} +3\gamma\big)-{\pi^2\over 2}\Big]
\label{loopintegralotvet1}
\end{eqnarray}
This gives us Eq. (\ref{evolrunpart1}).

\subsection{The light-cone expansion of propagators \label{sect:lico}}

In this appendix we derive the light-cone expansion of various propagators in the first order in the background field with one (quark) loop 
accuracy. First, we present necessary formulas for quark propagators.
The typical integral appears as
\begin{eqnarray}
&&\hspace{-0mm}
(x|{\Gamma(a)\over (-p^2-\ie)^a}\Phi{\Gamma(b)\over (-p^2-\ie)^b}|0)
~=~i^{a+b}\int_0^\infty\! ds~s^{a+b-1}\!\int_0^1\! du~\baru^{a-1}u^{b-1}(x|e^{is\baru p^2}\Phi e^{isup^2}|0)
\end{eqnarray}
where $\Phi(z)$ is some operator, such as $\Psi(z)$ or $\calf_{\mu\nu}(z)$. 
Using expansion in powers of proper time $s$ \cite{Balitsky:1987bk},
\begin{eqnarray}
&&\hspace{-0mm}
(x|e^{is\baru p^2}\Phi e^{isup^2}|0)~=~(x|e^{isp^2}|0)\Big[\!\int_0^1\! du~\Phi(ux)-is\!\int_0^1\! du~\baru u\partial^2\Phi(ux)~+~O(s^2)\Big]
\label{licoexpanshen}
\end{eqnarray}
we get the light-cone expansion 
\begin{eqnarray}
&&\hspace{-0mm}
(x|{\Gamma(a)\over (-p^2)^a}\Phi{\Gamma(b)\over (-p^2)^b}|0)
~=~(x|{\Gamma(a+b)\over (-p^2)^{a+b}}|0)\!\int_0^1\! du~\baru^{a-1}u^{b-1}\Phi(ux)
-(x|{\Gamma(a+b+1)\over (-p^2)^{a+b+1}}|0)\!\int_0^1\! du~\baru^{a}u^{b}\partial^2\Phi(ux)~+~...
\label{licoexp}
\end{eqnarray}
For our background fields that depend only on $x^+$ we need only the first term of this expansion since $\partial^2\Phi(x^+)=0$
so 
\begin{eqnarray}
&&\hspace{-0mm}
(x|{\Gamma(a)\over (-p^2)^a}\Phi{\Gamma(b)\over (-p^2)^b}|0)
~=~(x|{\Gamma(a+b)\over (-p^2)^{a+b}}|0)\!\int_0^1\! du~\baru^{a-1}u^{b-1}\Phi(ux)
\label{licomaster}
\end{eqnarray}
This is our master formula for light-cone expansions. In this appendix we discuss only Feynman propagators so $p^2$  always means $p^2+\ie$.

Let us start from the light-cone expansion of Eq. (\ref{klico1}).
Using standard trick  
\begin{eqnarray}
&&\hspace{-0mm}
(x|{\ln{\tmu^2\over -p^2}\over p^2}\Phi {1\over p^2}|0)~=~\bigg[(x|{\tmu^{2\lambda}\over (-p^2)^{1+\lambda}}\Phi {1\over (-p^2)}|0)
\bigg]_\lambda
\end{eqnarray}
 where $\big[\dots\big]_\lambda$ denotes the first nontrivial term in the expansion in powers of $\lambda$, we obtain 
\begin{eqnarray}
&&\hspace{-0mm}
(x|{\ln{\tmu^2\over -p^2}\over p^2}\Phi {1\over p^2}|0)~=~(x|{\ln{\tmu^2\over -p^2}\over p^4}|0)\!\int_0^1\! du~\Phi(ux)
+(x|{1\over p^4}|0)\!\int_0^1\! du~(1+\ln\baru)\Phi(ux)
\nonumber\\
&&\hspace{-0mm}
=~{i\Gamma\big(\ve)\over 16\pi^{d\over 2}(-x^2)^\ve}
\Big(\big[\ln{-\tmu^2x^2\over 4}+{1\over \ve}-\psi(1+\ve)+\gamma \big]\!\int_0^1\! du~\Phi(ux)
+\!\int_0^1\! du~\ln\baru\Phi(ux)\Big)
\label{clicon1}
\end{eqnarray}
where we used Eq. (\ref{licomaster}). 

Restoring the end point $y$ we obtain
\begin{eqnarray}
&&\hspace{-0mm}
(x|{1\over p^2}\Phi {\ln{\tmu^2\over -p^2}\over p^2}|y)~=~(x|{\ln{\tmu^2\over -p^2}\over p^4}|y)\!\int_0^1\! du~\Phi(x_u)
+(x|{1\over p^4}|y)\!\int_0^1\! du~(1+\ln u)\Phi(x_u)
\nonumber\\
&&\hspace{-0mm}
(x|{\ln{\tmu^2\over -p^2}\over p^2}\Phi {1\over p^2}|y)~=~(x|{\ln{\tmu^2\over -p^2}\over p^4}|y)\!\int_0^1\! du~\Phi(x_u)
+(x|{1\over p^4}|y)\!\int_0^1\! du~(1+\ln \baru)\Phi(x_u)
\label{clico2}
\end{eqnarray}
where $x_u=ux+\baru y$ as usual. 

We will also need formulas for differentiation with respect to the point-splitting cutoff.
The master formula is
\begin{eqnarray}
&&\hspace{-1mm}
-\delta^-{d\over d\delta^-}\!\int_{-\infty}^0\! dz^+~f(x_\perp^2-2z^+\delta^-)\!\int_0^t\! du~\Phi(uz^+)
~\stackrel{1>t>0}=~\int_{-\infty}^0\! dz^+\Phi(z^+)f\big(x_\perp^2-{2\over t}z^+\delta^-\big)
\label{glavormula}
\end{eqnarray}
and corollaries
\begin{eqnarray}
&&\hspace{-1mm}
\delta^-{d\over d\delta^-}\!\int_{-\infty}^0\! dz^+~f(x_\perp^2-2z^+\delta^-)\!\int_0^1\! du~\ln u~\Phi(uz^+)
~=~\int_{-\infty}^0\! dz^+\Phi(z^+)\int_0^1\!{dt\over t}~f\big(x_\perp^2-{2\over t}z^+\delta^-\big)
\label{glavormulas}\\
&&\hspace{-0mm}
\delta^-{d\over d\delta^-}\!\int_{-\infty}^0\! dz^+~f(x_\perp^2-2z^+\delta^-)\!\int_0^1\! du~\ln \baru~\Phi(uz^+)
~=~\int_{-\infty}^0\! dz^+\Phi(z^+)\int_0^1\!{dt\over 1-t}~\Big[f\big(x_\perp^2-2z^+\delta^-\big)-f\big(x_\perp^2-{2\over t}z^+\delta^-\big)\Big]
\nonumber
\end{eqnarray}

Next, let us present formulas relevant for  the gluon propagator (\ref{gluproplicoe}) (see Fig. \ref{fig:1loopgprop}):
\begin{eqnarray}
&&\hspace{-0mm}
\langle A^a_\mu(x) A^b_\nu(y)\rangle_{\rm quark~loop}~=~{g^2\over 24\pi^2}\Big\{ig_{\mu\nu}(x|{\ln -\tmu^2/ P^2\over P^2}|y)
-i(x|P_\mu{\ln -\tmu^2/ P^2\over P^4}P_\nu|y)
+2(x|{1\over p^2}\big\{\calf_{\mu\nu},\ln{\tmu^2\over -p^2}\big\}{1\over p^2}|y)
\nonumber\\
&&\hspace{5mm}
-~\!\int\! dz_1dz_2\!\int_0^1\! du ~\Big[u(x|{1\over p^2}]z_1)\calf_{\mu \xi}(z_u)(z_1|{p^\xi\over p^2}|z_2)(z_2|{p_\nu\over p^2}|y)
-\baru (x|{p_\mu\over p^2}]z_1)\calf_{\nu \xi}(z_u)(z_1|{p^\xi\over p^2}|z_2)(z_2|{1\over p^2}|y)
\nonumber\\
&&\hspace{10mm}
-~(x|{1\over p^2}|z_1)\Big(
 2i\baru u(z_1|{p^\xi\over p^2}|z_2)(D_\mu \calf_{\nu \xi}(z_u)+\mu\leftrightarrow\nu)
 +(z_1|2\ln{\tmu^2\over -p^2}-{5\over 2}|z_2)\calf_{\mu\nu}(z_u)\Big)(z_2|{1\over p^2}|y)\Big]
\Big\}
\label{gluproplico1}
\end{eqnarray}

To get the light-cone expansion of the gluon propagator we need a formula
\begin{eqnarray}
&&\hspace{-0mm}
(x|e^{isP^2}|y)~=~(x|e^{isp^2}|y)\Big([x,y]+s\!\int_0^1\! du~\baru u[x,x_u]D^\mu \calf_{\mu x}(x_u)[x_u,y]
\nonumber\\
&&\hspace{0mm}
+~2i\!\int_0^1\! du\!\int_0^u\! dv~\baru v[x,x_u]\calf_{\mu x}(x_u)[x_u,x_v]\calf^{\mu}_{~x}(x_v)[vx,y]~+~O(s^2)\Big)
~=~(x|e^{isp^2}|y)[x,y]~+~O(D\calf,\calf^2)
\end{eqnarray}
(where $ \calf_{\mu x}\equiv  x^\xi\calf_{\mu \xi}$) and therefore
\begin{equation}
(x|f(P^2)|y)~=~[x,y](x|f(p^2)|y)
\end{equation}
in our approximation. Hereafter $+~O(D\calf,\calf^2)$ is assumed in all equations.
By differentiation of gauge link $[x,y]$ using formulas
\begin{eqnarray}
&&\hspace{-11mm}
i{\partial\over\partial x_\mu}[ux+\baru y,vx+\barv y]
~=~-uA_\mu(x_u)[x_u,x_v]+[x_u,x_v]vA_\mu(x_v)+\!\int_v^u\! dt ~t[x_u,x_t]\calf_{x\mu}(x_t)[x_t,x_v]
\nonumber\\
&&\hspace{-11mm}
[ux+\baru y,vx+\barv y]\big(-i\stackrel{\leftarrow}{\partial\over \partial y_\nu}\big)~=~
\baru A_\nu(x_u)[x_u,x_v]-[x_u,x_v]\barv A_\nu(x_v)-\!\int_v^u\! dt ~\bart[x_u,x_t]\calf_{x\nu}(x_t)[x_t,x_v]
\end{eqnarray}
(where $x_t=tx+\bart y$ as usual) one obtains
\begin{eqnarray}
&&\hspace{-1mm}
(x|g_{\mu\nu}{\Gamma(a)\over(-P^2)^{a-1}}+P_\mu{\Gamma(a)\over (-P^2)^{a}} P_\nu|y)
~=~~[x,y](x|g_{\mu\nu}{\Gamma(a)\over(-p^2)^{a-1}}+p_\mu{\Gamma(a)\over (-p^2)^a} p_\nu|y)
\nonumber\\
&&\hspace{-1mm}
+~\!\int_0^1\! du ~\Big(u[x,x_u]\calf_{x\mu}(x_u)[x_u,y](x|{p_\nu\Gamma(a)\over(-p^2)^a}|y)
-\baru[x,x_u]\calf_{x\nu}(x_u)[x_u,y](x|{p_\mu\Gamma(a)\over(-p^2)^a}|y)
\nonumber\\
&&\hspace{-1mm}
-~{i\over 2}(x|{\Gamma(a)\over(-p^2)^a}|y)\!\int_0^1\!du~[x,x_u]\Big(\calf_{\mu\nu}(x_u)-\baru u (D_\mu \calf_{\nu x}(x_u)+\mu\leftrightarrow\nu)\Big)[x_u,y]~+~O(D\calf,\calf^2)
\nonumber\\
&&\hspace{5mm}
=~[x,y](x|(p_\mu p_\nu-p^2g_{\mu\nu}){\Gamma(a)\over (-p^2)^{a}}|y)
-{2\Gamma\big({d\over 2}-a+1\big)\over 4^{a}\pi^{d\over 2}(-\Delta^2)^{{d\over 2}-a+1}}
\!\int_0^1\! du~[x,x_u]\big(u\Delta_\nu \calf_{\Delta\mu}(x_u)-\baru \Delta_\mu \calf_{\Delta\nu}(x_u)\big)[x_u,y]
\nonumber\\
&&\hspace{11mm}   
+~{\Gamma\big({d\over 2}-a\big)\over 4^a\pi^{d\over 2}(-\Delta^2)^{{d\over 2}-a}}\half\!\int_0^1\! du
[x,x_u]\big(\calf_{\mu\nu}(x_u)-\baru u (D_\mu \calf_{\nu \Delta}(x_u)+\mu\leftrightarrow\nu)\big)[x_u,y]~+~O(D\calf,\calf^2)
\label{firsterm}
\end{eqnarray}
where $\Delta\equiv x-y$. Using this formula it is easy to get the first term in Eq. (\ref{gluproplico1}),
\begin{eqnarray}
&&\hspace{-0mm}
ig_{\mu\nu}(x|{\ln -\tmu^2/ P^2\over P^2}|y)-i(x|P_\mu{\ln -\tmu^2/ P^2\over P^4}P_\nu|0)
~=~i(x|g_{\mu\nu}{\ln -\tmu^2/ p^2\over p^2}-p_\mu p_\nu{\ln -\tmu^2/ p^2\over p^4}|y)
\nonumber\\
&&\hspace{-0mm}
+~{i\over 8\pi^{2+\ve}}{\Gamma(1+\ve)\over (-\Delta^2)^{1+\ve}}\big[\ln{-\tmu^2 \Delta^2\over 4}-\psi(1+\ve)-\psi(2)\big]
\!\int_0^1\! du~\big(u\Delta_\nu \calf_{\Delta\mu}(x_u)-\baru \Delta_\mu \calf_{\Delta\nu}(x_u)\big)
\nonumber\\
&&\hspace{-0mm}
-~{i\Gamma\big(\ve)\over 32\pi^{d\over 2}(-\Delta^2)^\ve}\Big[ {1\over\ve}+\ln{-\tmu^2 \Delta^2\over 4}-\psi(1+\ve)-\psi(2)\Big]
\!\int_0^1\! du~
\big(\calf_{\mu\nu}(x_u)-\baru u (D_\mu \calf_{\nu \Delta}(x_u)+\mu\leftrightarrow\nu)\big)
\end{eqnarray}
Hereafter we drop gauge links for brevity.

The last term in the first line of Eq. (\ref{gluproplico1}) follows easily from Eq. (\ref{clico2}):
\begin{eqnarray}
&&\hspace{-1mm}
(x|{1\over p^2}\big\{\calf_{\mu\nu},\ln{\tmu^2\over -p^2}\big\}{1\over p^2}|y)
~=~2(x|{\ln{\tmu^2\over -p^2}\over p^4}|y)\!\int_0^1\! du~\calf_{\mu\nu}(x_u)
+(x|{1\over p^4}|y)\!\int_0^1\! du~(2+\ln \baru u)\calf_{\mu\nu}(x_u)
\nonumber\\
&&\hspace{-0mm}
=~{i\Gamma\big(\ve)\over 8\pi^{d\over 2}(-\Delta^2)^\ve}
\Big(\big[\ln{-\tmu^2\Delta^2\over 4}+{1\over \ve}-\psi(1+\ve)+\gamma \big]\!\int_0^1\! du~\calf_{\mu\nu}(x_u)
+\half\!\int_0^1\! du~\ln\baru u~\calf_{\mu\nu}(x_u)\Big)
\label{clico1}
\end{eqnarray}

To calculate terms in the second and third lines of Eq. (\ref{gluproplico1}) we use 
formulas
\begin{eqnarray}
&&\hspace{-1mm}
(x|{1\over p^2}|z_1)(z_1|{\Gamma(a)\over(-p^2)^a}|z_2)\!\int_0^1\! du ~\calf_{\alpha\beta}(uz_1+\baru z_2)(z_2|{1\over p^2}|y)
=~-i^{a}\!\int_0^\infty\! ds_1ds_2\!\int_0^{s_1}\!\int_0^{s_2}{dt_1 dt_2\over  (t_1+t_2)^{2-a}}
\nonumber\\
&&\hspace{11mm}
\times~(x|e^{is_1p^2}\calf_{\alpha\beta}e^{is_2p^2}p_\beta|y)~=~-{i^a\over a(a-1)}\!\int_0^\infty\! ds_1ds_2\big[(s_1+s_2)^a
-s_1^a-s_2^a\big](x|e^{is_1p^2}\calf_{\alpha\xi}e^{is_2p^2}p_\beta|y)
\nonumber\\
&&\hspace{22mm}
=~{1\over a(a-1)}(x|{\Gamma(a+2)\over (-p^2)^{a+2}}|y)\!\int_0^1\! du\big[1-\baru^a-u^a\big]\calf_{\alpha\beta}(ux+\baru y)
\label{fla85}
\end{eqnarray}
and
\begin{eqnarray}
&&\hspace{-1mm}
(x|{p^\xi\over p^2}|z_1)(z_1|{\Gamma(a)\over(-p^2)^a}|z_2)\!\int_0^1\! du ~u\calf_{\alpha\xi}(uz_1+\baru z_2)(z_2|{p_\beta\over p^2}|y)
\\
&&\hspace{-1mm}
=~-i^{a}\!\int_0^\infty\! ds_1ds_2\!\int_0^{s_1}\!\int_0^{s_2}{t_2dt_1 dt_2\over (t_1+t_2)^{3-a}}
(x|p^\xi e^{is_1p^2}\calf_{\alpha\xi}e^{is_2p^2}p_\beta|y)
\nonumber\\
&&\hspace{-1mm}
=~{i^a\over a(a-1)(a-2)}\!\int_0^\infty\! ds_1ds_2\big[(s_1+s_2)^a-as_2(s_1+s_2)^{a-1}
-s_1^a-s_2^a+as_2^a\big](x|p^\xi e^{is_1p^2}\calf_{\alpha\xi}e^{is_2p^2}p_\beta|y)
\nonumber\\
&&\hspace{-1mm}
=~{1\over a(2-a)}
\!\int_0^1\! du\Big[-u+u^a+\baru{1-\baru^{a-1}\over a-1}\Big]
\Big[\calf_{\alpha\xi}(ux+\baru y)(x|{\Gamma(a+2)p^\xi p_\beta\over (-p^2)^{a+2}}|y)
-i\baru D_\beta \calf_{\alpha\xi}(ux+\baru y)(x|{\Gamma(a+2)p^\xi \over (-p^2)^{a+2}}|y)\Big]
\nonumber
\end{eqnarray}
where we neglected $D^\xi \calf_{\alpha\xi}$ as usual. 
Similarly
\begin{eqnarray}
&&\hspace{-1mm}
(x|{p_\alpha\over p^2}|z_1)(z_1|{\Gamma(a)\over(-p^2)^a}|z_2)\!\int_0^1\! du ~\baru \calf_{\beta\xi}(uz_1+\baru z_2)(z_2|{p^\xi\over p^2}|y)
\label{fla134}\\
&&\hspace{-1mm}
=~{1\over a(2-a)}
\!\int_0^1\! du\big[-\baru+\baru^a+u{1-u^{a-1}\over a-1}\big]\Big[\calf_{\beta\xi}(ux+\baru y)
(x|{p_\alpha p^\xi\Gamma(a+2)\over (-p^2)^{a+2}}|y)+iuD_\alpha \calf_{\beta\xi}(ux+\baru y)(x|{p^\xi\Gamma(a+2)\over (-p^2)^{a+2}}|y)\Big]
\nonumber
\end{eqnarray}
and 
\begin{eqnarray}
&&\hspace{-11mm}
(x|{1\over p^2}|z_1)(z_1|{p^\xi\Gamma(a)\over(-p^2)^a}|z_2)\!\int_0^1\! du ~\baru uD_\alpha \calf_{\beta \xi}(uz_1+\baru z_2)(z_2|{1\over p^2}|y)
\nonumber\\
&&\hspace{-11mm}
=~-{1\over a(2-a)(3-a)}(x|{p^\xi\Gamma(a+2)\over (-p^2)^{a+2}}|y)\!\int_0^1\! du\big[1-\baru^a-u^a-a\baru u\big]D_\alpha \calf_{\beta \xi}(ux+\baru y)
\label{flasdg}
\end{eqnarray}
Using the above formulas, we obtain
\begin{eqnarray}
&&\hspace{0mm}
-\!\int\! dz_1dz_2\!\int_0^1\! du ~\Big[u(x|{1\over p^2}]z_1)\calf_{\mu \xi}(z_u)(z_1|{p^\xi\over p^2}|z_2)(z_2|{p_\nu\over p^2}|y)
-\baru (x|{p_\mu\over p^2}]z_1)\calf_{\nu \xi}(z_u)(z_1|{p^\xi\over p^2}|z_2)(z_2|{1\over p^2}|y)
\nonumber\\
&&\hspace{10mm}
 +2i\baru u(z_1|{p^\xi\over p^2}|z_2)(D_\mu \calf_{\nu \xi}(z_u)+\mu\leftrightarrow\nu)\Big)(z_2|{1\over p^2}|y)\Big]
\Big\}
\nonumber\\
&&\hspace{-11mm}
=~
-{i\over 16\pi^2\Delta^2}\!\int_0^1\! du~ \big[\baru\ln\baru~
\Delta_\nu \calf_{\mu \Delta}(x_u)-u\ln u~ \Delta_\mu \calf_{\nu \Delta}(x_u))\big]
+{i\over 32\pi^{d\over 2}}{\Gamma(\ve)\over (-\Delta^2)^{\ve}}\!\int_0^1\! du~ (u\ln u+\baru\ln\baru)
\calf_{\mu\nu}(x_u)
\nonumber\\
&&\hspace{-1mm}
-~{i\over 32\pi^{d\over 2}}{\Gamma(\ve)\over (-\Delta^2)^{\ve}}
\!\int_0^1\! du~ \big[u^2\ln uD_\mu \calf_{\nu \Delta}(x_u)
+\baru^2\ln\baru  D_\nu \calf_{\mu \Delta}(x_u)+\baru u\big(D_\mu \calf_{\nu \Delta}(x_u)+\mu\leftrightarrow\nu\big)\big]
\end{eqnarray}
and 
\begin{eqnarray}
&&\hspace{-0mm}
\!\int\! dz_1dz_2\!\int_0^1\! du~(x|{1\over p^2}|z_1)(z_1|2\ln{\tmu^2\over -p^2}-{5\over 2}|z_2)\calf_{\mu\nu}(z_u)
 (z_2|{1\over p^2}|y)
\nonumber\\
&&\hspace{11mm}
=~{i\Gamma\big(\ve)\over 8\pi^{d\over 2}(-\Delta^2)^\ve}
\Big(\big[\ln{-\tmu^2\Delta^2\over 4}+{1\over \ve}-\psi(1+\ve)+\gamma -{5\over 4}\big]\!\int_0^1\! du~\calf_{\mu\nu}(x_u)
+\!\int_0^1\! du~\ln\baru u~\calf_{\mu\nu}(x_u)\Big)
\label{gluproplico8}
\end{eqnarray}

Let us present the final formula for the one-loop correction to the gluon propagator in the background field  ($\ve={d\over 2}-2$):
\begin{eqnarray}
&&\hspace{-0mm}
\langle A^a_\mu(x) A^b_\nu(y)\rangle_{\rm quark~loop}^{ab}~=~{g^2\over 24\pi^2}\bigg\{
i(x|g_{\mu\nu}{\ln -\tmu^2/ p^2\over p^2}-p_\mu p_\nu{\ln -\tmu^2/ p^2\over p^4}|y)
\nonumber\\
&&\hspace{-0mm}
+~{i\over 8\pi^2}{1\over \Delta^2}\big[\ln{-\tmu^2 \Delta^2\over 4}-1+2\gamma \big]
\!\int_0^1\! du~\big(u\Delta_\nu \calf_{\mu\Delta}(x_u)-\baru \Delta_\mu \calf_{\nu\Delta}(x_u)\big)
\nonumber\\
&&\hspace{-0mm}
-~
{i\over 16\pi^2\Delta^2}\!\int_0^1\! du~ \big[\baru\ln\baru~
\Delta_\nu \calf_{\mu \Delta}(x_u)-u\ln u~ \Delta_\mu \calf_{\nu \Delta}(x_u))\big]
\nonumber\\
&&\hspace{-0mm}
+~{i\Gamma\big(\ve)\over 32\pi^{d\over 2}(-\Delta^2)^\ve}
\!\int_0^1\! du~\calf_{\mu\nu}(x_u)\Big(-{1\over\ve}-\ln{-\tmu^2 \Delta^2\over 4}+\psi(1+\ve)-\gamma 
+6-4\ln\baru u+u\ln u+\baru\ln\baru\Big)
\nonumber\\
&&\hspace{-0mm}
+~{i\Gamma\big(\ve)\over 32\pi^{d\over 2}(-\Delta^2)^\ve}\Big[ {1\over\ve}+\ln{-\tmu^2 \Delta^2\over 4}-\psi(1+\ve)-2+\gamma \Big]
\!\int_0^1\! du~
\baru u (D_\mu \calf_{\nu \Delta}(x_u)+\mu\leftrightarrow\nu)
\nonumber\\
&&\hspace{-1mm}
-~{i\over 32\pi^{d\over 2}}{\Gamma(\ve)\over (-\Delta^2)^{\ve}}
\!\int_0^1\! du~ \big[u^2\ln uD_\mu \calf_{\nu \Delta}(x_u)
+\baru^2\ln\baru  D_\nu \calf_{\mu \Delta}(x_u)\big]\bigg\}^{ab}~+~O\big(D^\mu \calf_{\mu \nu}, \calf_{\mu \nu}\calf^{\alpha\beta}\big)
\label{gproponlico}
\end{eqnarray}
As usual, the lightlike gauge links are implied.

\subsection{Formulas for the light-cone expansions in Sect. \ref{sec:qloop} \label{formulas-lighcone}}

In principle, we can use Eq. (\ref{gproponlico}) to find e.g. Eq. (\ref{gtmdael1}),
but calculations are greatly simplified by using some intermediate results such as Eq. (\ref{gluproplicog}) 
since many of the terms in Eq. (\ref{gproponlico}) cancel after differentiation. 
To use Eq. (\ref{gluproplicog}),  we need some additional formulas listed here.

First, similar to Eq. (\ref{fla134}) one obtains
\begin{eqnarray}
&&\hspace{0mm}
\!\int\! dz_1dz_2\!\int_0^1\! du ~
\baru (x|{1\over p^2}]z_1)\calf_{\nu \xi}(z_u)(z_1|{p^\xi\over p^2}|z_2)(z_2|{p_\mu\over p^2}|y)
~-~(\mu\leftrightarrow\nu)
=~{i\Delta_\mu\over 16\pi^{d\over 2}\Delta^2}\!\int_0^1\! du ~u\ln u~\calf_{\nu\Delta}(x_u)
\nonumber\\
&&\hspace{-0mm}
+~{i\over 32\pi^{d\over 2}}{\Gamma(\ve)\over (-\Delta^2)^{\ve}}\!\int_0^1\! du ~ u\ln u~\calf_{\mu\nu}(x_u)
+{i\over 32\pi^{d\over 2}}{\Gamma(\ve)\over (-\Delta^2)^{\ve}}\!\int_0^1\! du ~ \baru u\ln u~D_\mu\calf_{\nu\Delta}(x_u)
~-~(\mu\leftrightarrow\nu)
\nonumber\\
&&\hspace{-0mm}
=~{i\Delta_\mu\over 16\pi^{d\over 2}\Delta^2}\!\int_0^1\! du ~u\ln u~\calf_{\nu\Delta}(x_u)~-~(\mu\leftrightarrow\nu)
+~{i\over 32\pi^{d\over 2}}{\Gamma(\ve)\over (-\Delta^2)^{\ve}}\!\int_0^1\! du ~ (\ln u+\baru u)~\calf_{\mu\nu}(x_u)
\label{vanishterm}
\end{eqnarray}
Second, we need formulas
\begin{eqnarray}
&&\hspace{-1mm}
\delta^-{d\over d\delta^-}\!\int_{-\infty}^0\! dz^+ ~{\Gamma(a)\over (x_\perp^2-2z^+\delta^-\big)^a}\!\int_0^1\! du~\baru ~\calo(uz^+)
~=~-\int_0^1\! du \!\int_{-\infty}^0\! dz^+\calo(z^+){\Gamma(a)\over (x_\perp^2-{2\over u}z^+\delta^-)^a}
\nonumber\\
&&\hspace{-1mm}
\delta^-{d\over d\delta^-}\!\int_{-\infty}^0\! dz^+ ~{\Gamma(a)\over (x_\perp^2-2z^+\delta^-\big)^a}\!\int_0^1\! du~\baru u~\calo(uz^+)
~=~\int_0^1\! dt~(1-2t)\!\int_{-\infty}^0\! dz^+\calo(z^+){\Gamma(a)\over (x_\perp^2-{2\over t}z^+\delta^-)^a}
\nonumber\\
&&\hspace{-0mm}
=~\!\int_{-\infty}^0\! dz^+\calo(z^+)\!\int_0^1\! du~{\baru\over u}{2z^+\delta^-\Gamma(a+1)\over (x_\perp^2-{2\over u}z^+\delta^-)^{a+1}}
\label{nado3}
\end{eqnarray}
which follow from Eq. (\ref{glavormula}), and
\begin{eqnarray}
&&\hspace{-1mm}
\!\int\! dz_1dz_2\!\int_0^1\! du ~(x|{1\over p^2}|z_1)(z_1|{p^j\over p^2}|z_2)\!\int_0^1\! du ~\baru uD^-\calf^{-,j}(uz_1+\baru z_2)
(z_2|{p_i\over p^2}|y)
\nonumber\\
&&\hspace{-1mm}
=~(x|{p_i p_j\over (p^2)^3}|y)\!\int_0^1\! du~\baru uD^- \calf^{-,j}(ux+\baru y)
=~i\!\int_0^1\! du~\baru u~D^- \calf^{-,j}(ux+\baru y)\Big({g_{ij}\Gamma(\ve)\over 64\pi^{d\over 2}(-\Delta^2)^\ve}
-{\Delta_i \Delta_j\over 32\pi^2x^2}\Big)
\label{flasdg1}
\end{eqnarray}
which is obtained by differentiation of Eq. (\ref{flasdg}). Also, from the general formula 
\begin{eqnarray}
&&\hspace{-1mm}
\!\int\! dz_1dz_2\!\int_0^1\! du ~(x|{1\over p^2}|z_1)(z_1|{p_\mu\Gamma(a)\over(-p^2)^a}|z_2)\!\int_0^1\! du ~\baru u~\calo(uz_1+\baru z_2)(z_2|{1\over p^2}|y)
\nonumber\\
&&\hspace{-1mm}
=~-{1\over a(2-a)(3-a)}(x|{p_\mu\Gamma(a+2)\over (-p^2)^{a+2}}|y)\!\int_0^1\! du\big[1-\baru^a-u^a-a\baru u\big]\calo(ux+\baru y)
\nonumber\\
&&\hspace{-1mm}
-~i(x{\Gamma(a+2)\over(-p^2)^{a+2}}|y)\!\int_0^1\! du ~
{(a-4)(\baru^{a+1}-u^{a+1})-(a-2)(\baru^a-u^a)+2(\baru-u)\over a(a-2)(a-3)(a-4)}\partial_\mu\calo(ux+\baru y)
\label{flagene}
\end{eqnarray}
we get
\begin{eqnarray}
&&\hspace{-0mm}
\!\int\! dz_1dz_2\!\int_0^1\! du ~(x|{1\over p^2}|z_1)(z_1|{p_\mu\over p^2}|z_2)\!\int_0^1\! du ~\baru u~\calo(uz_1+\baru z_2)(z_2|{1\over p^2}|y)
~=~(x|{p_\mu\over p^6}|y)\!\int_0^1\! du~\baru u~\calo(ux+\baru y)
\label{flagen}
\end{eqnarray}
in agreement with Eq. (\ref{flasdg}). Finally, we used
\begin{eqnarray}
&&\hspace{-0mm}
\!\int_0^1\! du~(x|{1\over p^2}|z_1)(z_1|f(p^2)|z_2)\calo(z_u)(z_2|{p^-\over p^2}|y)
\nonumber\\
&&\hspace{0mm}
=~
 \!\int_0^1\! du~(x|{p^-\over p^2}|z_1)(z_1|f(p^2)|z_2)\calo(z_u)(z_2|{1\over p^2}|y)
 - \!\int_0^1\! du~(x|{1\over p^2}|z_1)(z_1|f(p^2)|z_2)i\partial^-\calo(z_u)(z_2|{1\over p^2}|y)
 \label{speminusom}
\end{eqnarray}

\subsection{Rapidity-only cutoff {\it vs} UV+rapidity regularization \label{sec:compare}}

\begin{figure}[htb]
\begin{center}
\includegraphics[width=77mm]{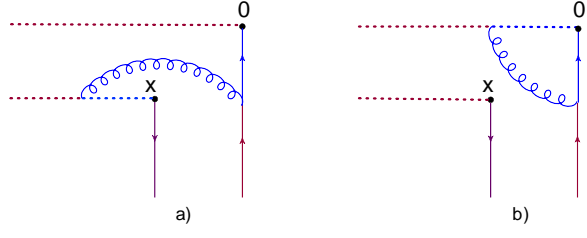}
\end{center}
\caption{Typical diagrams for one-loop evolution of the quark TMD operator.\label{fig:compare}}
\end{figure}

In this appendix we discuss the comparison between the small-$x$ inspired rapidity-only cutoff used in this paper and the 
 combination of  UV and rapidity cutoffs characteristic for the CSS approach.  Consider the typical contribution
to the quark TMD operator shown in Fig. \ref{fig:compare} at $x^+=0$ and $p_{B_\perp}=0$.  As discussed above, at such
a separation we can use Feynman diagrams instead of cut diagrams,

%
\begin{eqnarray}
&&\hspace{-0mm}
\langle{\rm T}[0^+,-\infty]_x[-\infty,0^+]_0\Gamma\psi(0)\rangle_\Psi^{\rm Fig. \ref{fig:compare}}
~=~g^2c_F\!\int\! \dhd\beta_B\dhd p_{B_\perp}\Gamma\Psi(\beta_B)I(\beta_B,x_\perp),
\nonumber\\
&&\hspace{-1mm}
I(\beta_B,x_\perp)~=~-i\!\int\!\dhd\alpha\dhd\beta\dhd p_\perp{1\over \beta+\ie}{e^{-i\alpha\vro\delta^-}\over \alpha\beta s-p_\perp^2+\ie}
{s(\beta-\beta_B)\over \alpha(\beta-\beta_B)s-p_\perp^2+\ie}\big(1-e^{i(p,x)_\perp}\big)\Gamma\Psi(\beta_B)
\end{eqnarray}
Without the cutoff in $\alpha$,  the integral
\begin{eqnarray}
&&\hspace{-11mm}   
I(\beta_B,x_\perp)~=~-i\mu^{-2\ve}\!\int\!\dhd\alpha\dhd\beta\dhd p_\perp{1\over \beta+\ie}
{1\over \alpha\beta s-p_\perp^2+\ie}
{s(\beta_B-\beta)\over \alpha(\beta_B-\beta)s+p_\perp^2-\ie}\big(1-e^{i(p,x)_\perp}\big)
\nonumber\\
&&\hspace{-11mm}
=~-\mu^{-2\ve}\!\int\!{\dhd p_\perp\over p_\perp^2}\big(1-e^{i(p,x)_\perp}\big)\!\int_0^{\beta_B}{\!\dhd\beta\over\beta_B}{\beta_B-\beta\over\beta+\ie}
~=~{1\over 8\pi^2}{\Gamma(\ve)\over (x_\perp^2\mu^2)^\ve}\!\int_0^{\beta_B}\!{d\beta\over\beta_B}
{\beta_B-\beta\over\beta}
\end{eqnarray}
diverges as $\beta\rightarrow 0$ even at $\ve\neq 0$. The so-called $\delta$-regularization 
with $A^-(z^+)\rightarrow A^-(z^+)e^{\pm\delta z^+}$ gives
\begin{equation}
[0^+,-\infty]_x[-\infty,0^+]_0~\rightarrow~
i\!\int_{-\infty}^0\!dz^+[-A^-(z^+,x_\perp)+A^-(z^+,0_\perp)]e^{\delta z^+}
\end{equation}
so that
\begin{equation}
I^\delta(\beta_B,x_\perp)~=~{1\over 8\pi^2}{\Gamma(\ve)\over (x_\perp^2\mu^2)^\ve}\!\int_0^{\beta_B}\! {d\beta\over\beta_B}{\beta_B-\beta\over \beta+i\delta}
~\simeq~-{1\over 8\pi^2}\Big(-{1\over\ve}+\ln{\mu^2x_\perp^2\over 4}+\gamma \Big)\big(\ln{\beta_B\over i\delta}-1\big)
\end{equation}
which gives
\begin{equation}
I^\delta(\beta_B,x_\perp)~=~{1\over 8\pi^2}{\Gamma(\ve)\over (x_\perp^2\mu^2)^\ve}\!\int_0^{\beta_B}\! {d\beta\over\beta_B}{\beta_B-\beta\over \beta-i\delta}
~\simeq~-{1\over 8\pi^2}\Big(\ln{\mu^2x_\perp^2\over 4}+\gamma \Big)\big(\ln{\beta_B\over i\delta}-1\big)
\label{idelta}
\end{equation}
after subtraction of the counterterm.

On the other hand, the rapidity-only cutoff $\delta^-={1\over\vro\sigma}$ gives [see Eq. (\ref{eiknapravotvet})]
\begin{eqnarray}
&&\hspace{-1mm}   
I^\sigma(\beta_B,x_\perp)~=~-i\!\int\!\dhd\alpha\dhd\beta\dhd p_\perp{1\over \beta+\ie}
{e^{-i{\alpha\over\sigma}}\over \alpha\beta s-p_\perp^2+\ie}
{s(\beta-\beta_B)\over \alpha(\beta-\beta_B)s-p_\perp^2+\ie}\big(1-e^{i(p,x)_\perp}\big)
\nonumber\\
&&\hspace{-1mm}
=~\!\int\!{\dhd p_\perp\over p_\perp^2}\big(1-e^{i(p,x)_\perp}\big)\!\int_0^{\infty}\!\dhd\alpha
{\beta_Bs\over\alpha\beta_Bs+p_\perp^2}e^{-i{\alpha\over\sigma}}
~=~-{1\over 16\pi^2}\ln^2\big(-i\beta_B\sigma s{x_\perp^2\over 4}e^{\gamma }\big)
\label{isigma}
\end{eqnarray}
The integrals (\ref{idelta}) and (\ref{isigma}) coincide when $\mu^2$ is two times BLM scale
$\mu^2=2\mu_\sigma^2=2x_\perp^{-1}\sqrt{\beta_B\sigma s}$ and $\delta={4\over \sigma s x_\perp^2}$. 
Hopefully, the double evolution \cite{Echevarria:2015usa} along the line $\mu^2x^2\sqrt\delta=4\sqrt \beta_B$ will produce results compatible 
with Eq. (\ref{ouresult2}).

\subsection{Rapidity-only evolution beyond Sudakov region at small and moderate $x$ \label{sec:beyond}}
As we demonstrated in this paper, the Sudakov double logs are universal and the evolution of quark and gluon TMDs 
is the same for low and moderate $x$ until $\sigma \beta_B s\sim b_\perp^{-2}\sim q_\perp^2$. From that point, the evolution (or the lack of it) depends on $\beta_B=x_B$ and $q_\perp^2$. There are three different scenarios. We will consider them for the case of gluon TMDs since we can use the explicit formulas for the leading-order rapidity evolution at arbitrary $\beta_B=x_B$ from Ref. \cite{Balitsky:2015qba}.
 
First, if $x_B\sim 1$ and $q_\perp^2\gtrsim m_N^2$, there is no room for any evolution and one should turn to
phenomenological models of TMDs such as the replacement of $b$ by $b_\ast$ in Refs. \cite{Collins:1984kg,Collins:2017oxh}.

If  $x_B\sim 1$ and $q_\perp^2\gg m_N^2$, there is room for DGLAP-type evolution summing logs 
$\big(\alpha_s\ln{q_\perp^2/m_N^2}\big)^n$. The rapidity evolution in this case has the form \cite{Balitsky:2015qba}
\footnote{We have omitted the term $\sim(2k_ik^j-\delta_i^j)\scrf^{i,a;\sigma}\scrf^{a;\sigma}_{~~i}$ 
from Eq. (3.25) from Ref. \cite{Balitsky:2015qba}.  This term is not essential for our discussion here.}
\begin{eqnarray}
&&\hspace{-0mm}
\sigma{d\over d\sigma}
\langle p_N|\scrf^{i,a;\sigma}(\beta_B,x_\perp)\scrf^{a;\sigma}_{~~i}(\beta_B,0_\perp)|p_N\rangle
\label{evoleqlico}\\
&&\hspace{2mm}
=~4\alpha_sN_c\!\int\! dk_\perp~
\bigg\{e^{i(k,x)_\perp}\langle p_N|\scrf^{i,a;\sigma}\big(\beta_B+{k_\perp^2\over\sigma s},x_\perp\big)
\scrf^{a;\sigma}_{~~i}\big(\beta_B+{k_\perp^2\over\sigma s},0_\perp\big)|p_N\rangle
\theta\big(1-\beta_B-{k_\perp^2\over\sigma s}\big)
\nonumber\\
&&\hspace{-0mm}
\times~\Big[{1\over k_\perp^2}-{2\over \sigma\beta_Bs+k_\perp^2}+{(\sigma\beta_Bs)^2\over (\sigma\beta_Bs+k_\perp^2)^4}\Big]-{\sigma\beta_Bs\over k_\perp^2(\sigma\beta_Bs+k_\perp^2)}
\langle p_N|\scrf^{i,a;\sigma}(\beta_B,x_\perp)
\scrf^{a;\sigma}_{~~i}(\beta_B,y_\perp)|p_N\rangle\bigg\}
\nonumber
\end{eqnarray}
Note that if $\sigma\beta_Bs\gg x_\perp^{-2}$ we get leading-order equation (\ref{gloevoleq}) at $\beta=\beta'$.
On the other hand, as demonstrated in Ref. \cite{Balitsky:2015qba}, if $\sigma\ll {q_\perp^2\over\beta_Bs}$ the factor $e^{i(k,x)_\perp}$ in the RHS of Eq. (\ref{evoleqlico}) can be neglected and we have the leading-order DGLAP equation with identification 
$\mu^2_{\rm DGLAP}=\sigma\beta_Bs$.  The result of this DGLAP evolution should be convoluted with Eq. (\ref{gevolution}) using full Eq. (\ref{evoleqlico}) for proper matching.

Similarly, if $x_B=\beta_B\ll 1$, even at $\beta_B\sigma s=q_\perp^2$ there is room for BFKL-type evolution 
from $\sigma ={q_\perp^2\over \beta_Bs}$ to $\sigma ={q_\perp^2\over s}$ which corresponds to summing logs 
$(\alpha_s\ln x_B)^n$. The leading-order rapidity equation at arbitrary $\beta_B$ has the form 
\cite{Balitsky:2015qba}
\begin{eqnarray}
&&\hspace{-0mm}
\sigma{d\over d\sigma}
\langle p_N|\scrf^{i,a;\sigma}(\beta_B,x_\perp)\scrf^{a;\sigma}_{~~i}(\beta_B,0_\perp)|p_N\rangle
\label{evoleqgen}\\
&&\hspace{2mm}
=~4\alpha_s\!\int\! dk_\perp~\bigg[\theta\big(1-\beta_B-{k_\perp^2\over\sigma s}\big)
\langle p_N|N_c\scrf^{i,a;\sigma}\big(\beta_B+{k_\perp^2\over\sigma s},x_\perp\big)
\scrf^{a;\sigma}_{~~i}\big(\beta_B+{k_\perp^2\over\sigma s},0_\perp\big)|p_N\rangle
{e^{i(k,x)_\perp}\over k_\perp^2}
\nonumber\\
&&\hspace{-1mm}
-4N_c{\sigma\beta_Bs\over k_\perp^2(\sigma\beta_Bs+k_\perp^2)}
\langle p_N|\scrf^{i,a;\sigma}(\beta_B,x_\perp)
\scrf^{a;\sigma}_{~~i}(\beta_B,y_\perp)|p_N\rangle
-\theta\big(1-\beta_B-{k_\perp^2\over\sigma s}\big)
\nonumber\\
&&\hspace{-0mm}
\times~\langle p_N|{\rm Tr}(x_\perp|U{p_j\over\sigma\beta_Bs+p_\perp^2} U^\dagger
\scrf^{i,a;\sigma}\big(\beta_B+{k_\perp^2\over\sigma s}\big)|k_\perp)
(k_\perp|\scrf^{a;\sigma}_{~~i}\big(\beta_B+{k_\perp^2\over\sigma s}\big)
U{p^j\over\sigma\beta_Bs+p_\perp^2}U^\dagger|y_\perp)|p_N\rangle+~ \dots
\nonumber
\end{eqnarray}
where $U(x_\perp)\equiv [x_\perp-\infty^+,x_\perp+\infty^+]$ is a Wilson line (infinite gauge link)
and dots stand for a number of nonlinear terms similar to the last one [see Eq. (5.5) from Ref. \cite{Balitsky:2015qba}]. 
The small-$x$ evolution is relevant from $\sigma={q_\perp^2\over\beta_Bs}$ to  $\sigma={q_\perp^2\over s}$. As demonstrated in Ref. \cite{Balitsky:2015qba}, at $\sigma\ll {q_\perp^2\over\beta_Bs}$ the evolution equation (\ref{evoleqgen}) reduces to the
BK equation which can be studied using standard small-$x$ methods. After that, the matching 
of  the double-log Sudakov evolution (\ref{evolog}) to a single-log BK evolution should be done using full nonlinear equation (\ref{evoleqgen}).


\begin{thebibliography}{10}
 
\bibitem{Brodsky:1982gc}
S.~J.~Brodsky, G.~P.~Lepage and P.~B.~Mackenzie,
Phys. Rev. D \textbf{28}, 228 (1983)
doi:10.1103/PhysRevD.28.228

\bibitem{Collins:1981uw}
J.~C.~Collins and D.~E.~Soper,
Nucl. Phys. B \textbf{194}, 445-492 (1982)
doi:10.1016/0550-3213(82)90021-9

\bibitem{Collins:1984kg}
J.~C.~Collins, D.~E.~Soper and G.~F.~Sterman,
Nucl. Phys. B \textbf{250}, 199-224 (1985)
doi:10.1016/0550-3213(85)90479-1

\bibitem{Ji:2004wu}
X.~d.~Ji, J.~p.~Ma and F.~Yuan,
Phys. Rev. D \textbf{71}, 034005 (2005)
doi:10.1103/PhysRevD.71.034005
[arXiv:hep-ph/0404183 [hep-ph]].

\bibitem{GarciaEchevarria:2011rb}
M.~G.~Echevarria, A.~Idilbi and I.~Scimemi,
JHEP \textbf{07}, 002 (2012)
doi:10.1007/JHEP07(2012)002
[arXiv:1111.4996 [hep-ph]].

\bibitem{Collins:2011zzd}
J.~Collins,
Camb. Monogr. Part. Phys. Nucl. Phys. Cosmol. \textbf{32}, 1-624 (2011)

\bibitem{Collins:2003fm}
J.~C.~Collins,
Acta Phys. Polon. B \textbf{34}, 3103 (2003)
[arXiv:hep-ph/0304122 [hep-ph]].

\bibitem{Balitsky:2015qba}
I.~Balitsky and A.~Tarasov,
JHEP \textbf{10}, 017 (2015)
doi:10.1007/JHEP10(2015)017
[arXiv:1505.02151 [hep-ph]].

\bibitem{Balitsky:2016dgz}
I.~Balitsky and A.~Tarasov,
JHEP \textbf{06}, 164 (2016)
doi:10.1007/JHEP06(2016)164
[arXiv:1603.06548 [hep-ph]].

\bibitem{Balitsky:2019ayf}
I.~Balitsky and G.~A.~Chirilli,
Phys. Rev. D \textbf{100}, no.5, 051504 (2019)
doi:10.1103/PhysRevD.100.051504
[arXiv:1905.09144 [hep-ph]].

\bibitem{Beneke:1994qe}
M.~Beneke and V.~M.~Braun,
Phys. Lett. B \textbf{348}, 513-520 (1995)
doi:10.1016/0370-2693(95)00184-M
[arXiv:hep-ph/9411229 [hep-ph]].

\bibitem{Brodsky:1998kn}
S.~J.~Brodsky, V.~S.~Fadin, V.~T.~Kim, L.~N.~Lipatov and G.~B.~Pivovarov,
JETP Lett. \textbf{70}, 155-160 (1999)
doi:10.1134/1.568145
[arXiv:hep-ph/9901229 [hep-ph]].

\bibitem{Balitsky:2006wa}
I.~Balitsky,
Phys. Rev. D \textbf{75}, 014001 (2007)
doi:10.1103/PhysRevD.75.014001
[arXiv:hep-ph/0609105 [hep-ph]].

\bibitem{Kovchegov:2006vj}
Y.~V.~Kovchegov and H.~Weigert,
Nucl. Phys. A \textbf{784}, 188-226 (2007)
doi:10.1016/j.nuclphysa.2006.10.075
[arXiv:hep-ph/0609090 [hep-ph]].

\bibitem{Collins:2014jpa}
J.~Collins and T.~Rogers,
Phys. Rev. D \textbf{91}, no.7, 074020 (2015)
doi:10.1103/PhysRevD.91.074020
[arXiv:1412.3820 [hep-ph]].

\bibitem{Balitsky:2017flc}
I.~Balitsky and A.~Tarasov,
JHEP \textbf{07}, 095 (2017)
doi:10.1007/JHEP07(2017)095
[arXiv:1706.01415 [hep-ph]].

\bibitem{Balitsky:2017gis}
I.~Balitsky and A.~Tarasov,
JHEP \textbf{05}, 150 (2018)
doi:10.1007/JHEP05(2018)150
[arXiv:1712.09389 [hep-ph]].

\bibitem{Balitsky:2020jzt}
I.~Balitsky,
JHEP \textbf{05}, 046 (2021)
doi:10.1007/JHEP05(2021)046
[arXiv:2012.01588 [hep-ph]].


\bibitem{Balitsky:2007feb}
I.~Balitsky and G.~A.~Chirilli,
Phys. Rev. D \textbf{77}, 014019 (2008)
doi:10.1103/PhysRevD.77.014019
[arXiv:0710.4330 [hep-ph]].

\bibitem{Balitsky:2009xg}
I.~Balitsky and G.~A.~Chirilli,
Nucl. Phys. B \textbf{822}, 45-87 (2009)
doi:10.1016/j.nuclphysb.2009.07.003
[arXiv:0903.5326 [hep-ph]].




\bibitem{Balitsky:1987bk}
I.~I.~Balitsky and V.~M.~Braun,
Nucl. Phys. B \textbf{311}, 541-584 (1989)
doi:10.1016/0550-3213(89)90168-5

\bibitem{Mulders:2000sh}
P.~J.~Mulders and J.~Rodrigues,
Phys. Rev. D \textbf{63}, 094021 (2001)
doi:10.1103/PhysRevD.63.094021
[arXiv:hep-ph/0009343 [hep-ph]].

\bibitem{Mueller:2012uf}
A.~H.~Mueller, B.~W.~Xiao and F.~Yuan,
Phys. Rev. Lett. \textbf{110}, no.8, 082301 (2013)
doi:10.1103/PhysRevLett.110.082301
[arXiv:1210.5792 [hep-ph]].

\bibitem{Mueller:2013wwa}
A.~H.~Mueller, B.~W.~Xiao and F.~Yuan,
Phys. Rev. D \textbf{88}, no.11, 114010 (2013)
doi:10.1103/PhysRevD.88.114010
[arXiv:1308.2993 [hep-ph]].

\bibitem{Gutierrez-Reyes:2017glx}
D.~Guti\'errez-Reyes, I.~Scimemi and A.~A.~Vladimirov,
Phys. Lett. B \textbf{769}, 84-89 (2017)
doi:10.1016/j.physletb.2017.03.031
[arXiv:1702.06558 [hep-ph]].

\bibitem{Li:2016axz}
Y.~Li, D.~Neill and H.~X.~Zhu,
Nucl. Phys. B \textbf{960}, 115193 (2020)
doi:10.1016/j.nuclphysb.2020.115193
[arXiv:1604.00392 [hep-ph]].

\bibitem{Gehrmann:2014yya}
T.~Gehrmann, T.~Luebbert and L.~L.~Yang,
JHEP \textbf{06}, 155 (2014)
doi:10.1007/JHEP06(2014)155
[arXiv:1403.6451 [hep-ph]].

\bibitem{Gutierrez-Reyes:2018iod}
D.~Gutierrez-Reyes, I.~Scimemi and A.~Vladimirov,
JHEP \textbf{07}, 172 (2018)
doi:10.1007/JHEP07(2018)172
[arXiv:1805.07243 [hep-ph]].

\bibitem{Echevarria:2015usa}
M.~G.~Echevarria, I.~Scimemi and A.~Vladimirov,
Phys. Rev. D \textbf{93}, no.1, 011502 (2016)
[erratum: Phys. Rev. D \textbf{94}, no.9, 099904 (2016)]
doi:10.1103/PhysRevD.93.011502
[arXiv:1509.06392 [hep-ph]].

\bibitem{Collins:2017oxh}
J.~Collins and T.~C.~Rogers,
Phys. Rev. D \textbf{96}, no.5, 054011 (2017)
doi:10.1103/PhysRevD.96.054011
[arXiv:1705.07167 [hep-ph]].
 
   \end{thebibliography}
\end{document}